\documentclass{aa}
\usepackage{natbib}
\bibpunct{(}{)}{;}{a}{}{,}
\usepackage{graphicx}
\graphicspath{{images_de_larticle/}}

\begin{document}


\title{A study of simulated histories of reionization with merger trees of HII regions}

   \author{Jonathan Chardin\thanks{\email{jonathan.chardin@astro.unistra.fr}}
          \and 
           Dominique Aubert
	  \and
	   Pierre Ocvirk 
	  \fnmsep
          }

   \institute{Observatoire Astronomique de Strasbourg, Universit\'e de Strasbourg, CNRS UMR 7550, 11 rue de l'Universit\'e, F-67000 Strasbourg, France\\
             }

   \date{Received ?? ??, 2011; Accepted ?? ??, 2011}

\titlerunning{Merger trees of HII regions}

\date{Accepted / Received}

\abstract  {} {We  describe  a  new methodology  to  analyze the  reionization
  process  in  numerical  simulations:  the  chronology and  the  geometry  of
  reionization is investigated by means  of merger histories of individual HII
  regions.}   {From the  merger tree  of ionized  patches, one  can  track the
  individual  evolution  of  the regions properties such  as  e.g.  their size,  or  the
  intensity of the  percolation process by looking at  the formation rate, the
  frequency of  mergers and the number  of individual HII  regions involved in
  the mergers.  We apply the  merger tree technique  to simulations of
  reionization with three
    different kinds of ionizing source  models and two resolutions. Two of them use star particles
    as ionizing  sources. In this  case we confront two  emissivity evolutions
    for the sources in order to reach  the reionization at $z \sim
    6$.  As  an alternative  we built a  semi-analytical model where  the dark
    matter halos  extracted from  the density fields  are assumed  as ionizing
    sources.}  {We then show how this methodology is a good candidate to
    quantify the  impact of the adopted star formation  on the  history of the
    observed  reionization.  The  semi-analytical  model shows  a  homogeneous
    reionization history with 'local' hierarchical growth steps for individual
    HII regions.
    On  the other  hand  auto-consistent  models for  star  formation tend  to
    present fewer  regions with  a dominant region  in size which  governs the
    fusion process  early in  the reionization at  the expense of  the 'local'
    reionizations.
    The  differences are attenuated when the  resolution of the
    simulation is increased.}  {}


\maketitle
 

\section{Introduction}
\label{intro}

The first generation  of stars appears after the period  of dark ages (between
$z \sim 1090$  and $z \sim 20$), creating a multitude  of distinct HII regions
in  the  Universe.  These  regions  eventually  merge  with other  ionized
regions, leading to a total reionization of the Universe. The diffusion of CMB
photons    on    the    electrons    released   during    reionization    (see
e.g. \citealt{2009ApJS..180..330K}) and the absorption features in the spectra
of   high   redshift    quasars   (see   e.g.   \citealt{2006AJ....132..117F},
\citealt{2007AJ....134.2435W}) suggest  that the end  of reionization occurred
at  $z\sim  11-6$ ,  with  an average  neutral  fraction  of hydrogen  between
$10^{-4}$ and $10^{-6}$ (see \citealt{2006AJ....132..117F}).

Many    models   (see
e.     g.     \citealt{2004ApJ...609..474B},    \citealt{2004ApJ...613....1F},
\citealt{2007ApJ...654...12Z},                   \citealt{2007ApJ...669..663M},
\citealt{2009MNRAS.394..960C})         and        simulation        techniques
(\citealt{1999ApJ...523...66A},                  \citealt{2001NewA....6..437G},
\citealt{2002ApJ...572..695R},                   \citealt{2006MNRAS.371.1057I},
\citealt{2007ApJ...671....1T},                   \citealt{2008MNRAS.387..295A},
\citealt{2009MNRAS.393.1090F}  and  \citealt{2009MNRAS.396.1383P})  have  been
proposed  to  investigate  the  impact   of  the  radiative  transfer  on  the
reionization epoch.  See \cite{2009arXiv0906.4348T}  for a complete  review of
these   models.    In    parallel,   semi-analytic   models   were   developed
(\cite{2004ApJ...609..474B},                        \cite{2004ApJ...613....1F},
\cite{2007ApJ...654...12Z},                         \cite{2007ApJ...669..663M},
\cite{2009MNRAS.394..960C}),  mostly based  on the  'excursion  set formalism'
method  (\cite{1991ApJ...379..440B}) where  both the  source  distribution and
ionization fields are derived from the density fields.

In  this context,  one challenge  is  to describe  the geometry  and the  time
sequence of the reionization as they  will be influenced by e.g. the formation
rate of  ionizing sources,  the distribution of  their formation sites  or the
size and growth  of HII regions. Aiming at  this description, many theoretical
studies have  thus been conducted focusing on the global evolution of physical
fields    such    as    the    ionized   fraction    or    temperature    (see
e.g.       \citealt{2010ApJ...724..244A},      \citealt{2009MNRAS.400.1049F}).
Alternately, many groups have  undertaken analyzes to characterize this period
but   by   focusing   on    the   properties   of   individual   HII   regions
(\citealt{2006MNRAS.369.1625I},                  \citealt{2007ApJ...654...12Z},
\citealt{2007MNRAS.377.1043M},                   \citealt{2008ApJ...675....8L},
\citealt{2008MNRAS.388.1501C},                   \citealt{2008ApJ...681..756S},
\citealt{2011MNRAS.413.1353F}).

With  the current article,  we also  aim at  investigating the  chronology and
geometry of the reionization  through numerical simulations. In particular, we
describe the  process through the  study of \textit{the merger  process} of
HII regions  and for  this purpose  we track the  individual histories  of the
ionized regions formed in  numerical simulations. Using this point-of-view, we
combine the  analyzis in  terms of  individual HII  regions  with the
temporal evolution of global fields. As a consequence, it results in a `local'
perspective with  \textit{a set of histories of reionization},  allowing to study  their scatter
throughout  a cosmological  volume. This  will be  an alternative  approach to
those already undertaken.

In this paper, we first focus on the procedure to separate the different HII regions  
and to track each of them along time. For this purpose, we follow many other investigations
(\citealt{2006MNRAS.369.1625I}, \citealt{2008ApJ...681..756S} and \citealt{2011MNRAS.413.1353F})
by using a \textit{Friends-of-friends algorithm} (FOF) to characterize the different 
ionized regions. Then, to follow the HII regions' properties with time or redshift, 
we build their \textit{merger tree}.
Using the merger tree and investigating its properties to infer the  
redshift evolution of the HII regions' formation number, their size evolution
and their merger history, we suggest
a new way to constrain the evolution of the reionization process.

We test the method on three models for the ionizing sources in order to 
quantify their impact on the simulated reionization history.
We also quantify the impact of the spatial resolution of the simulations on the history by performing
simulations with sizes of 200 and 50 Mpc/h boxes. 
By comparing the resulting reionization histories, we show how different source prescriptions lead to different 
properties on the evolution of HII regions as reionization progresses.  

This paper is organized as follows. 
In Section \ref{merger_tree1}, we present the tools developed to investigate the history of the reionization 
in simulations such as the FOF algorithm and the merger tree of HII regions.
In Section \ref{Simu}, we detail the properties of the simulations that we study.
In Section \ref{global_quantities}, first, we present our results in terms of general quantities such as the ionization 
fraction or the optical depth evolution.
Then, in section \ref{sources_topology}, we compare the different histories of each model enlightened with the merger tree methodology.  
We finally discuss these results and their limits in section \ref{Discussion} and get our conclusions.

\section{Analysing merger trees of HII regions}
\label{merger_tree1}

\subsection{Rationale}

Merger trees have been used for 20 years in e.g. the theoretical studies of galaxy formation 
(see e.g. \cite{1993MNRAS.262..627L} and \cite{1993MNRAS.261..921K}). 
More recently seminal works such as \cite{2004MNRAS.354..695F} and subsequent analytical and semi-analytical 
investigations have used excursion set formalisms to construct histories of HII regions during reionization. 
Even tough it is strongly connected to the aforementioned works, the current paper aims at using merger trees differently. 
The trees \textit{are extracted from simulations}, are used and \textit{analyzed as objects worthy of studies for their own properties}: 
in other words we analyze the simulated HII region merger histories from the merger tree they produce. 
The latter is not an intermediate to provide higher levels prediction such as e.g. statistics on 21cm signal, 
but is itself a probe of the properties of the simulation. 
On a broader perspective, the methodology presented here could be compared to genus or skeleton 
calculations as a reproductive and quantitative mean to discuss the simulated process. 
It is thus complementary to power spectra or time-evolution of average quantities for instance. 
In practice, in addition to tracking individual HII regions, we are thus led to analyze the "graph" 
properties of tree-based data as shown hereafter and draw conclusions on the simulated physics. 
Hereafter, we present briefly our implementation of the classic FOF technique to identify 
HII regions and the procedure to build merger trees. 
Additional details of these classic tools can be found in the appendix.

\subsection{FOF identification of HII regions}
\label{fofHII}

We have developed a FOF algorithm which is able
to separate the different HII regions at a given instant in snapshots of
simulations and allocate them an identification number.
The following description of this \textit{Friends-of-friends method} is similar to 
that used in previous studies (\citealt{2006MNRAS.369.1625I}, \citealt{2008ApJ...681..756S}, 
\citealt{2011MNRAS.413.1353F}). The ionization state of the gas is
sampled on a regular grid.

First we define the status of the box cells 
in terms of ionization fraction. 
The rule we have adopted is that a cell is ionized if it has an ionization fraction $ x \ge 0.5 $ 
(as often used in other works e.g. \citealt{2006MNRAS.369.1625I},
\citealt{2008ApJ...681..756S}, \citealt{2011MNRAS.413.1353F}). 
An individual HII region is defined by an identification number (ID hereafter). 
Then if a cell is ionized and has an ID, then all its ionized nearest
neighbors belong to the same ionized region and thus share the same ID. 
Our definition of nearest neighbors is limited to 6 cells along the main
directions. One single isolated cell is then considered as a single ionized region.
More details about the FOF implementation can be found in appendix
\ref{fof}.

We have to note that the FOF procedure is not without caveats. 
Thereby we could have adopted greater values for the ionization fraction
threshold for example. We can then imagine that HII regions would be more divided and more numerous
with smaller sizes in this case, possibly leading to different reionization
histories. In appendix, we show that the statistics of the regions depend weakly on
this specific value of the ionization threshold in the 0.3-0.9 range, except for the smallest regions with recombining episodes.
The FOF technique can also connect distant regions with ionized
bridges, an effect that might be undesirable. 
Alternately we could have chosen other methods for the identification such as the \textit{spherical average} 
method described by \cite{2007ApJ...654...12Z}.
In this case the distribution of HII region sizes would have been smoother (see \citealt{2011MNRAS.413.1353F}).
We have chosen the FOF method because it allows to follow regions from snapshot to snapshot in order to 
build the merger tree, which is a feature that the \textit{spherical average} method does not allow.
We thus have to keep in mind the features of the method when we will give our conclusion in the next.

\subsection{Merger tree}
\label{merger_tree2}

The identification  of the  ionized regions in  each snapshot of  a simulation
being known, the merger tree of HII regions can be built.  Such a tree aims at
tracking the identification number of an ionized region in order to follow the
evolution of its properties along  time (see appendix \ref{merger} for details
in the merger tree implementation).

\begin{figure}
\includegraphics[width=9cm]{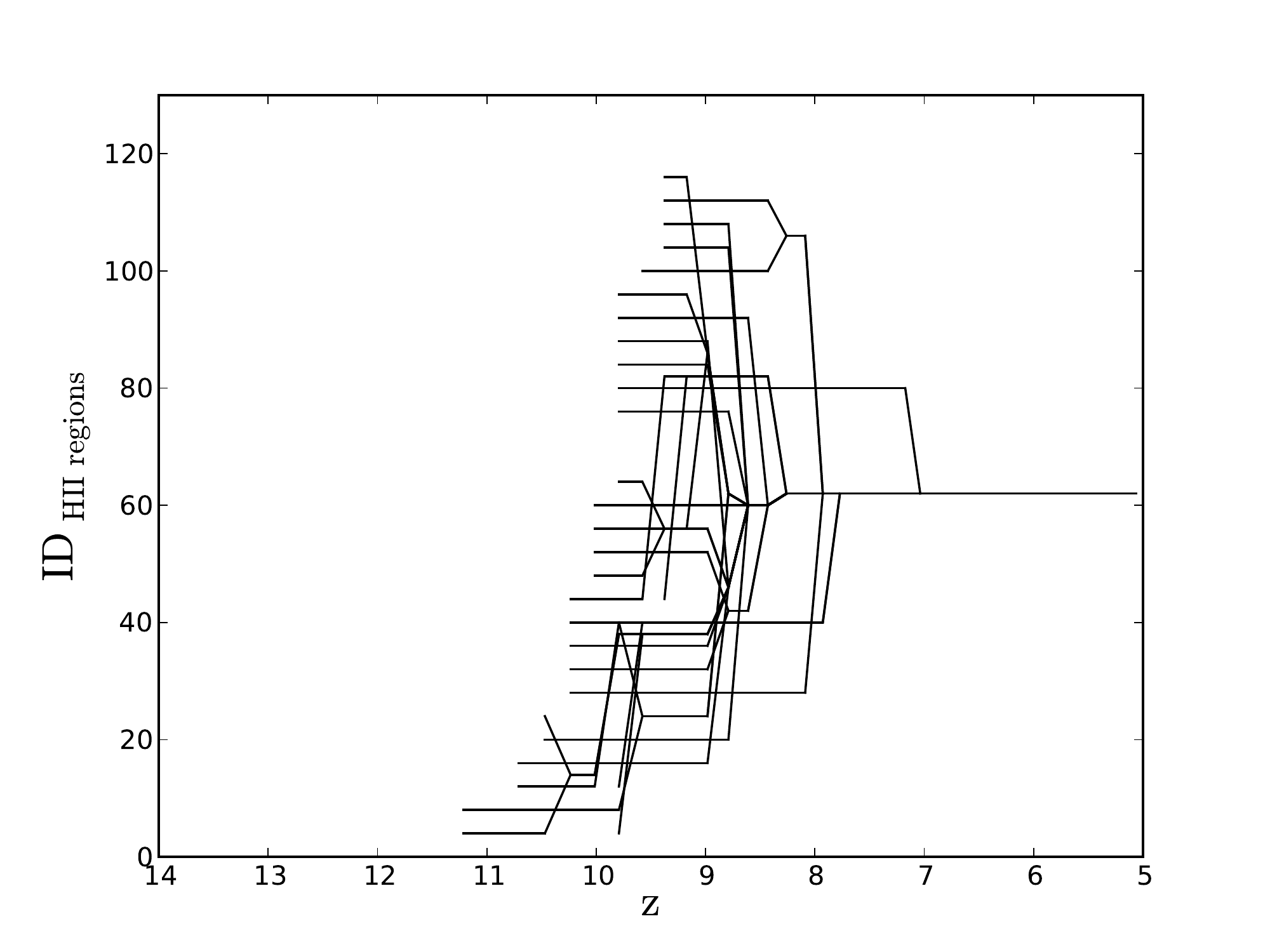}
\caption{Illustration of the merger tree of HII regions. 
Each black line represents an ID evolution with the redshift for a distinct HII region.
For clarity, we represent here only the ID evolution for 30 regions.
Mergers occur when several branches of the tree lead to the same ID.}
\label{a_view_of_the_merger_tree}
\end{figure}

In  Figure \ref{a_view_of_the_merger_tree}  an  example of  a  merger tree  is
represented where  each black line represents  an ID evolution  for a distinct
HII region.  As reionization progresses,  there is a growing number of ionized
regions to  follow, as represented  by the growing  number of branches  in the
tree.  We can also follow the  merger process between HII regions when several
branches of the tree lead to the same ID.  We observe a decrease in the number
of regions until there remains a single one at the end of reionization.

From  the  merger  tree thus  constructed,  we  will  be  able to  follow  the
properties of ionized regions during the entire simulated reionization.
Indeed,  it will  be possible  to  study the  temporal evolution  of each  HII
region,  their  individual merging  history,  their  geometric properties  and
enlighten in a different way the scenario of reionization.

Let us note  that this technique can be sensitive to  the snapshot sampling of
the  simulations. Non-regularly  spaced  snapshots or  a  sparse sampling  can
weaken the  conclusions obtained with a merger  tree. Preliminary experiments,
by taking e.g. half  of the snapshots, have shown that our  results seem to be
robust, however the quantitative conclusions may be prone to variations.

\begin{figure}
\begin{center}
\includegraphics[width=8cm]{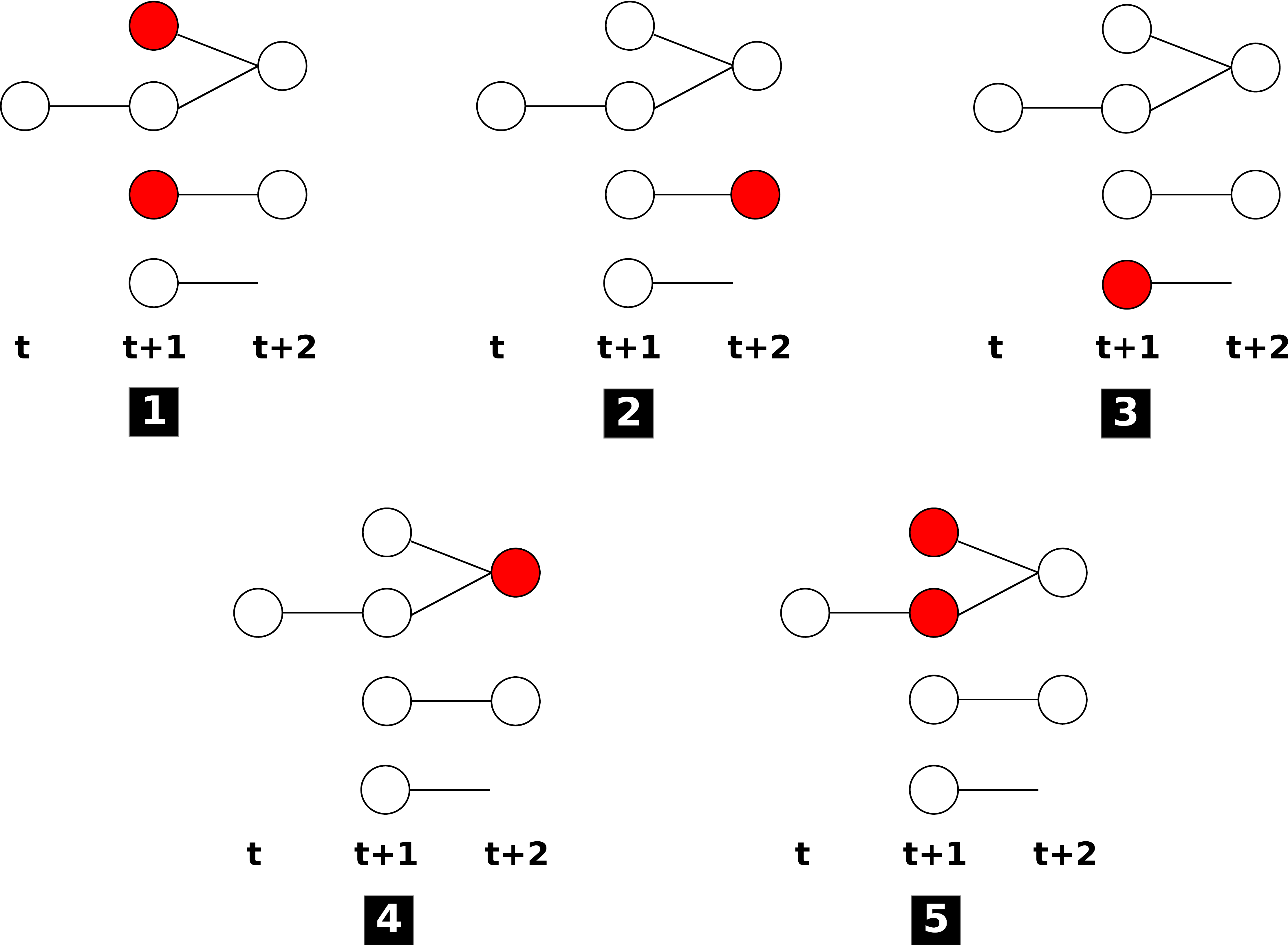}
\caption{Illustration of the properties that we can follow with the merger tree. In each diagram,
red items symbolise the kind of properties that we follow. 1: the number of new HII regions between two snapshots, 
2: the number of growing ionized regions , 
3: the number of HII regions which recombine,
4: the number of HII regions resulting from mergers
and 5: the number of parents involved for an HII region resulting from mergers }
\label{comprehension_plot}
\end{center}
\end{figure}

In figure \ref{comprehension_plot} we illustrate the typical properties that
can be investigated through the merger tree~:
\renewcommand{\labelitemi}{\textbullet}
\begin{itemize}
 \item the number of new HII regions between two snapshots. 
 \item the number of growing ionized regions.
 \item the number of HII regions which recombine.
 \item the number of HII regions resulting from mergers.
 \item the number of parents involved for a HII region resulting from mergers
\end{itemize}

\section{Simulations}
\label{Simu}

\subsection{Gas Dynamics and Post-processed Radiative transfer}
\label{radiative_transfer}

In this section we aim at describing the cosmological simulations of
reionization performed and analyzed with the merger tree methodology. These
simulations were part of a set of experiments at several resolutions and fully
described in \cite{2010ApJ...724..244A} and only a brief summary of the
methodology will be given here.

The evolution of the gas and sources distribution were provided by outputs of
the RAMSES cosmological code (\citealt{2002A&A...385..337T}) that handles the
co-evolution of dark matter, gas and star particles using an adaptive mesh
refinement strategy. The hydrodynamical equations are solved thanks to a second
  order Godunov Scheme with an HLLC Riemann Solver. The gas is assumed to be
  perfect with a 5/3 polytropic index.  Metals are included and taken
  in account in the cooling of the gas. The star formation is
  included following the prescription of \citet{2006A&A...445....1R} as well
  as supernovae feedback.

Initial conditions were
generated on $1024^3$ grids, according to the WMAP 5 cosmology (\cite{2009ApJS..180..330K}) using
the MPgrafic package and white noise statistics from the Horizon collaboration (\cite{2008ApJS..178..179P}).
Two box sizes were considered, 50 and 200 Mpc/h, both with a $1024^3$ coarse
grid resolution+ 3 levels of refinement and simulations were conducted down
to z$\sim 5.5$. The refinement strategy is quasi-lagrangian with finer levels
being triggered when the mass within a cell is 8 times the mass resolution. The large box is less subject to finite-volume variance effects
but lacks resolution whereas the 50 Mpc/h box better resolve small scales
physics (such as star formation) but is more prone to cosmic variance and its
percolation process during reionization may be already affected by periodic boundary conditions

The radiative transfer was included as a post-processing step using the ATON code (\citealt{2008MNRAS.387..295A}) in all simulations.
It tracks the propagation of radiation on a $1024^3$ coarse grid using a
moment-based description of the radiative transfer equation. 
It can reconcile the high resolution and the intensive time-stepping of the
calculation thanks to Graphics Processing Units (GPUs)
acceleration obtained through NVidia's CUDA extension to C (see \cite{2010ApJ...724..244A} for further details of the
implementation). ATON also tracks atomic hydrogen processes such as heating/cooling
and photo-ionization and can handle multiple group of frequencies, even though
in the current work only a single group of photon has been dealt with. The
typical photon energy is 20.26 eV, corresponding to the average energy of
an hydrogen ionizing photon emitted by a 50 000 K black body: such spectrum is
a good approximation of a salpeter IMF integrated over the stellar particles
mass assumed here (see e.g. \cite{2009A&A...495..389B}). Calculations were run
on 64-128 M2068 NVidia GPUs on the hybrid sections of the Titane and Curie supercomputers
hosted by the CCRT/CEA facilities. Having post-processed the simulations,
  we applied the merger tree methodology described in the previous section to
  the ionisation fraction field degraded at a $512^3$ resolution, making them
  easier to analyse.

\subsection{Ionising source models}
\label{ionizing_sources_models}
At each resolution, three different models were considered for the sources: two considered the
self-consistent stellar particles spawned by the cosmological simulation
code. The third one uses dark matter halos as proxies for the sources. All
the models were tuned by trial and error to provide a complete reionization by
$z\sim 6.5-5.5$ and a half-reionization at $z\sim 7$. Sources were included in the radiative transfer calculation following the same procedure as in \citet{2010ApJ...724..244A}.

\begin{figure}
\begin{center}
\includegraphics[width=0.8\columnwidth,height=0.7\columnwidth]{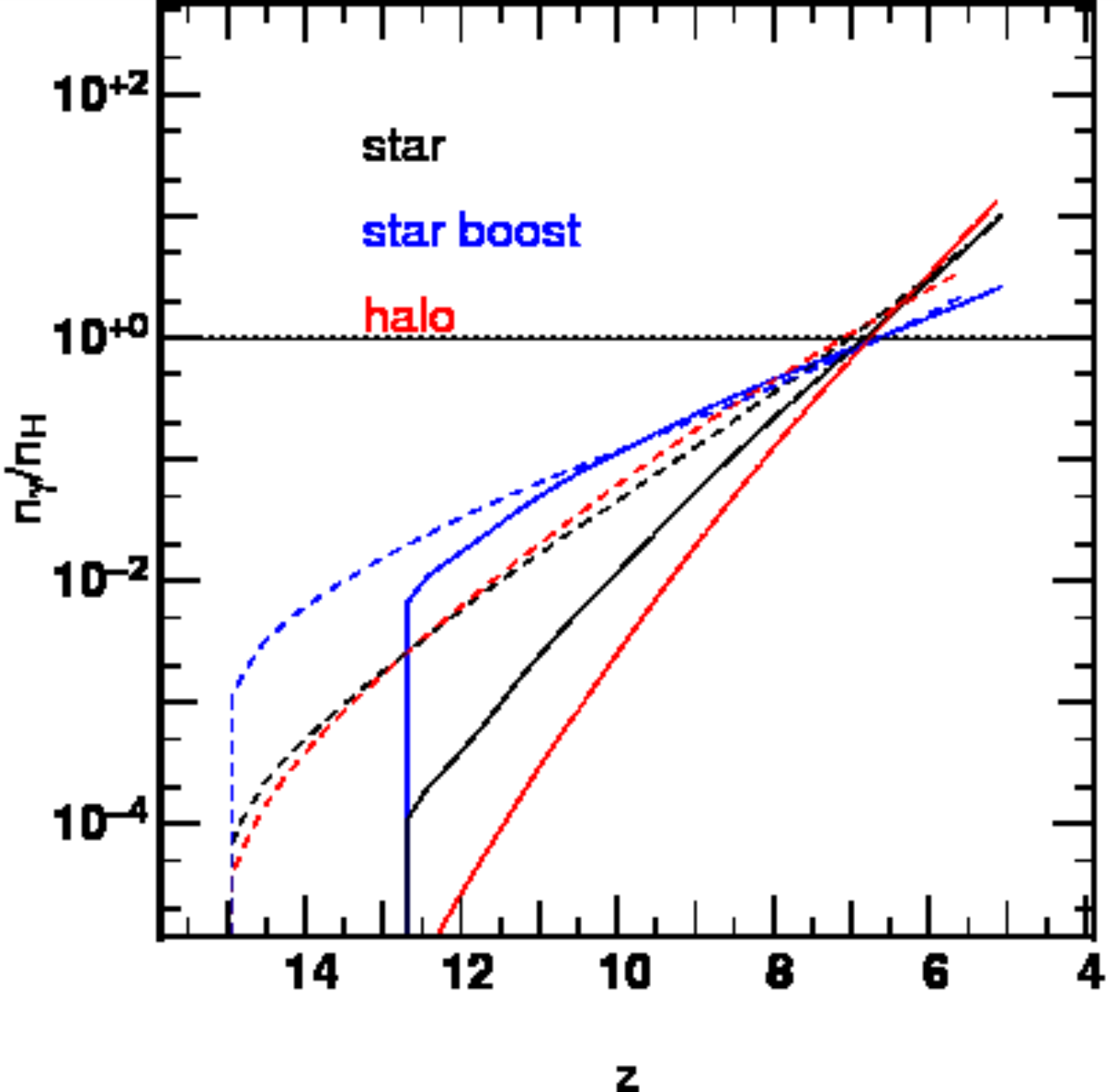}
\end{center}
\caption{Redshift evolution of the cumulative number of ionizing photons
  emitted by the sources relative to the number of hydrogen atoms. The
  horizontal line stand for one photon per atom. Solid (resp. dashed) lines stand for the
  200 (resp. 50) Mpc/h simulations. Black, blue and red curve stand
  respectively for the 'Star', 'Boosted Star' and 'Halo' models of ionizing sources.}
\label{source}
\end{figure}

\subsubsection{`Star' models}
\label{Ramses}
The star formation recipe is described in \citet{2006A&A...445....1R} and
assumes that above a given baryon over-density ($\delta \sim 5$ in our case), gas
transforms into constant mass stars ($1 \times 10^{6}$ M$_\odot$and $2 \times 10^{4}$ M$_\odot$ in 200/50
Mpc/h boxes) with a given efficiency ($\epsilon=0.01$). 

The number of stellar
particles found at $z\sim 8.5$ (that corresponds to the peak of HII regions
number as seen hereafter) is 8500 (resp. 35 500) in the 200 Mpc/h (resp. 50
Mpc/h) simulations.
Even though it is now standard, the modelisation of the formation of `stellar' particles in cosmological simulations remains a
complex and subtle matter. Among other effects, it depends strongly on the
resolution: the growth of non-linearities is scale-dependent and as a
consequence the simulated star formation rate depends on the ability of the
simulation to resolve high-density peaks (see
e.g. \citet{2003MNRAS.339..289S}, \citet{2006A&A...445....1R} and references
therein). Therefore poorly resolved simulations tend to develop a population of stars at later times and at a
slower rate than more resolved ones. This limitation is further emphasized
during the reionization epoch at large z. As a direct consequence the amount
of ionizing photons is usually under-evaluated if taken directly from star
particles even though the situation improves as the non-linearities evolves
in the simulations. For instance, with the number of resolution elements and refinement
strategy used in the current work, a 12.5 Mpc/h box would be necessary to
achieve a convergence in the number of emitted photons
(\citet{2010ApJ...724..244A}). Otherwise, the star particles may even be too
scarce to reionize the cosmological box. 
Furthermore, these sources are subject
to stochasticity: they contribute to the UV flux only during the lifetime of
strong UV-emitting stars (that we chose to be 20 Myrs). As this component fade away in a given `stellar
particle', the latter may not be replaced by a neighbour. Until a stationary regime of
source renewal is installed, these numerical artefacts due to source
  discretness and lack of convergence may lead to `blinking' emitters of numerical nature, 
and therefore to artificial recombining HII regions unsustained by inner sources, without any relation to actual starburst episodes.

On the other hand, these `stellar' sources are directly
generated from the simulation, are simulated self-consistently with the gas
physics and do not require an additional processing to compute the sources locations and emissivities. One may suggest
that a simple correction may be enough to correct from the low resolution
effects. The simplest is to consider that stellar particles sample correctly
the location of the main UV sources and a constant fudge factor would
provide the adequate number of ionizing photons to reionize at $z\sim
7-6$. Such a model will be named as 'SXX' hereafter, where `XX' stands for
200 or 50 depending on the box size. A more sophisticated
one requires that not only the adequate number of photons is generated at
$z=6$ but also at each instant. Since the lack of resolution is more critical
at early times, it implies a stronger correction at high z than later
on. \cite{2010ApJ...724..244A} have shown that the correction is exponential
with characteristic times depending on the resolution. A drawback of this type
of correction is the fact that sources end up as being individually 'decaying'
with time: if the renewal rate of stellar particles is not high enough, this
decay could impact the simulated reionization. In this model, `stellar
particles' have \textit{boosted} emissivities per baryon at early times and will be
named 'SBXX' hereafter, XX standing for the box size. 
It should be noted that the density contrast chosen here to trigger star formation can be seen as low and is
  chosen to allow star formation at our moderate spatial resolution and at
  redshifts where non-linearities are weak at our scales. The value chosen
  here is similar to the one used by e.g. \citet{2000ApJ...541...25N} at the
  same resolution and is slightly more permissive than most of the thresholds reviewed e.g. in \citet{2002MNRAS.330..113K} with typical values of
  $\delta\sim 10$. However, even with such criteria, significant factors should be applied to the
  luminosity of these stellar particles to produce enough photons to obtain 'standard reionizations'
  histories. As shown hereafter, the ionized 'patchiness' of these models tend to be
  less structured on small scales than halo-based models while nevertheless
  having their overlap process sharing the same global properties. Overall it
  indicates that star particles are not overproduced.

In Fig. \ref{source}, one can see that the S200 and S50  'stellar' models show
different slopes for the cumulative number of photons. It results from the
fact that the same number of photons must be produced from a smaller number of
star particles and on a shorter period for the large box. At both resolutions, 2 ionizing photons per
hydrogen atom have been produced when reionization is completed at $z\sim
6.5$. The SB models are as expected `converged': as the photon
production become stationary, the same number of photons is produced in both
box sizes. Compared to the 'Star models' emissivities, the 'Boosted' ones are greater
at early times then smaller for $z<7.5$, to achieve the reionization at the
same epoch. On a more general note, one can see that `stellar' particles appear earlier at
high resolution as expected.

Quantitatively, we
  followed the simple modelisation of \citet{2009A&A...495..389B} to assign an
  emmissivity of 90 000 photons per stellar baryons over the lifetime of a
  source. For S50 and S200 the enhancement factors to account for convergence
  issue are respectively 3.8 and 30, while the boost temporal evolution of the
  SB50 and SB200 is equal to $\mathrm{max} (1,a \exp(k/t))$ with $(a,k\mathrm{[Myr]})$=(1.2,1500) for
  SB50 and (1.2,3000.) for SB200, $t$ being the cosmic time.

\subsubsection{Halo sources}
\label{halos_sources}

As a second choice we adopt a semi-analytical model for the production of ionizing sources based on dark matter halos.
Each halo is assumed as a star formation site with an emissivity proportional to the halo mass. This procedure is largely inspired by the work of
\cite{2006MNRAS.369.1625I} where a constant mass-to-light ratio is assumed such as the ionizing flux of each halo has the following expression:

\begin{equation}
\dot{N}_{\gamma}=\alpha M
\end{equation}

where $\alpha$ is the emissivity coefficient. Here we have chosen values of $\alpha = 5.9 \times 10^{43}$ and $\alpha = 3.5 \times 10^{42}$ 
photons/s/$M_{\odot}$ for both 200 and 50 Mpc/h boxes respectively. These
values were chosen in order to reach a reionization at $z \sim 6.5-6$ similar
to the `Star' and `Boosted Star' models. Halos were detected using the
parallel FoF finder of \cite{2011MNRAS.410.1911C}, with a linking length of
b=0.2 and a minimal mass of 10 particles. This minimal mass corresponds to
$9.8\times 10^7 M_\odot$ for the 50 Mpc/h box and $6.3\times 10^9 M_\odot$ for
the 200 Mpc/h box, implying that all halos can be considered as emitters. The
number of halos found at $z\sim 8.5$ is 33 100 (resp. 285 000) in the 200
Mpc/h (resp. 50 Mpc/h) simulations. It should also
be emphasized that, contrary to 'Star' and 'Boosted Star' models, a whole
spectrum of mass, and thus emissivities, is available for halos.
No lifetime has been assigned to these sources and a halo would produce photons as soon as it is being detected until
it disappears or merge. In the following sections we will refer to this
semi-analytical star formation simulations with the following acronyms: "H200" or "H50" for both box sizes of 200 and 50
Mpc/h. 

In Fig. \ref{source}, one can see that 'Halo 50' follows closely 'Star 50',
implying that provided the appropriate correction, stars can mimic halos
emission history. A greater difference can bee seen in the 200 Mpc/h box,
where the 'Halo 200' model shows a greater slope than the 'Star 200' model:
there exists a greater difference in the buildup of halos compared to the
star population at lower resolution, due mostly to the difficulty to achieve
star formation density thresholds in large scale simulations. At $z\sim 6.5$,
roughly two photons per hydrogen atom were produced in all cases.


\section{Global features}
\label{global_quantities}

\subsection{Ionisation fraction}
\label{ionization_fraction}

\begin{figure}
\includegraphics[width=9cm]{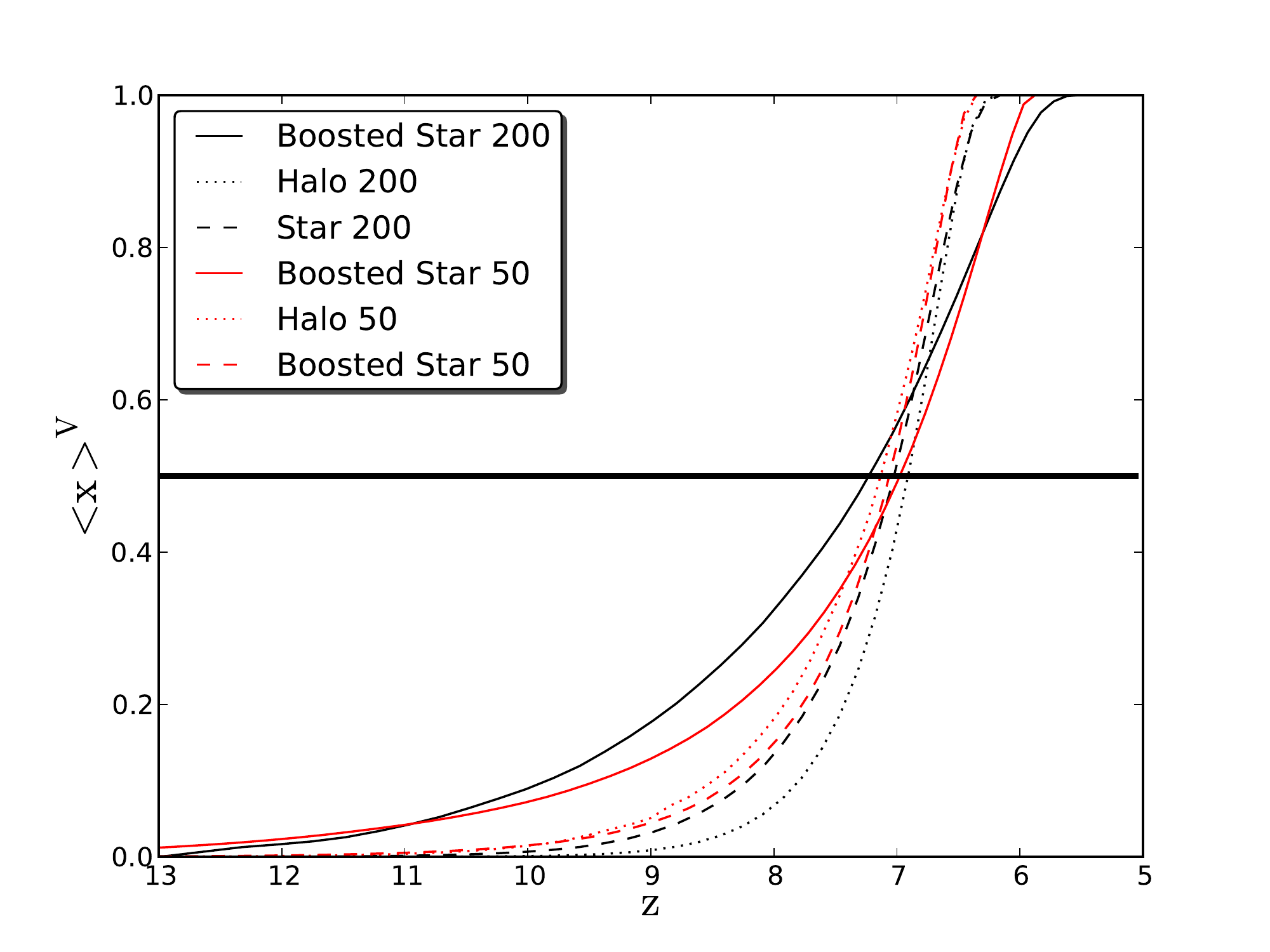}
\caption{Evolution of the volume weighted average ionization fraction with redshift for the six simulations.}
\label{xmoy_vs_z}
\end{figure}

Figure \ref{xmoy_vs_z} shows the evolution of the volume weighted average ionization fraction as a function of redshift for the
three models and for both 200 and 50 Mpc/h boxes. 
Average ionization curves show a similar value of $<x>\sim0.5$ at a redshift of $z\sim7$.
At first glance we see that every models are comparable in terms of ionization fraction. 
The boosted star model just shows an earlier "take off" in the ionization curve than in both others models for the two box sizes.
Conversely in the Star and Halo models the ionization curves "take off" later ($z \sim 10-9$) but the reionization
is achieved earlier than in the boosted Star model. 
These differences are attenuated in the 50 Mpc/h box where the ionization histories become very close from one model to another.

\subsection{Optical depth}
\label{optical_depth}

\begin{figure}
\includegraphics[width=9cm]{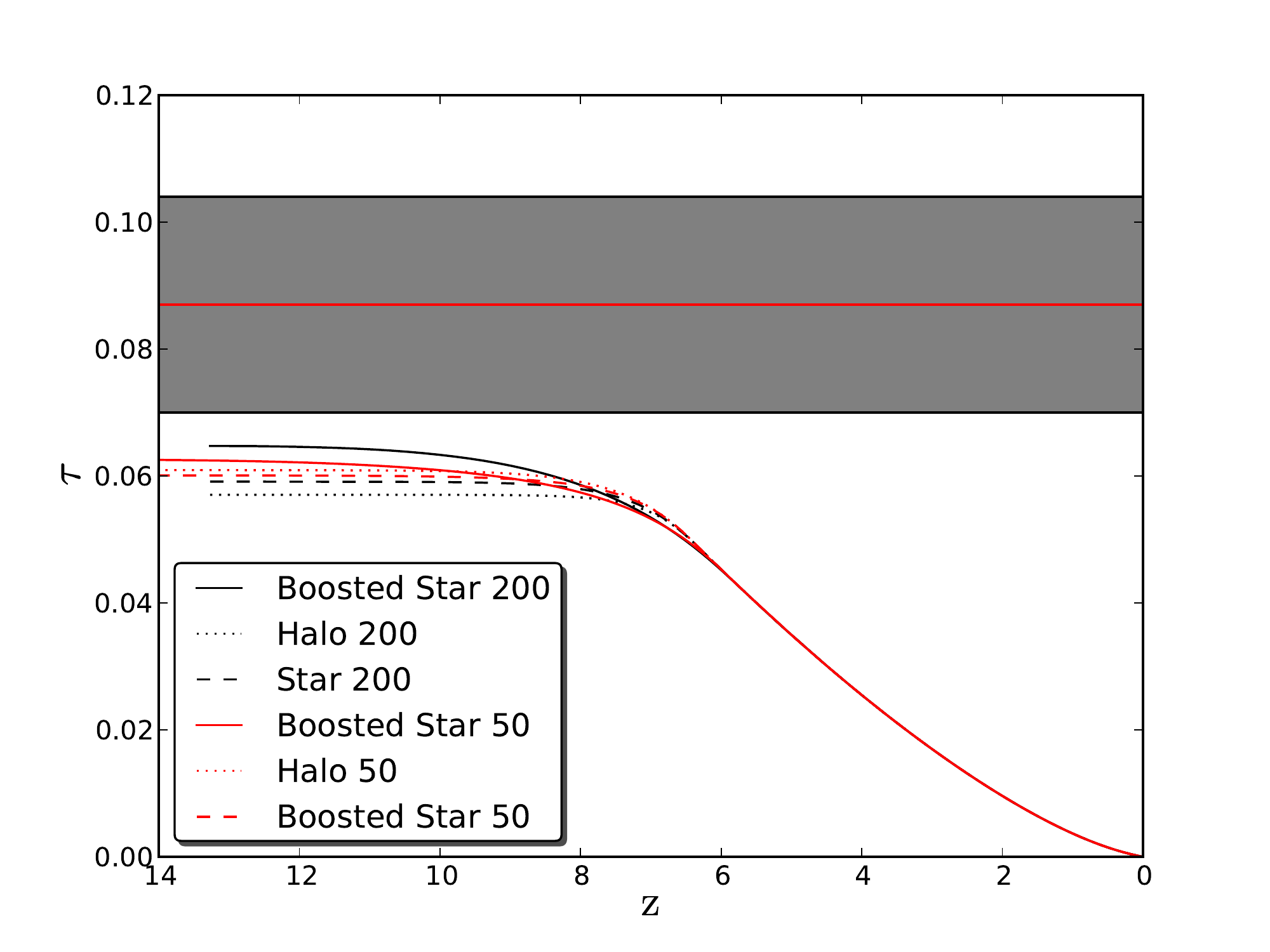}
\caption{Evolution of the Thomson optical depth with redshift for the six simulations.}
\label{tau_vs_z}
\end{figure} 

Figure \ref{tau_vs_z} shows the evolution of the optical depth as a function of redshift for the
six simulations given by

\begin{equation}
\tau(z)=c\sigma_t \int_{z}^0 n_e(z)\frac{\mathrm dt}{\mathrm dz}\, \mathrm dz
\end{equation}

\noindent where $\sigma_t$ is the Thompson cross section of the electron and $n_e(z)=<x(z)>n_H(z)$ is the density of electrons
released by ionized hydrogen atoms at redshift z.
We also represent the constraint range obtained from the five year release of CMB measurement
made by the \textit{WMAP} collaboration (\cite{2009ApJS..180..330K}) at the 1 $\sigma$ level.

We immediately see that all the simulations converge in terms of optical depth.
The 200 Mpc/h box simulations reach a same value at $z \sim 8$.
The only difference being that the optical depth is slightly greater in the boosted Star model before $z \sim 8$.
This is naturally explained by the fact that the ionization history is more extended in this model as seen with the 
ionization curve in figure \ref{xmoy_vs_z}.   
In the 50 Mpc/h box, the curves become almost superimposed from $z \sim 10$.
Again this comforts us in saying that all the simulations are comparable.

\subsection{Ionisation fields}
\label{ionization_fields}

\begin{figure*}
\begin{center}
\includegraphics[width=18cm]{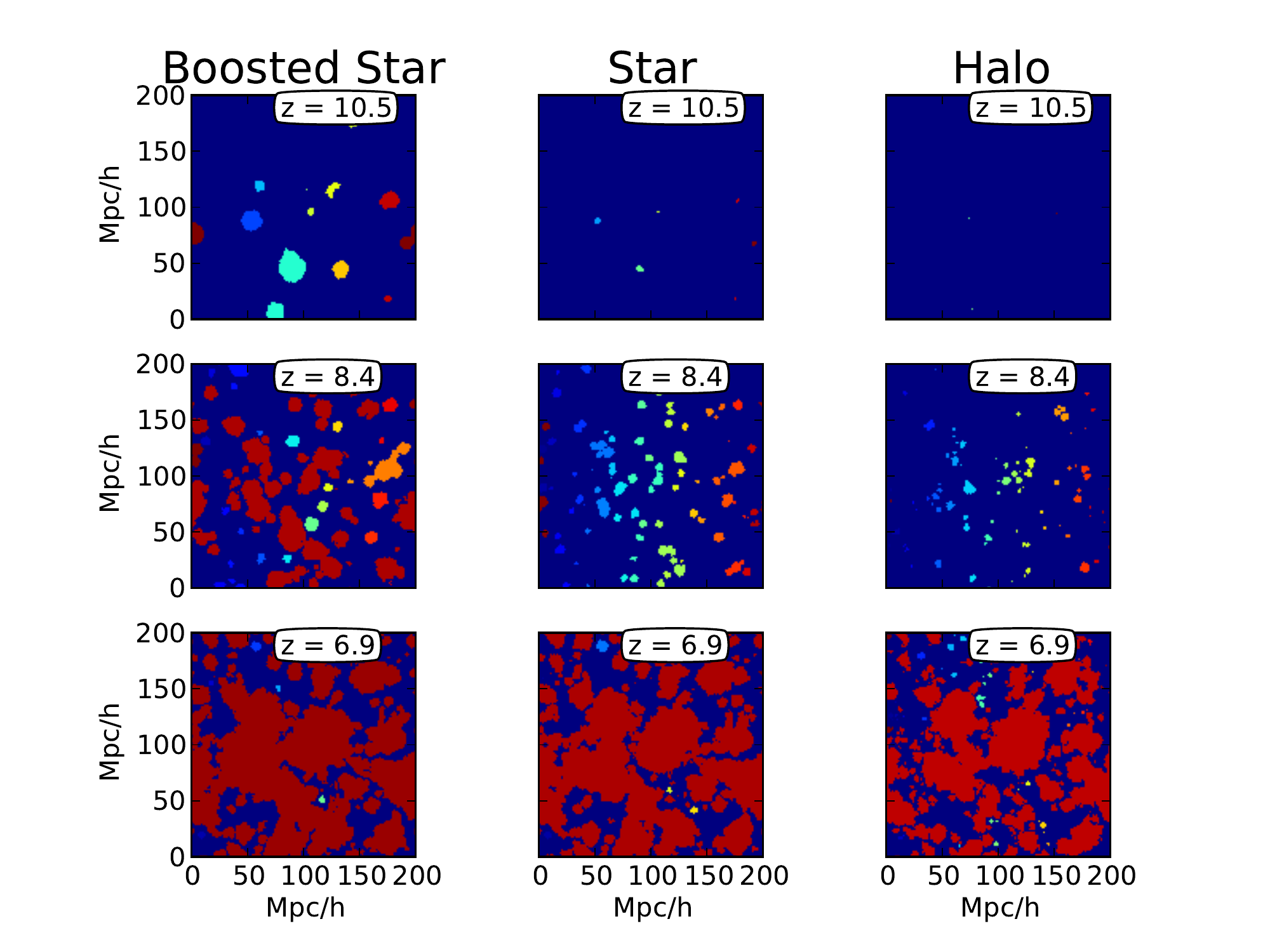}
\caption{Ionisation map for the three ionizing sources models for three distinct redshift for the 200 Mpc/h size box.
The colours encode the different identification number allocated to the HII regions by the FOF algorithm}
\label{champs_identifies200}
\end{center}
\end{figure*}

\begin{figure*}
\begin{center}
\includegraphics[width=18cm]{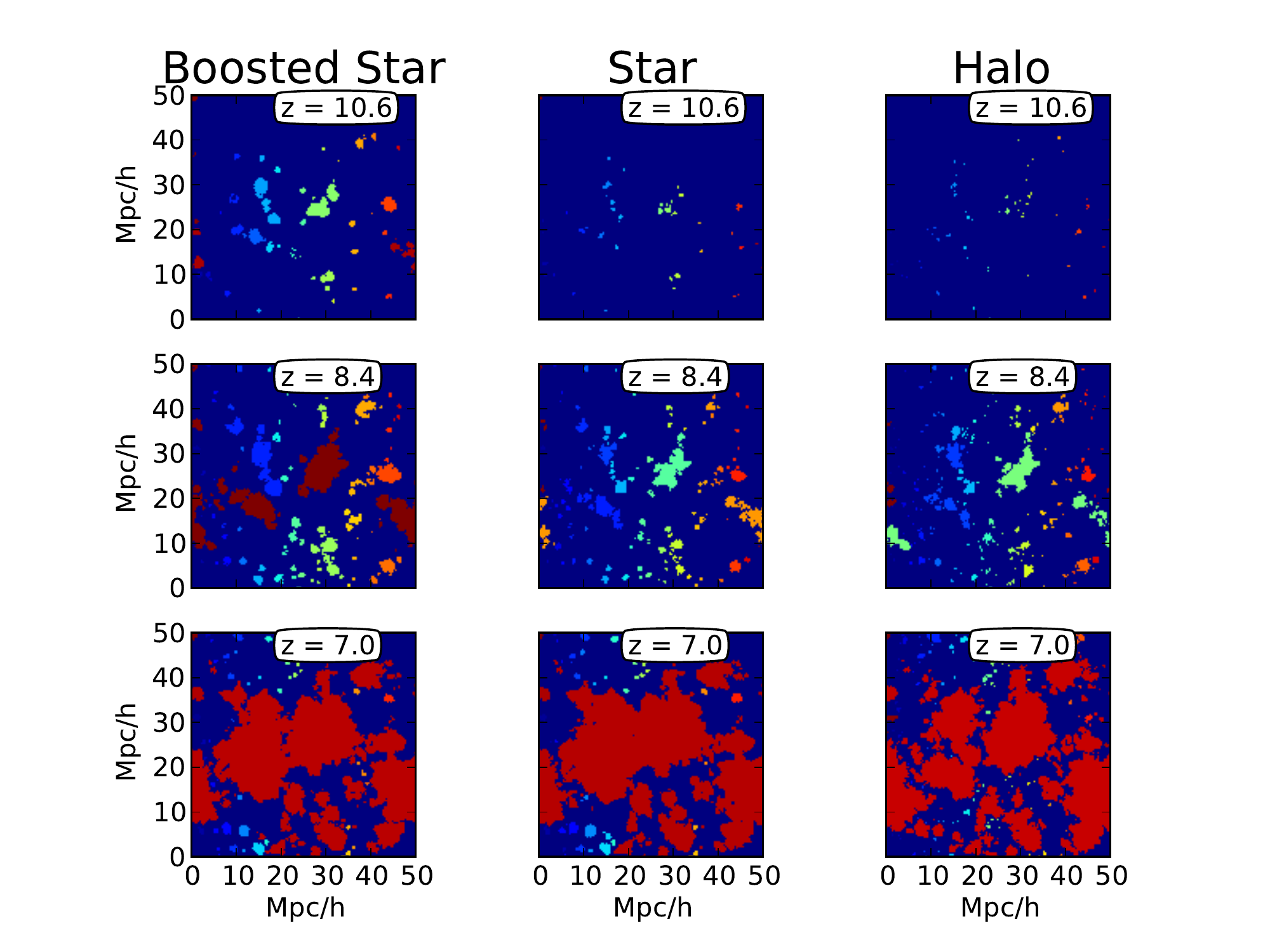}
\caption{Ionisation map for the three ionizing sources models for three distinct redshift for the 50 Mpc/h size box.
The colours encode the different identification number allocated to the HII regions by the FOF algorithm}
\label{champs_identifies50}
\end{center}
\end{figure*}

The maps of figures \ref{champs_identifies200} and \ref{champs_identifies50} show the ionization field maps for the three
models for both 200 and 50 Mpc/h boxes.
In each figure, colour encodes an individual HII region detected by the FOF procedure.

At $z \sim 10.5$ and $z \sim 8.4$, the basic features of the 3 type of
reionization can be spotted: SB models exhibit a few regions with large radii,
whereas H models have smaller and numerous individual regions. S models are
intermediate with larger ionized patch than detected in the halo-based
simulations but more individual regions than found in the SB models. As such
it reflects the differences in source modeling where halos are more numerous
than stellar particles and therefore share ionizing photons over a larger
number of weaker sources. On the other hand, SB models produce large regions
at the earliest redshifts because of the strong initial correction to the
source emissivity and which favors large regions  and potentially early overlaps.


At the end of the reionization at $z \sim 6.9$ and in all models, a single
large HII region is detected by the FOF method that results from a connected
network of multiple ionized regions. Nonetheless, the Halo model presents a
greater resilience to percolation as it presents a map more structured with more individual HII regions.

When the spatial resolution is increased with the 50 Mpc/h box, the figure \ref{champs_identifies50} shows the same tendencies as in the 200 Mpc/h box. 
The H50 map presents the most important number of regions with the smallest
sizes until late phases of the reionization and the SB50 still presents the
lowest number of regions with the largest sizes. S50 still represents an
intermediate case. Overall and as expected by the greater spatial resolution, ability to capture the clustering and larger number of sources in a 50
Mpc/h box, these maps present a greater level of
granularity and more individual regions than the 200 Mpc/h versions.

\section{Merger tree properties}
\label{sources_topology}

Our main interest is to investigate the impact of the ionizing source models on the reionization history.
Typically, we will see what quantities are retained from one model to another and how these models imprint 
differences on the history observed.

\subsection{Evolution of the number density of HII regions}
\label{number_density}

\begin{figure*}
   \begin{center}
    \begin{tabular}{ccc}
       \includegraphics[width=6cm,height=5cm]{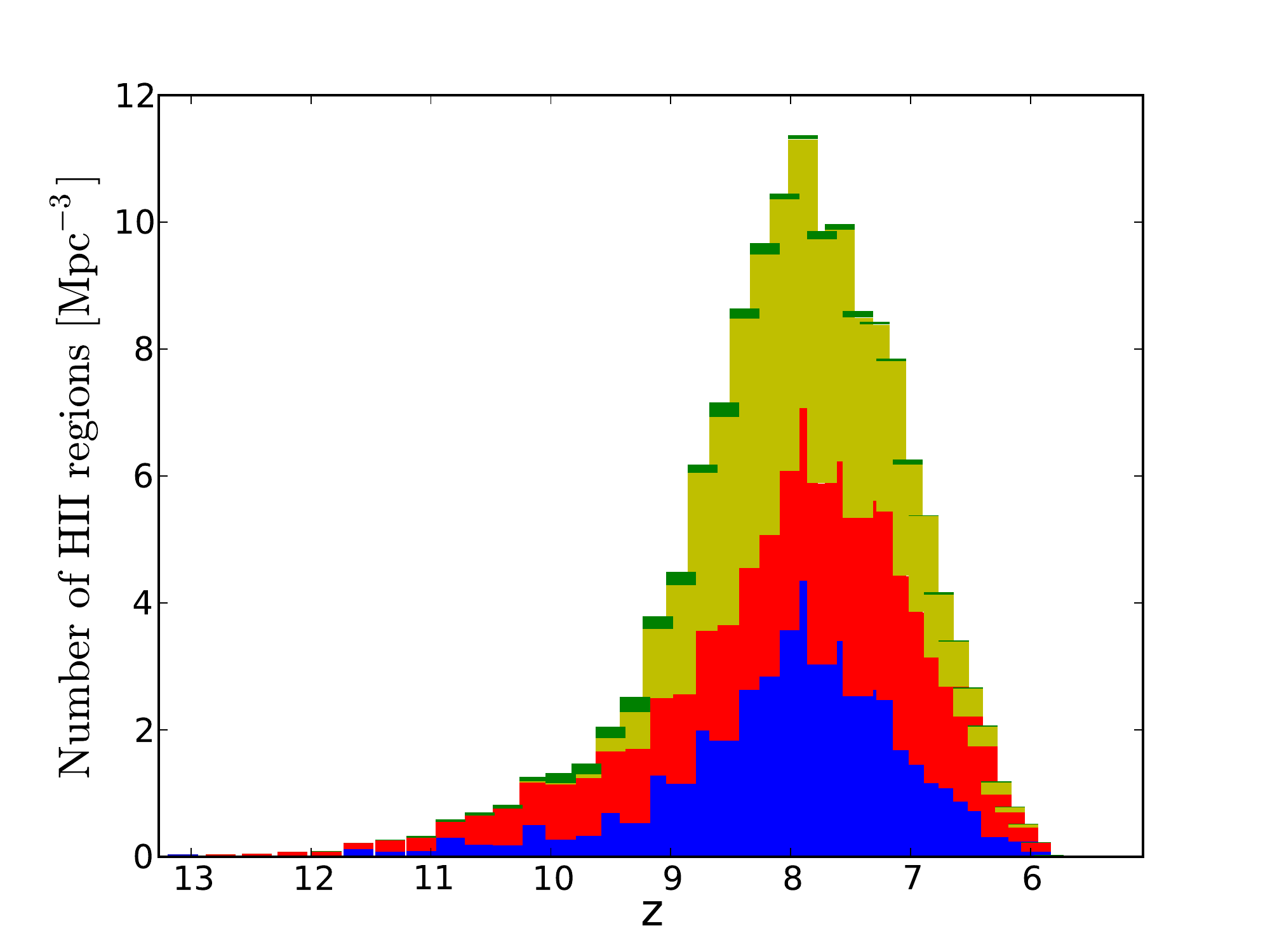} &
	\includegraphics[width=6cm,height=5cm]{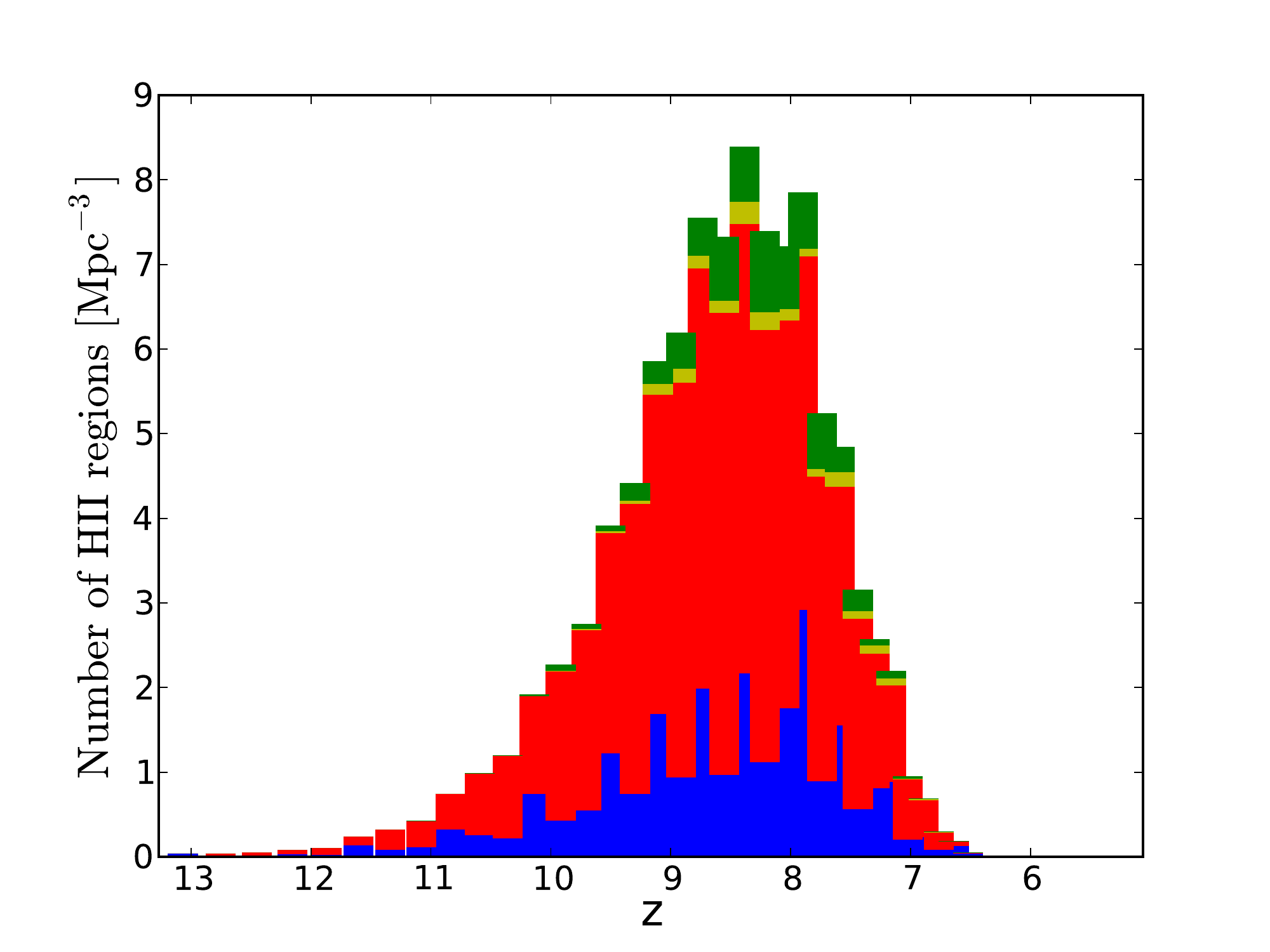} &
	\includegraphics[width=6cm,height=5cm]{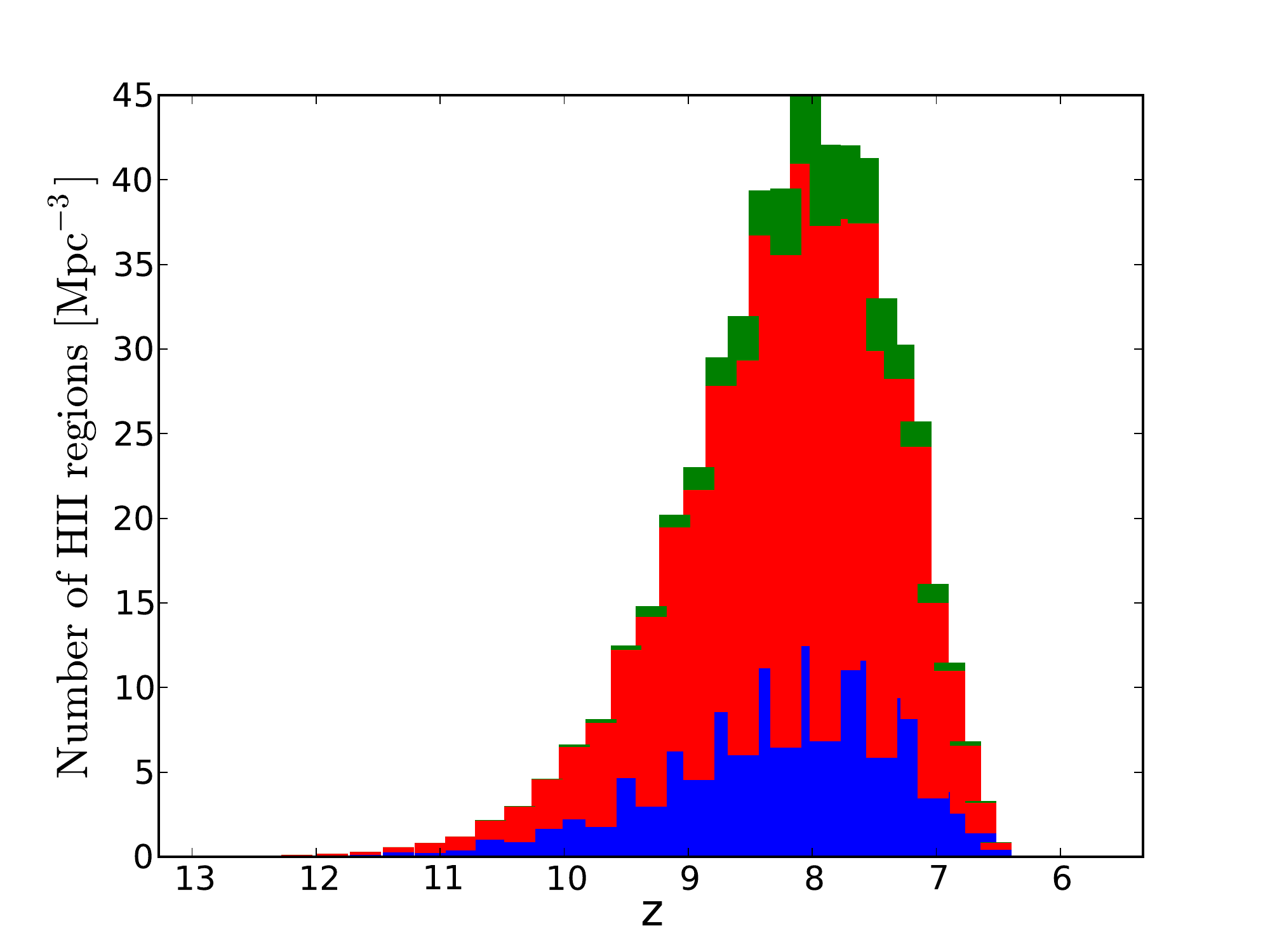} \\
    (a) Boosted Star 200 & (b) Star 200 & (c) Halo 200 \\
       \includegraphics[width=6cm,height=5cm]{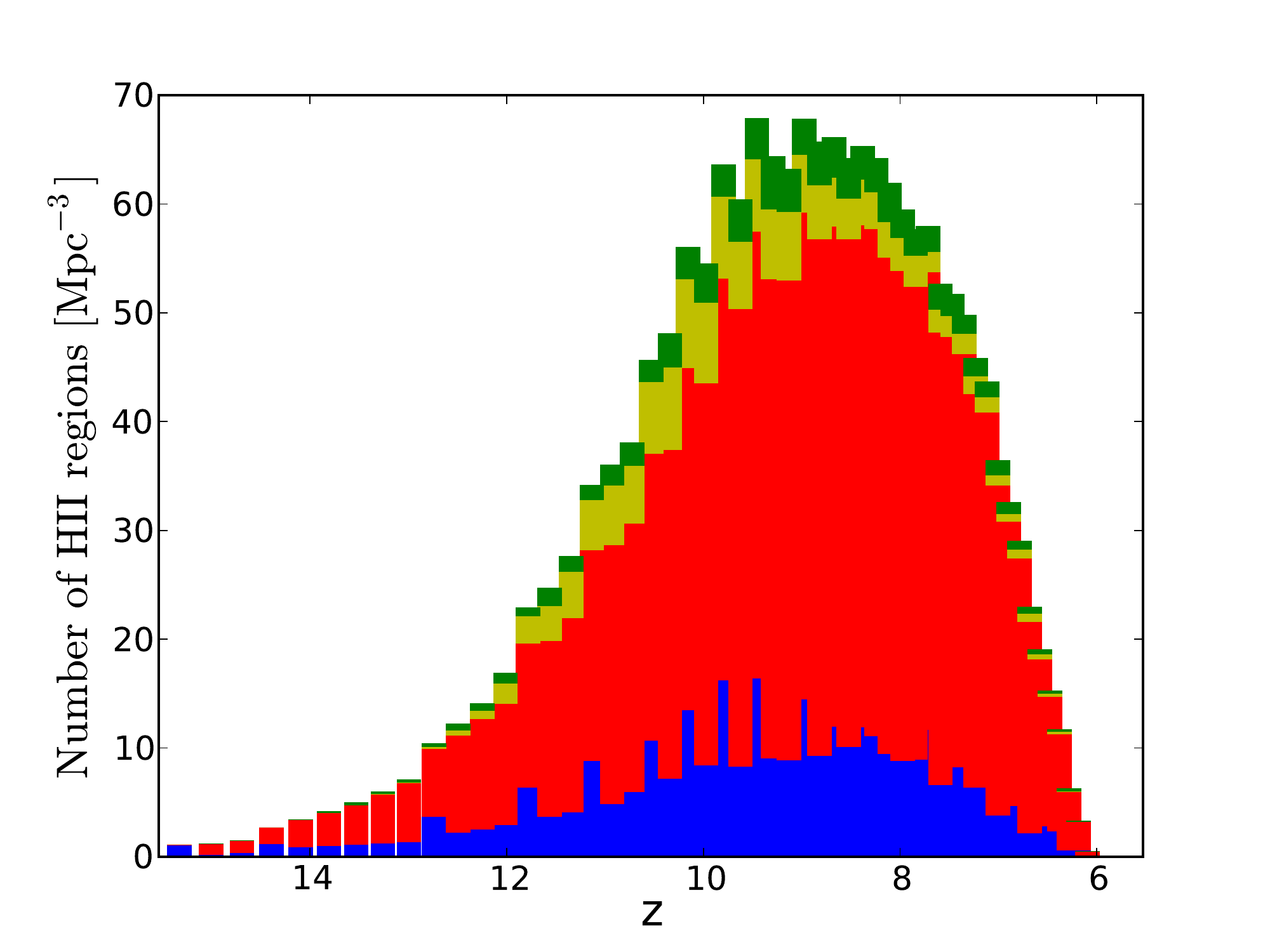} &
	\includegraphics[width=6cm,height=5cm]{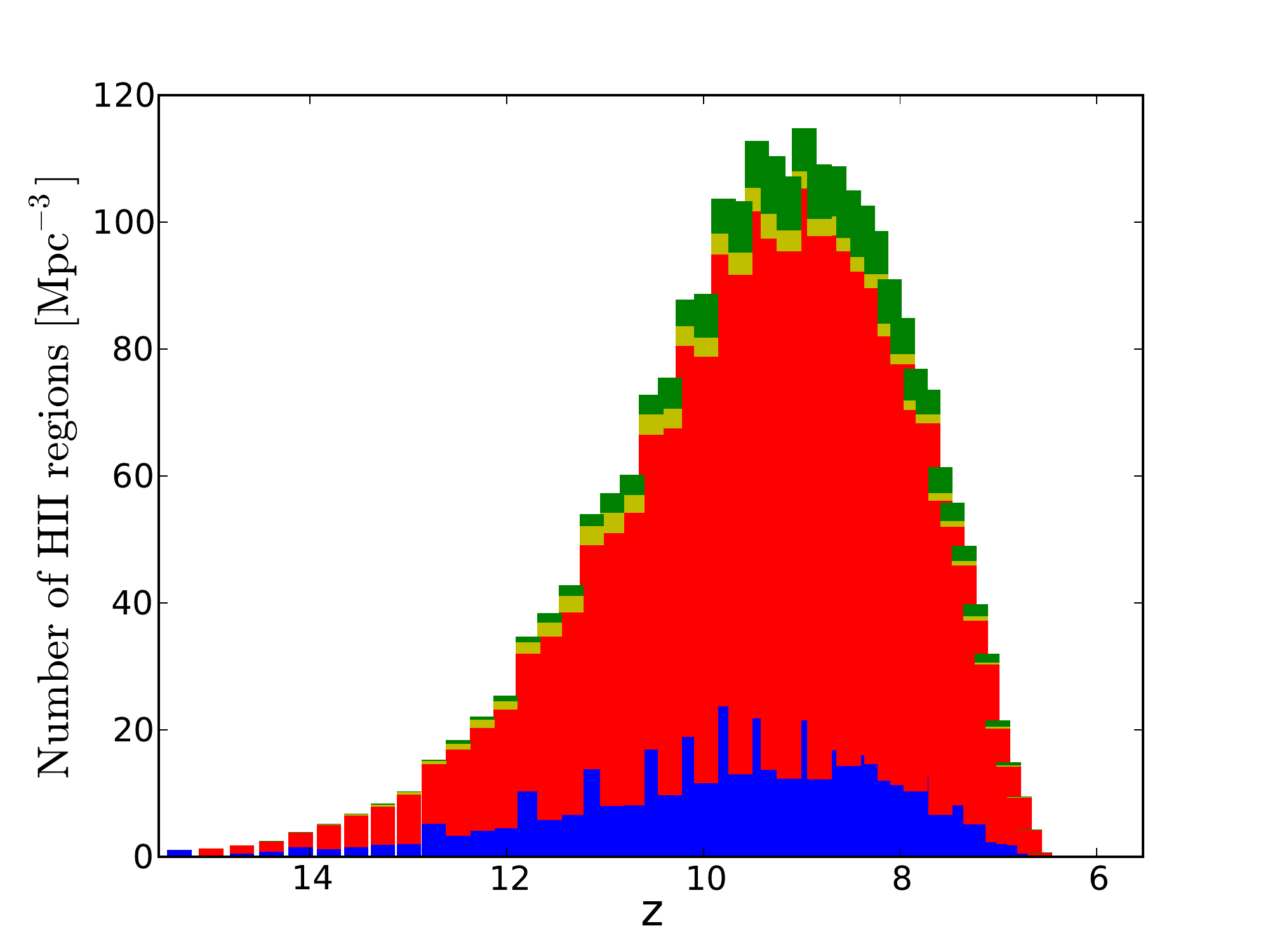} &
	\includegraphics[width=6cm,height=5cm]{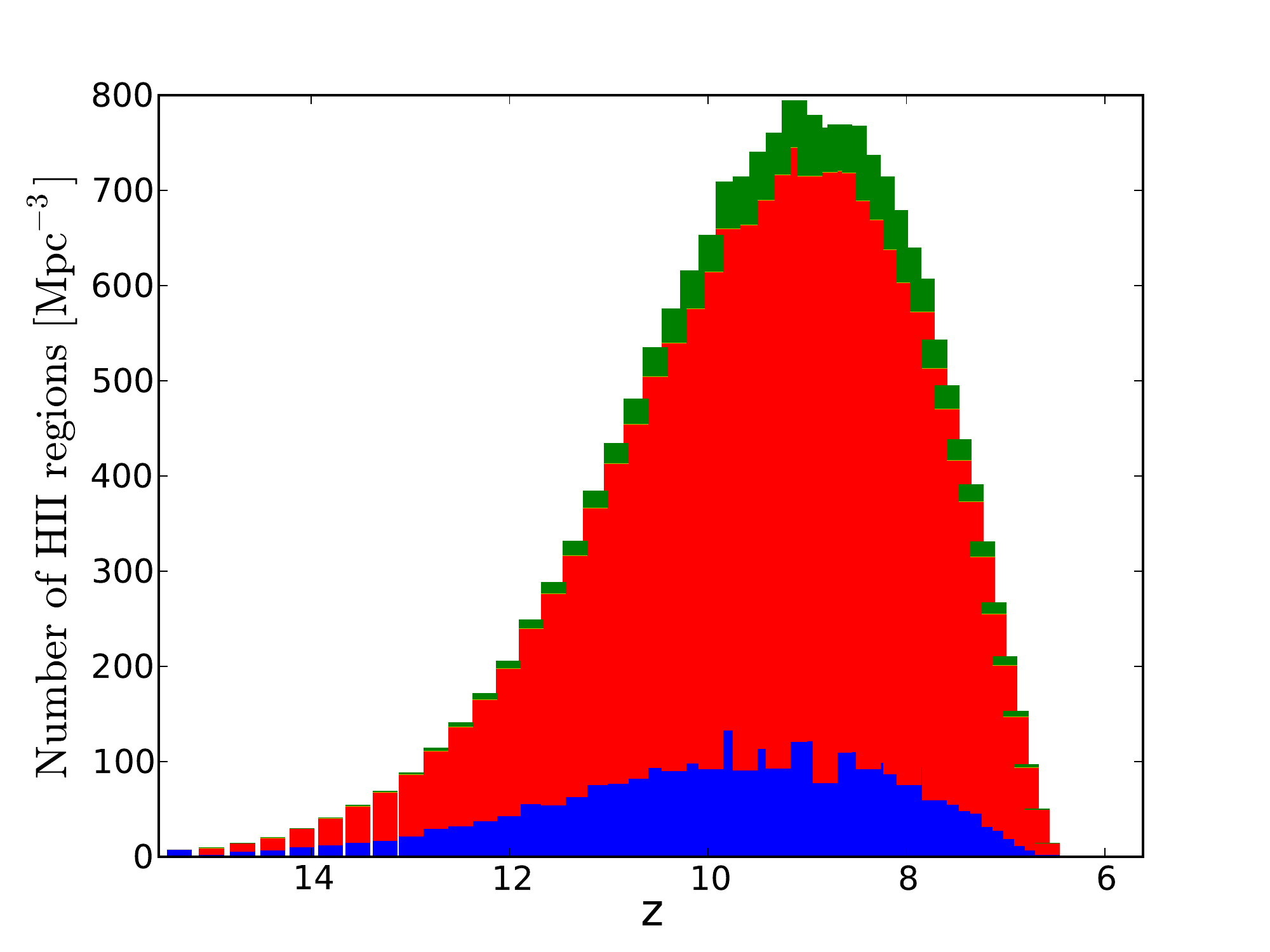} \\
     (d) Boosted Star 50 & (e) Star 50 & (f) Halo 50 \\
\end{tabular}    
   \caption{Evolution of the number density of each kind of HII regions as a function of redshift for the three models of ionizing sources
and for both boxes of 200 and 50 Mpc/h. Panels (a), (b) and (c) respectively represents the distribution for the Boosted Star, 
the Star and the Halo model for the 200 Mpc/h box, while the panels (d), (e) and (f) are for the same models but for the 50 Mpc/h box.
The colours stand as follow: Blue for the new HII regions, Red for the expanding regions, 
Yellow for the regions that will recombine and Green for the regions resulting from mergers.}
    \label{absolute_number}
  \end{center}
 \end{figure*}

The figure \ref{absolute_number} shows the evolution of the number density of ionized regions
as a function of redshift for the three models of ionizing sources and for both box sizes.
We distinguish in addition the number of each type of HII regions: new
regions, those resulting from mergers, those expanding without merging and
those recombining are respectively shown in blue, green, red and yellow.
We note that the distributions present the same general shape regardless of
the source model. They all show their maximum number of regions at broadly the same redshift
$\mathrm{z_{\,peak}}$ for both spatial resolution respectively at $z \sim 8$ and $z \sim 9$ for 200 Mpc/h and 50 Mpc/h.
Before $\mathrm{z_{\,peak}}$, HII regions appear and expand, populating the box with more and more regions.
We can refer to this period as the "pre-overlap" period where the reionization is dominated by
the expansion and the birth of new HII regions. Conversely after $\mathrm{z_{\,peak}}$, HII regions 
begin to merge intensively and decrease in number during an "overlap period", that continues until the reionization
is completed. Even though the shape of the distributions are similar, the
absolute number of HII regions is much greater in the Halo model than in the
two `stellar ones' and with a greater discrepancy in the 50 Mpc/h box.
This absolute number remains within the same order of magnitude between the Star and the boosted Star model
for both boxes. This difference is not surprising and typically
reflects the difference in the relative number of ionizing sources produced from one model to another.
Indeed the number of dark matter halos assumed as ionizing sources in the semi analytical model is greater than the number of 
auto-consistent ionizing sources generated in both Star models~: at z=8.5 there are
$\sim$ 4 (resp. 8) times more halos than stellar particles in the 200 Mpc/h
(resp. 50 Mpc/h) and it corresponds to the factor measured in terms of detected HII regions.
Finally, when splitting these distribution in different types of ionized regions
(new, expanding, merging and recombining), they follow the overall
pattern of increase and decline after $\mathrm{z_{\,peak}}$ with differences
detailed in the next section \ref{proportion_HII_regions}. Let us only mention the case of new regions that
also follow the same pattern even though at face value mergers only affect
pre-existing ionized patches. It should be reminded that as reionization progresses the
amount of neutral volume decreases, limiting the possibility of having new
sources in a neutral area. Hence, new regions are also affected by the
overlapping process, albeit indirectly.

\subsection{Number and volume fraction of the different types of HII regions}
\label{proportion_HII_regions}

\begin{figure*}
   \begin{center}
    \begin{tabular}{ccc}
      \includegraphics[width=6cm,height=5cm]{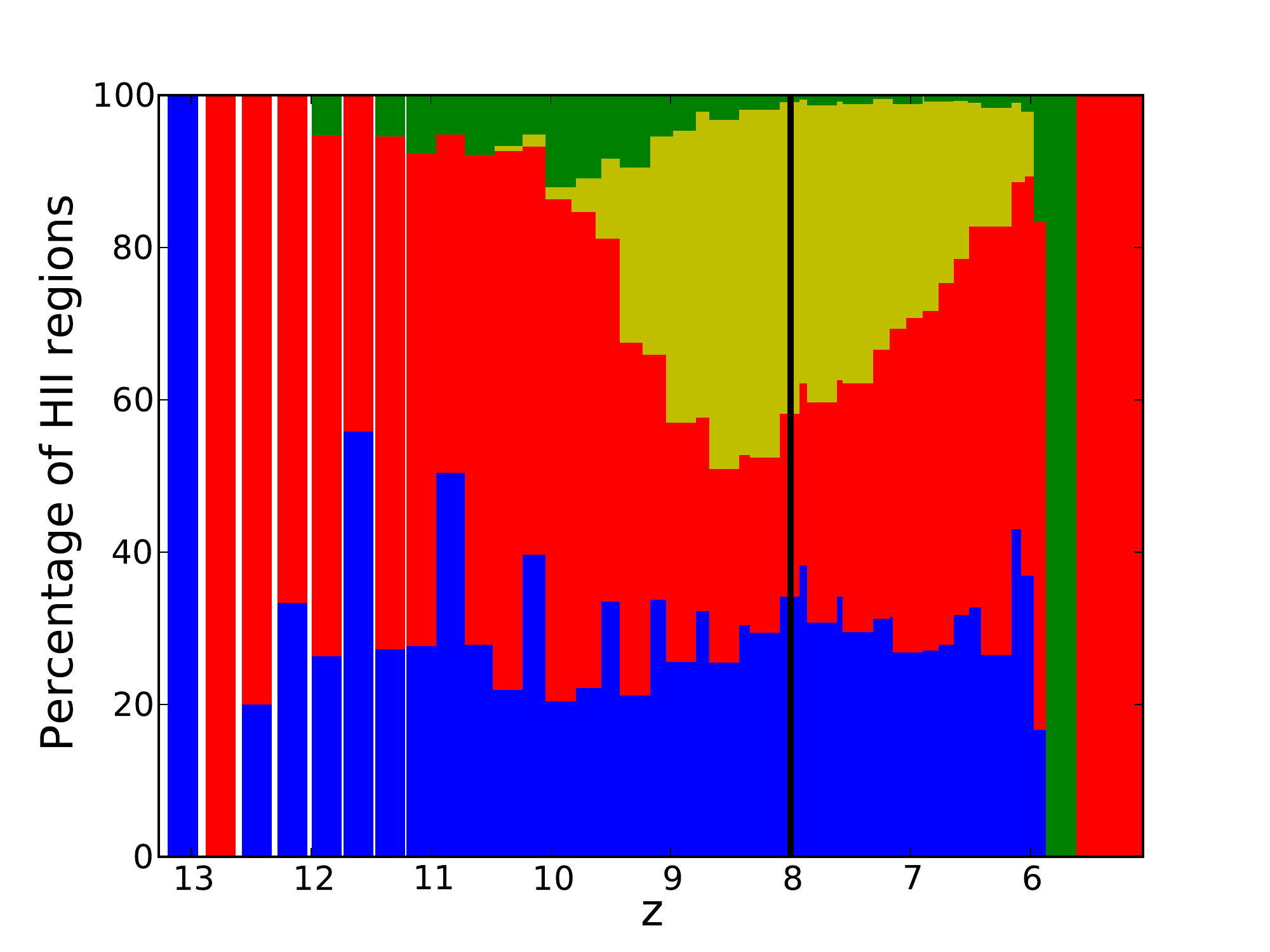} &
	\includegraphics[width=6cm,height=5cm]{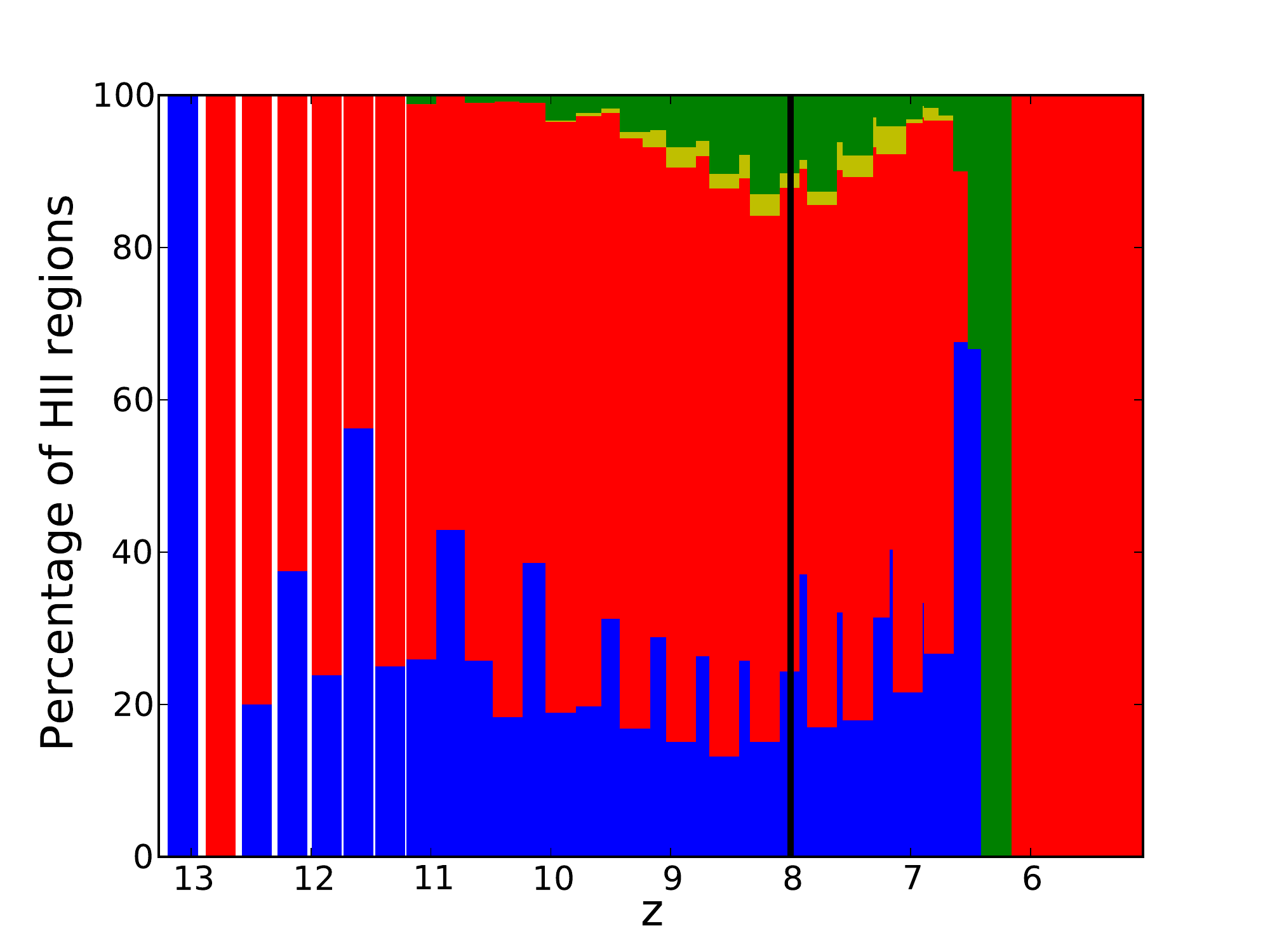} &
	\includegraphics[width=6cm,height=5cm]{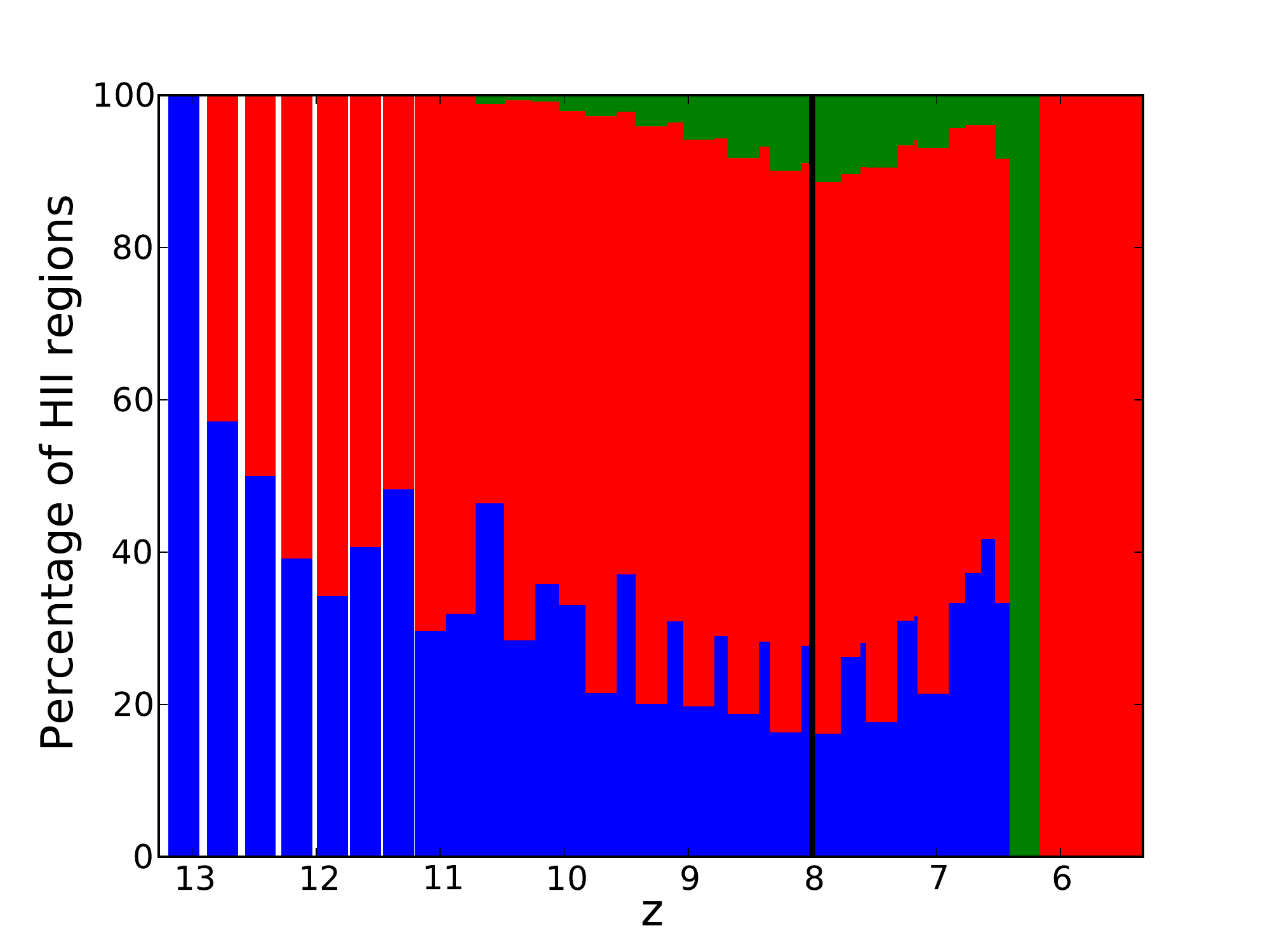} \\
    (a) Boosted Star 200 & (b) Star 200 & (c) Halo 200 \\
       \includegraphics[width=6cm,height=5cm]{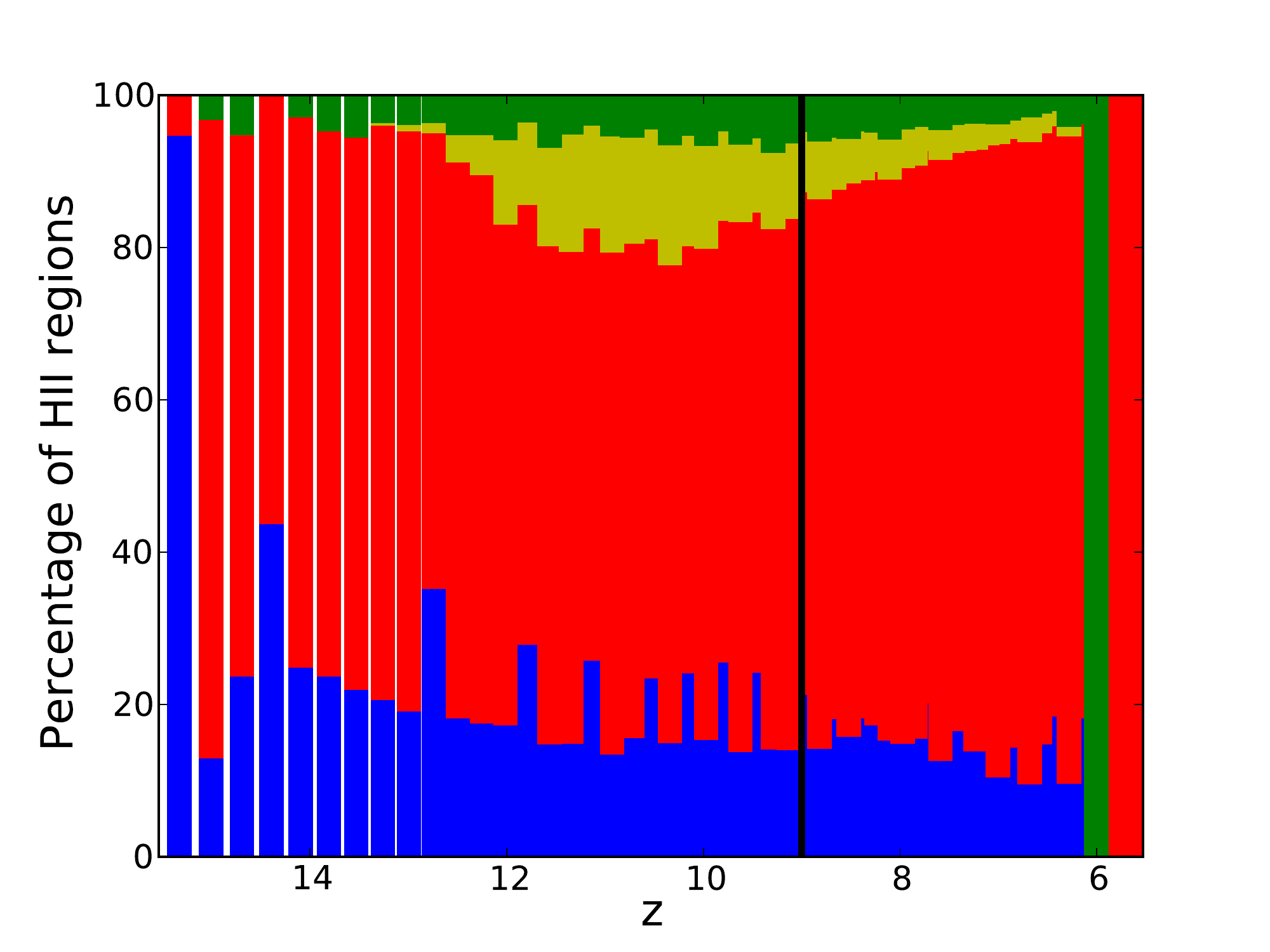} &
	\includegraphics[width=6cm,height=5cm]{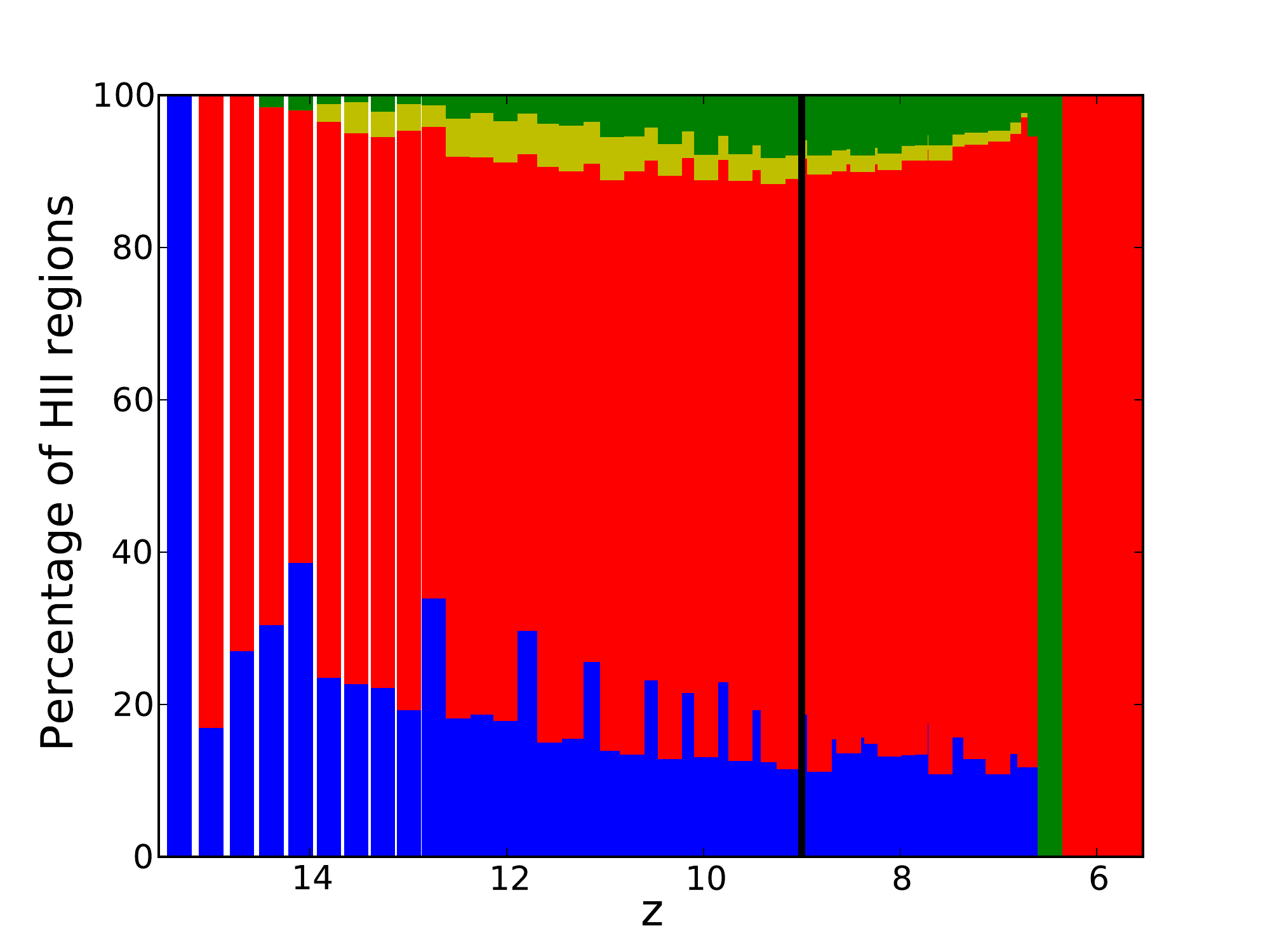} &
	\includegraphics[width=6cm,height=5cm]{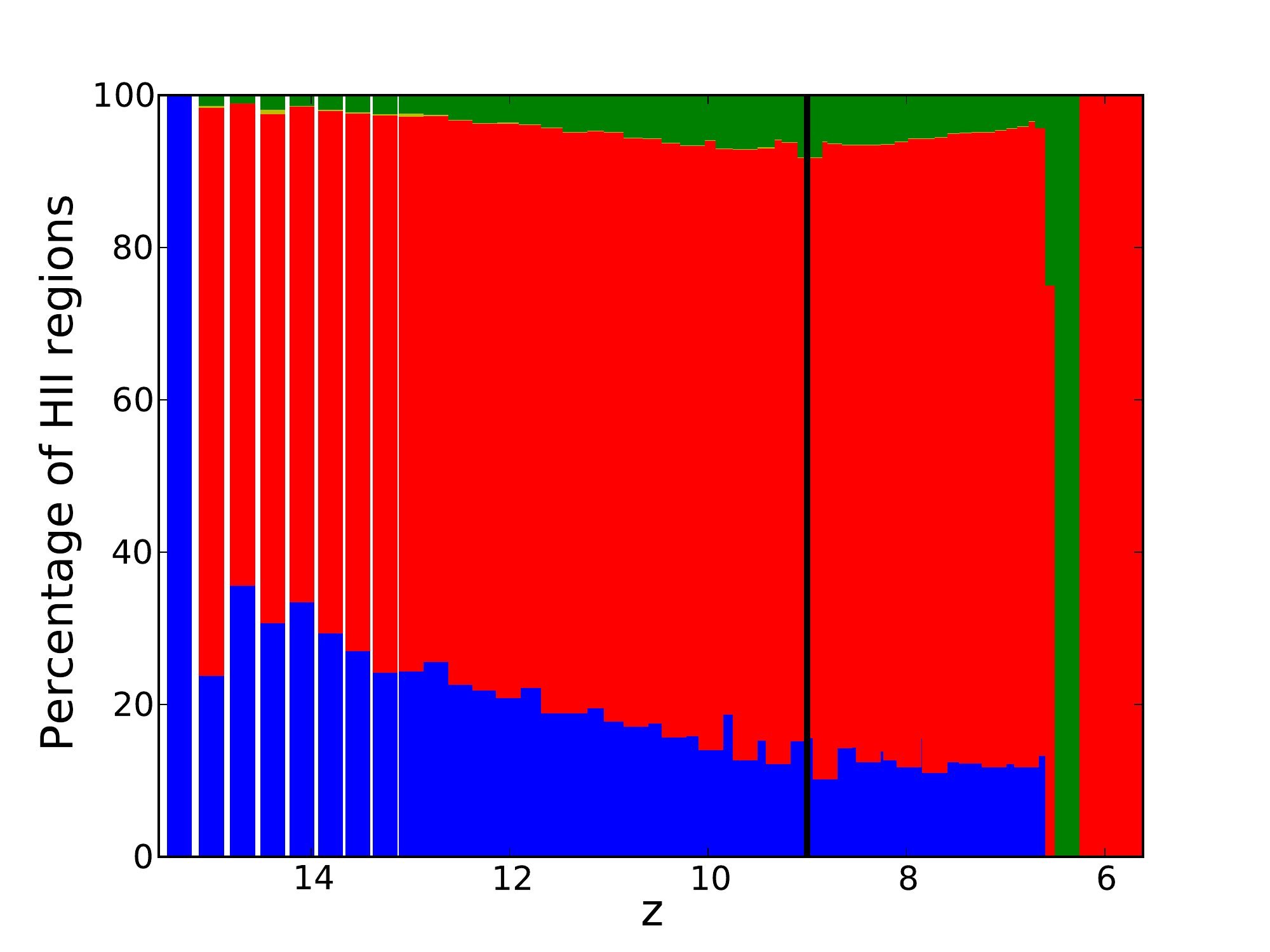} \\
      (d) Boosted Star 50 & (e) Star 50 & (f) Halo 50 \\
\end{tabular}    
  \caption{Evolution of the proportion of each kind of HII regions as a function of redshift for the three models of ionizing sources
and for both boxes of 200 and 50 Mpc/h. Panels (a), (b) and (c) respectively represents the distribution for the Boosted Star, 
the Star and the Halo model for the 200 Mpc/h box, while the panels (d), (e) and (f) are for the same models but for the 50 Mpc/h box.
The colours stands as follow: Blue for the new HII regions, Red for the expanding regions, 
Yellow for the regions that will recombine and Green for the regions resulting from mergers. The black vertical line shows the peak 
of the absolute number of HII regions: $\mathrm{z_{\,peak}}$.}
    \label{proportion}
  \end{center}
 \end{figure*}

\begin{figure*}
   \begin{center}
    \begin{tabular}{ccc}
      \includegraphics[width=6cm,height=5cm]{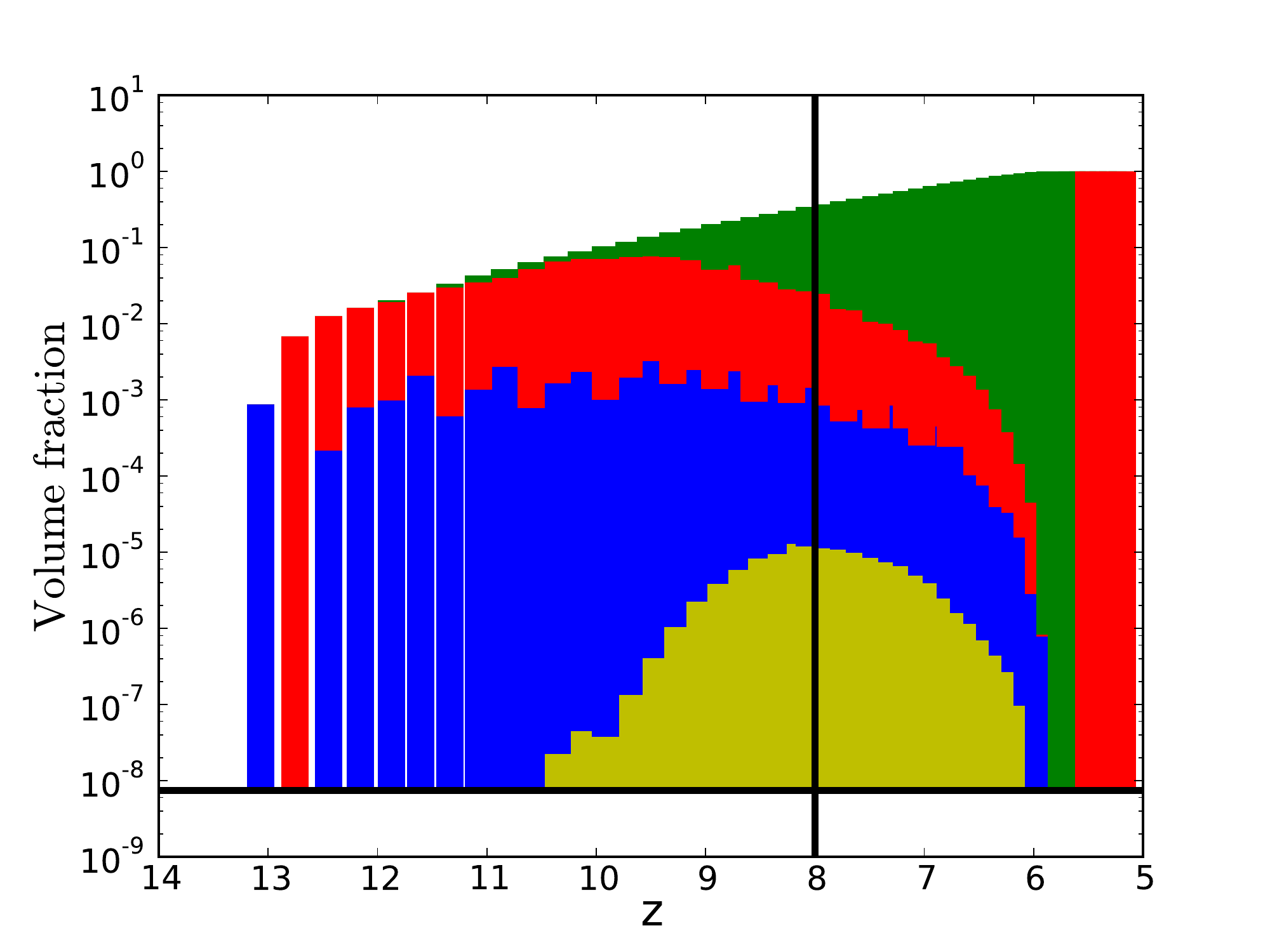} &
	\includegraphics[width=6cm,height=5cm]{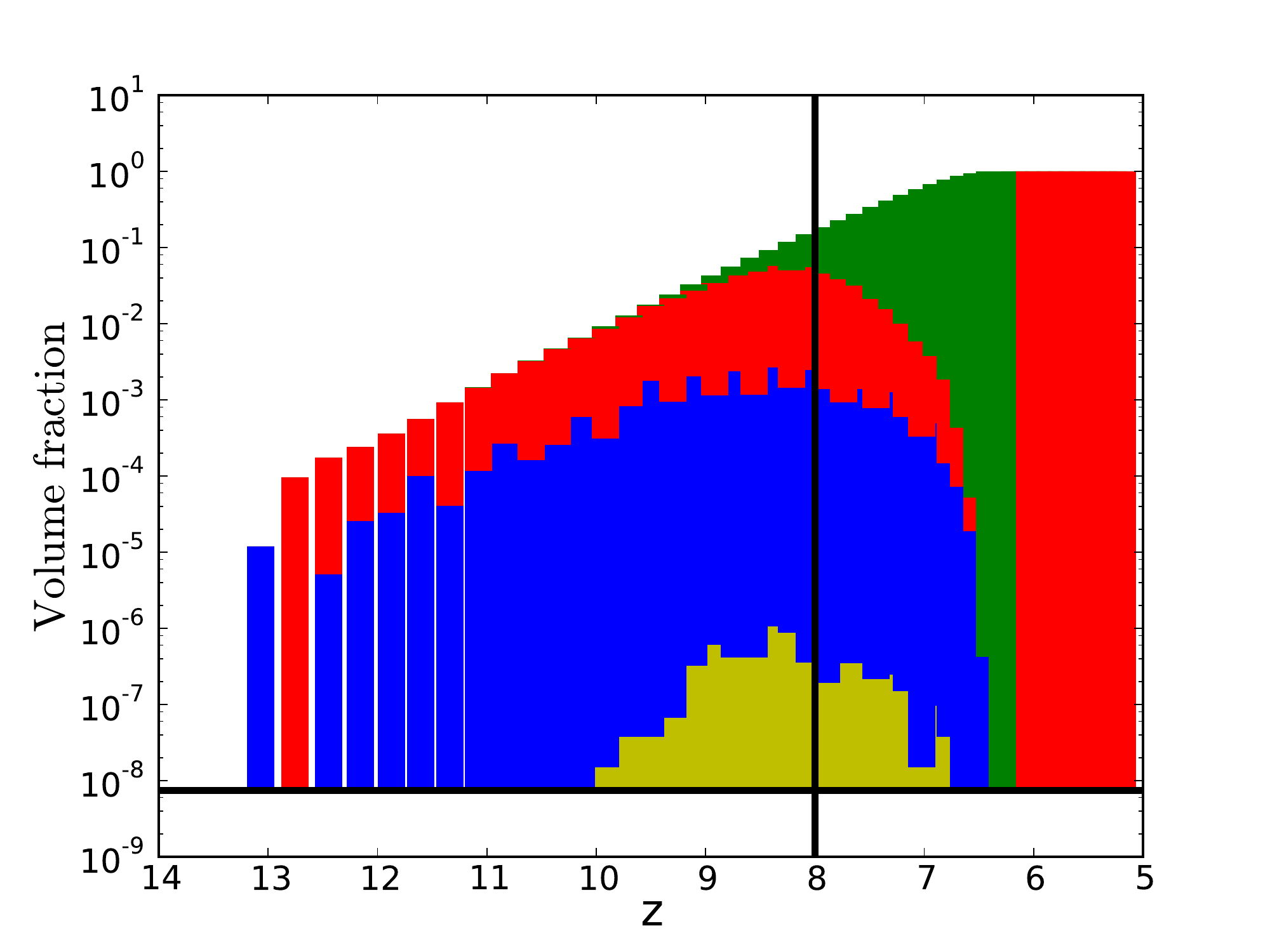} &
	\includegraphics[width=6cm,height=5cm]{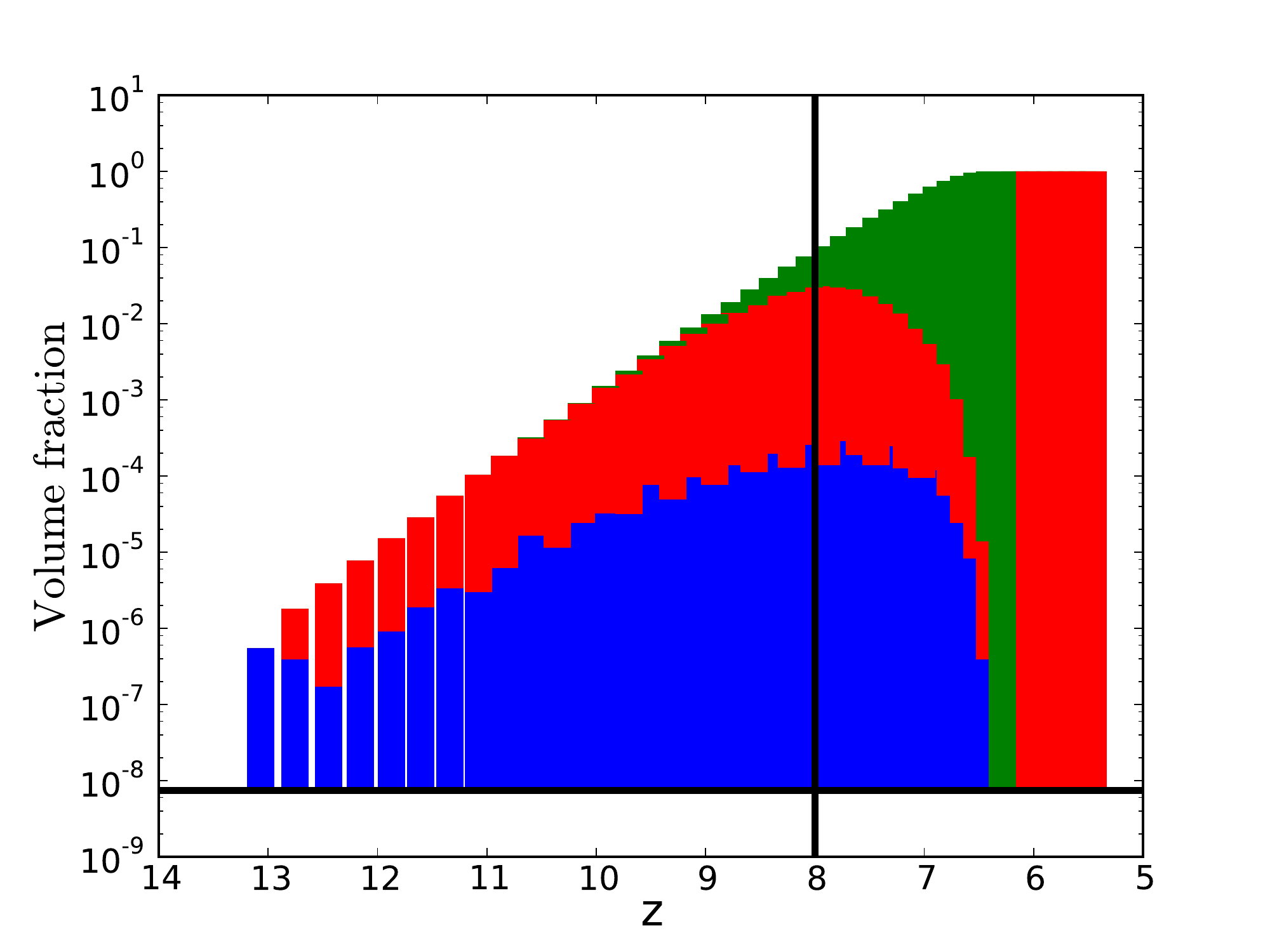} \\
      (a) Boosted Star 200 & (b) Star 200 & (c) Halo 200 \\
       \includegraphics[width=6cm,height=5cm]{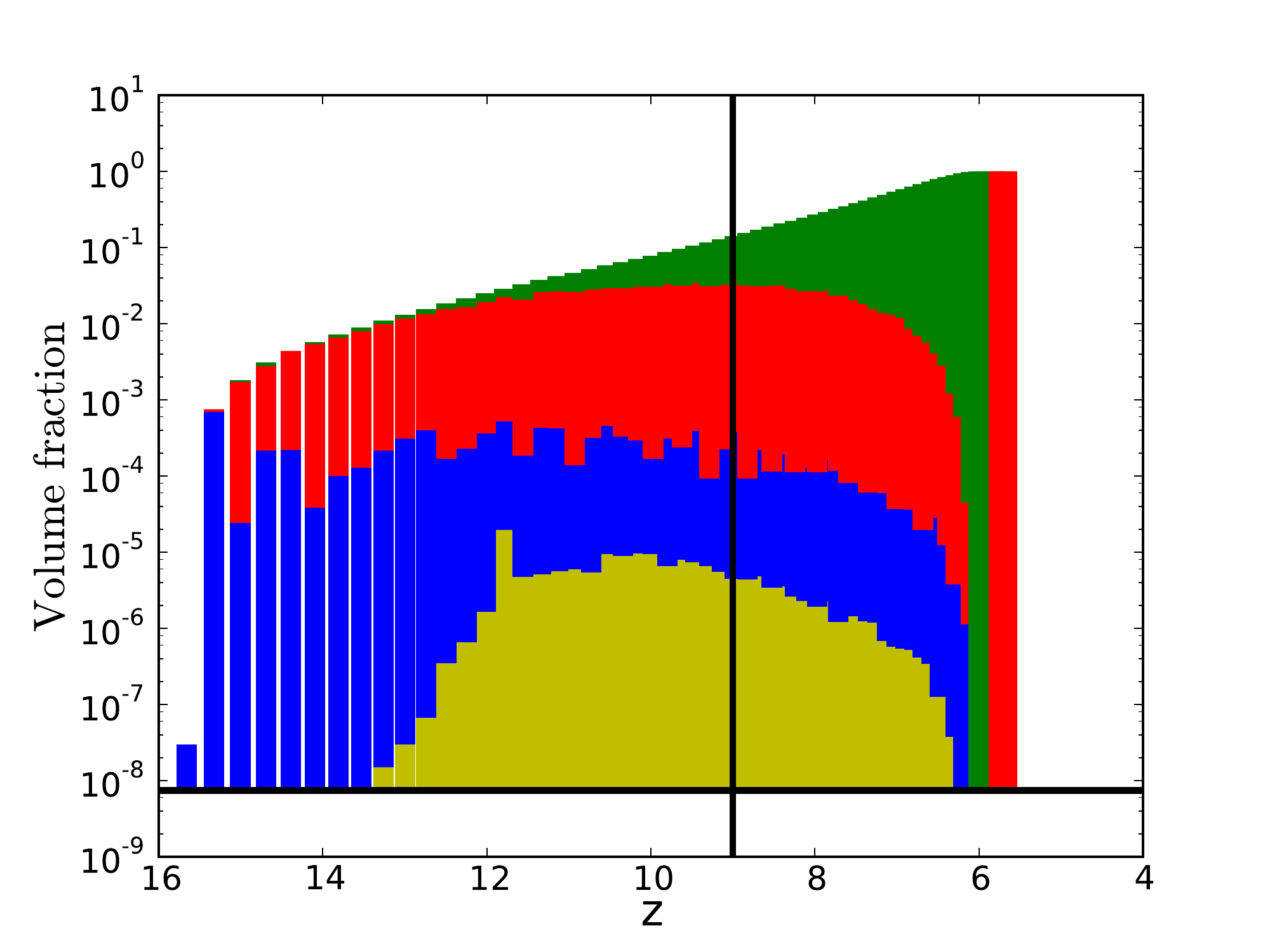} &
	\includegraphics[width=6cm,height=5cm]{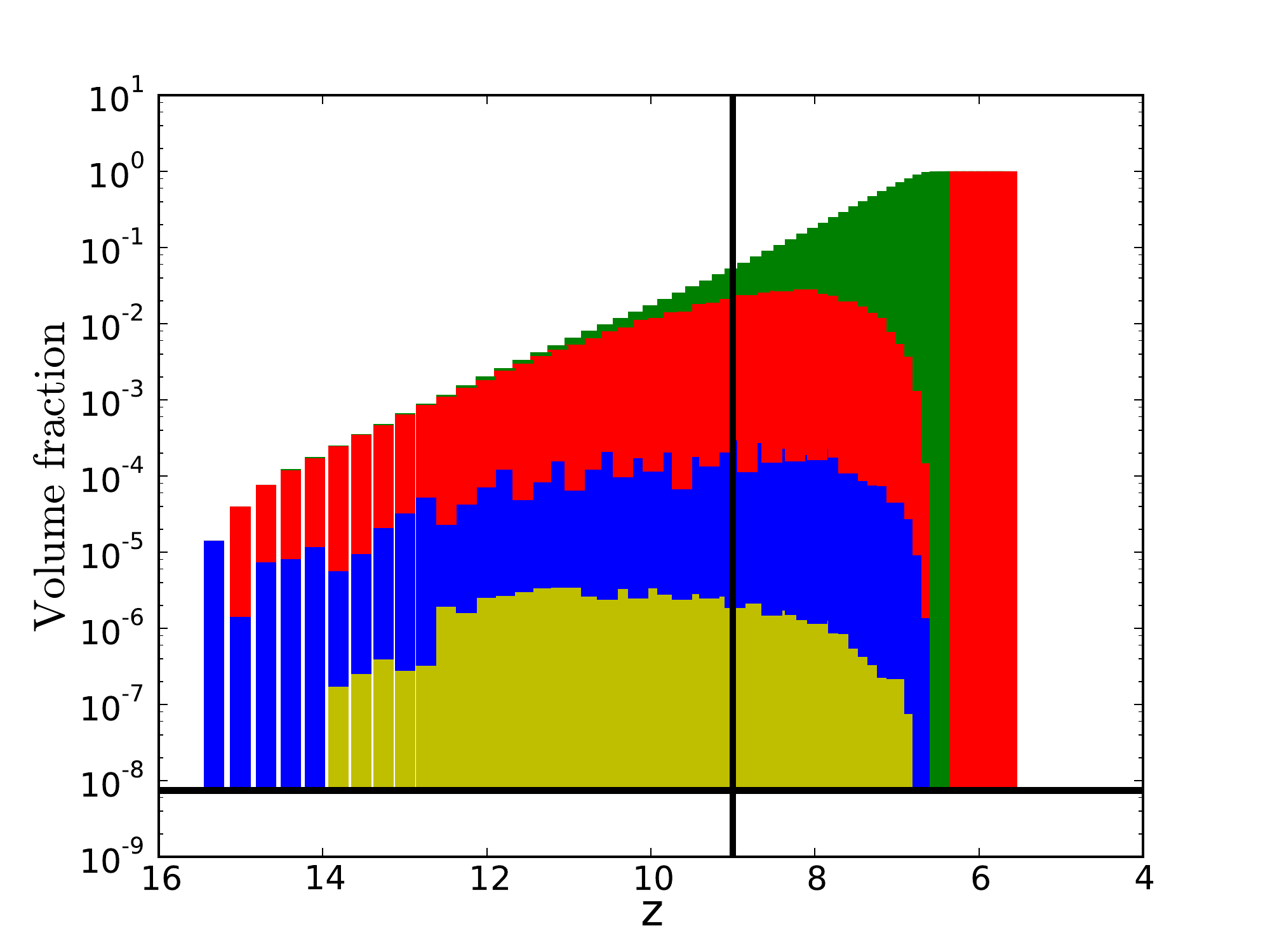} &
	\includegraphics[width=6cm,height=5cm]{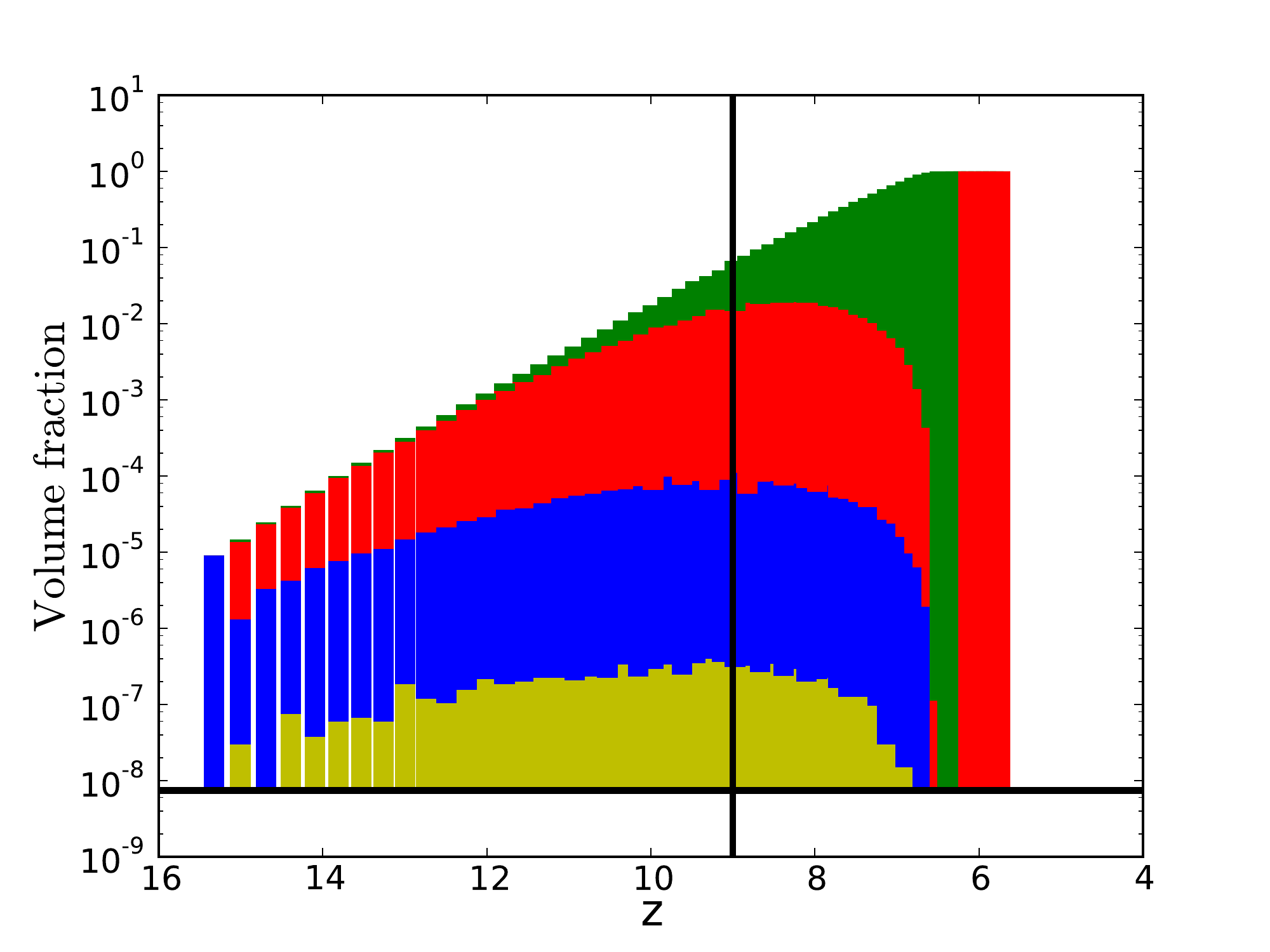} \\
      (d) Boosted Star 50 & (e) Star 50 & (f) Halo 50 \\
\end{tabular}    
  \caption{Evolution of the volume fraction of each kind of HII regions as a function of redshift for the three models of ionizing sources
and for both boxes of 200 and 50 Mpc/h. Panels (a), (b) and (c) respectively represents the distribution for the Boosted Star, 
the Star and the Halo model for the 200 Mpc/h box, while the panels (d), (e) and (f) are for the same models but for the 50 Mpc/h box.
The colours stands as follow: Blue for the new HII regions, Red for the expanding regions, 
Yellow for the regions that will recombine and Green for the regions resulting from mergers. The black vertical line shows the peak 
of the absolute number of HII regions: $\mathrm{z_{\,peak}}$ and the black horizontal line shows the volume fraction of one cell of the grid.}
    \label{volume_fraction}
  \end{center}
 \end{figure*}

To emphasize differences in the histories of the various kind of regions,
Fig. \ref{proportion} shows the evolution of their relative proportion (rather
than their number density) in terms of number and Fig. \ref{volume_fraction}
presents the volume fraction of the whole simulated boxes occupied by each
type of HII region.

\subsubsection{Recombining regions}
\label{recombination_number}
The most striking feature of SB models (and to a lesser extent S models)
  is the presence of recombining HII regions, tracked as connection-less
  branches in the merger tree. Recombination can be driven by two main factors:
  high density features or evolving sources. The evolution of the density is
  common to all experiments and the lack of such regions in H models indicates
  that it is unlikely to be the main origin of these peculiar
  regions. Nevertheless, a hint for this `density' effect can be seen in the larger amount of such
  regions in the S50 model compared to S200, or their marginal presence in
  H50: higher resolution leads to higher contrasts and thus higher
  recombination rates. On the other hand, stellar sources are more prone to variation than
  `halo' sources. First they exhibit some level of stochasticity, hence
  sources that turn off may not be replaced through an efficient renewal
  process in order to sustain some regions, whereas
  halo sources have by construction a continuous emission. It explains the
  lack of such regions in H models and their detection in S and SB
  models. Furthermore SB models have sources with a decreasing individual
  emissivity, leading to a typical scenario where strong early sources produce
  large regions that see their inner ionizing engine becoming progressively
  weaker, or stars being replaced by weaker ones, eventually leading to recombinations. Combined to stochasticity, it can lead to
  regions that are being `dissolved', potentially at several places within one
  region. Increasing the resolution tend to diminish the contribution of this type of regions in SB50
  models: the differential boost applied to source is weaker than in SB200,
  hence decreasing the effect of inner exhausted engine and the stellar renewal is
  more efficient, leading to a production of ionized regions closer to a stationary regime.
Looking at Fig. \ref{volume_fraction}, it is interesting to note that in all the
  cases where they are detected, recombining regions stand for a small fraction of the total ionized volume: between $\sim 10^{-7}$ and $\sim 10^{-5}$ of the total volume of the box for the H50 and the SB200 model respectively.
 Therefore, even if recombining regions are good indicators of the behaviour of the
 source models, it can be seen that they correspond only to an almost negligible fraction in terms of
 volume and can be considered to some extent as marginal.

\subsubsection{Regions in expansion}
\label{expansion_number}

Expanding regions are detected as having one `parent' only and
distinguish HII regions that expand without additional event. In absolute
number (see Fig. \ref{absolute_number}) their number peaks at
$z_\mathrm{peak}$ and their proportion is always dominant until the very end
of the reionization. Incidentally, it shows that the temporal sampling is high enough
to generally track the initial expansion of individual regions. Also, the fact that
such regions are still detected at the later stages suggests that even at the later stages of
the percolation, not all the regions are involved in a global merging
process: some regions still have room to experience stages of `quiet growth'
before eventually be part of the global ionized background.
In terms of volume (see Fig. \ref{volume_fraction}), expanding regions
  are the main contributors to the total ionized volume in every models until
  typically $z_\mathrm{peak}$. This is not surprising since the expanding regions
  are naturally much greater in size than the new regions and are
  always dominant in number (see Fig. \ref{absolute_number}). Once the overlap
start to be efficient, merged HII regions superseed the expanding ones in
volume, as expected.


\subsubsection{New regions}
\label{new_regions_number}

New regions (without parents in the tree) track the formation of new
sites of emission, i.e. star or halos forming in areas distant enough of
pre-existing ionized volumes. Of course, any source that appears within an
ionized region because of clustering won't contribute to this population. 
First, their proportion (and their absolute number) present regular `spikes'. The tree is constructed using a time sampling
different than the sampling used to include sources in the radiative transfer
calculation: the tree sampling frequency is low enough that between two
snapshots at least a new generation of source has been included and new regions
are thus detected. But this frequency is also low enough that regularly, two
generation of new sources were included in the RT calculation between two snapshots, creating bursts
of new regions. 
The presence of these spikes does not affect our analysis and conclusions.

Globally all experiments show an initial gradual decrease of the proportion of
new regions. In the 50 Mpc/h experiments, this decline stays until the end of
the reionization. It is related to a pile-up effect of pre-existing HII
regions: until $z_\mathrm{peak}$ their number increase at a higher rate than
the one of new regions. For instance, regions keep growing without efficient merging
over several generation of new HII regions: it increases the number of
pre-existing regions and  decrease mechanically the proportion of the new ones at a given moment.
After $z_\mathrm{peak}$, the proportion of new regions increases in 200 Mpc/h
experiments and arguably remains constant in 50 Mpc/h ones. By construction
the overall number of ionized sites decreases in this `post-overlap' period, indicating that
among these sites, new regions suffer to a lesser extent of the percolation
process. Hence while the destruction rate of pre-existing HII regions is
important, new stars or halos manage to appear in neutral areas, sustaining
(in the 50 Mpc/h box) or even increasing (in the 200 Mpc/h box) the relative
contribution of the new regions. This increase of the contribution of new
regions seen in S200 and H200 is indicative that the destruction of
pre-existing regions is more sudden than at higher resolution: it was hinted
in Fig. \ref{absolute_number} where the post $z_\mathrm{peak}$ evolution is
smoother at high resolution or in Fig. \ref{xmoy_vs_z} where the average
ionized fraction exhibits a sharper evolution at reionization. Regarding this
sharper evolution, let us emphasize that a more accurate tuning of the source
emissivity could have increased the matching of the reionization history of
the H and S models and may be the origin of this difference between high and
low resolution. It may also be the result of a different behaviour in the
propagation of fronts at low resolution (with faster fronts and weaker
shielding), leading to a more radical percolation in the 200 Mpc/h simulations
than in the 50 Mpc/h boxes. Finally, let us mention that the presence of a
strong recombining component in the SB200 model lead to a totally different
evolution of the new regions contribution: it remains broadly constant at all
redshifts with a weak dip at $z\sim10$. Without this recombining regions, it
would also exhibit a rising relative weight of new regions, albeit starting
much earlier than $z_\mathrm{peak}$.

Regarding their volume fraction (see Fig. \ref{volume_fraction}), these newly
  detected regions only represent a maximum of $\sim 10^{-4}-10^{-3}$ of the
  total volume even if there is a sizable number of this kind of
  regions detected. This is understandable because of their very nature: new
  regions are by definition expected to be small, since they just appeared and
their number cannot compensate this effect. Interestingly, the total volume
occupied by the new regions remains broadly constant until overlap~: it
indicates that even though sources increase in number with time(and thus the potential
number of new regions) it competes with the gradual decrease of available
neutral volume, these two effects `conspiring' to keep constant the volume imprint of
the new patches on the network of HII regions.

\subsubsection{Regions resulting from mergers}
\label{merger_number}

These regions have more than one parent and result therefore from the merger
of several regions. They track the coalescence of pre-existing regions into
larger ones and eventually track the final overlap into a single large HII
region at the end of reionization. This last stage can bee seen in all models
where there exist a late snapshot where $100\%$ of the detected region result
from mergers, an indication of the ultimate merger. S and H model present
similar evolutions for the merger populations. Their proportion is
`coincidentally' maximal at $\mathrm{z_{\,peak}}$: as the number of ionized patches gets larger, a larger
fraction of them is involved in mergers, indicating a `crowding' or clustering
effect where a smaller volume is available for expansion as the number of HII
regions increase. Later on, for $z<\mathrm{z_{\,peak}}$, the proportion of mergers
decreases indicating that only a subset of the ionized regions do actually
merge, potentially only one that would phagocyte the others, reducing the
overall number of individual regions until the end. Hence the peak of mergers
fraction could be seen as the rise of one or several dominant regions and this
rise appears when the number of individual region is maximal.

Comparing the two
resolutions, it can be noted that mergers are concentrated over a smaller
range of redshifts at low resolution whereas merger can be detected at a
significant level during the whole experiment in the S50, H50 and SB50 models.
Firstly, it can be the consequence of the slightly more extended history of
reionization in 50 Mpc/h simulations, already mentioned regarding the $\langle
x (z)\rangle$ trends (see Fig. \ref{xmoy_vs_z}). Furthermore, sources and individual regions are more
numerous at any time and in a smaller volume, in favor of a more generalized
contribution of mergers over a large range of redshifts.

Finally, it should be noted that the SB200 model presents an early peak in the
fraction of merger, at $z\sim 10$ instead of $z\sim 8$ for all the others
experiment, with an almost zero contribution later on, even though the
absolute number of regions manage to decrease at some point. It suggests that the onset of the global percolation started
earlier, due to early large HII regions induced by the large initial boost of
this model, and created a single region that monopolize the merger process. 
Interestingly, in this case $\mathrm{z_{\,peak}}$ occurs later, indicating
that while this main region grows, there remains a significant amount of neutral volume
to host the apparition of new regions. 

Finally, Fig. \ref{volume_fraction} clearly indicates that these merger regions
 dominate the fraction of the ionized volume for $z\le \mathrm{z_{\,peak}}$. It was not obvious given that their number
  density is always the lowest contribution (cf. Fig. \ref{absolute_number}).
It is indicative that few of these regions produce a network that dominate the ionized volume and/or that such regions are
individually very large, as one would expect from them being the result of
several mergers. This point is assessed in greater detail in the next sections.


\subsection{Sizes of HII regions}
\label{size_HII_regions}

\subsubsection{Sizes distribution with redshift}
\label{size_distribution}  

\begin{figure*}
   \begin{center}
    \begin{tabular}{ccc}
      \includegraphics[width=6cm,height=5cm]{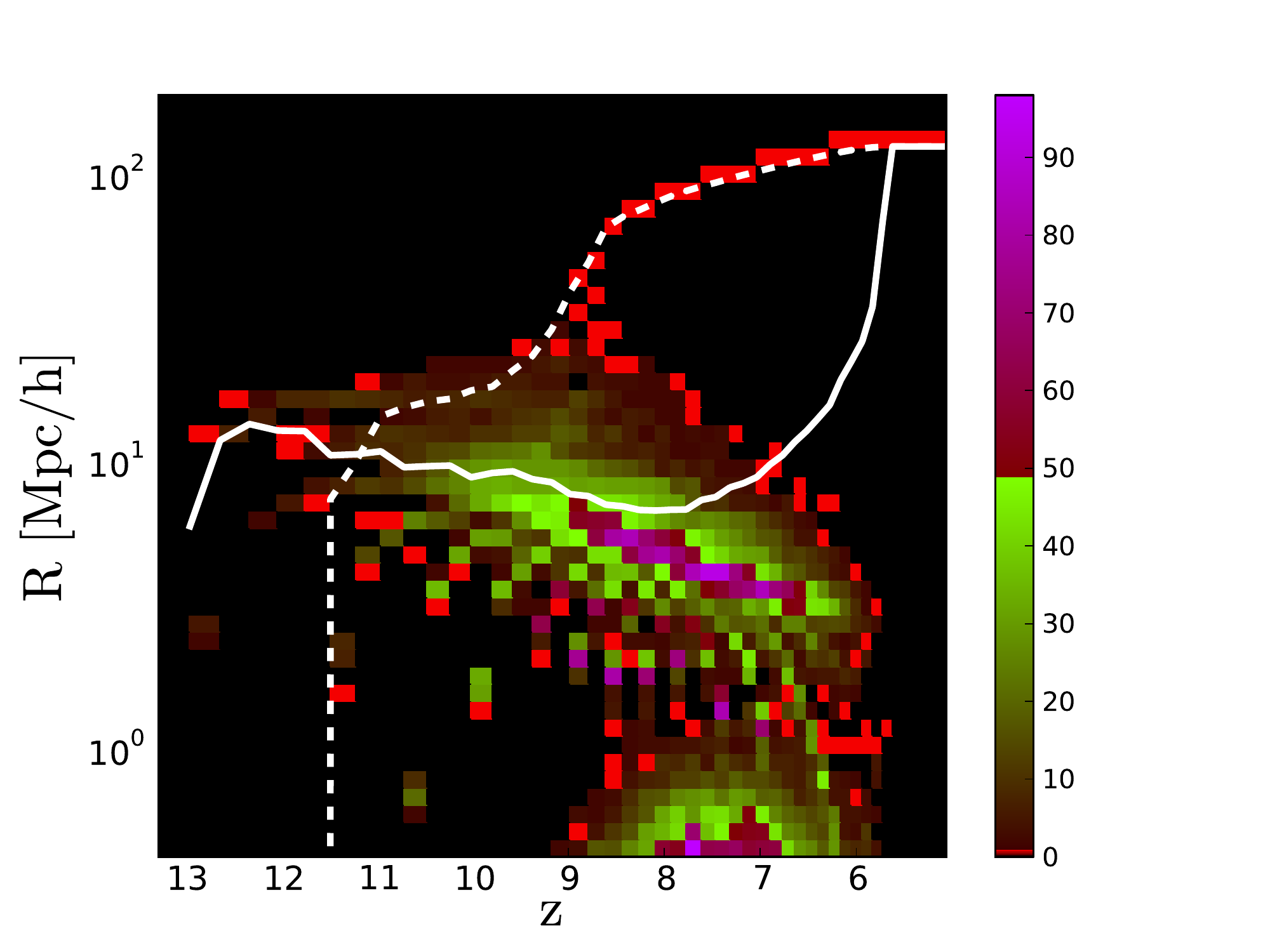} &
      \includegraphics[width=6cm,height=5cm]{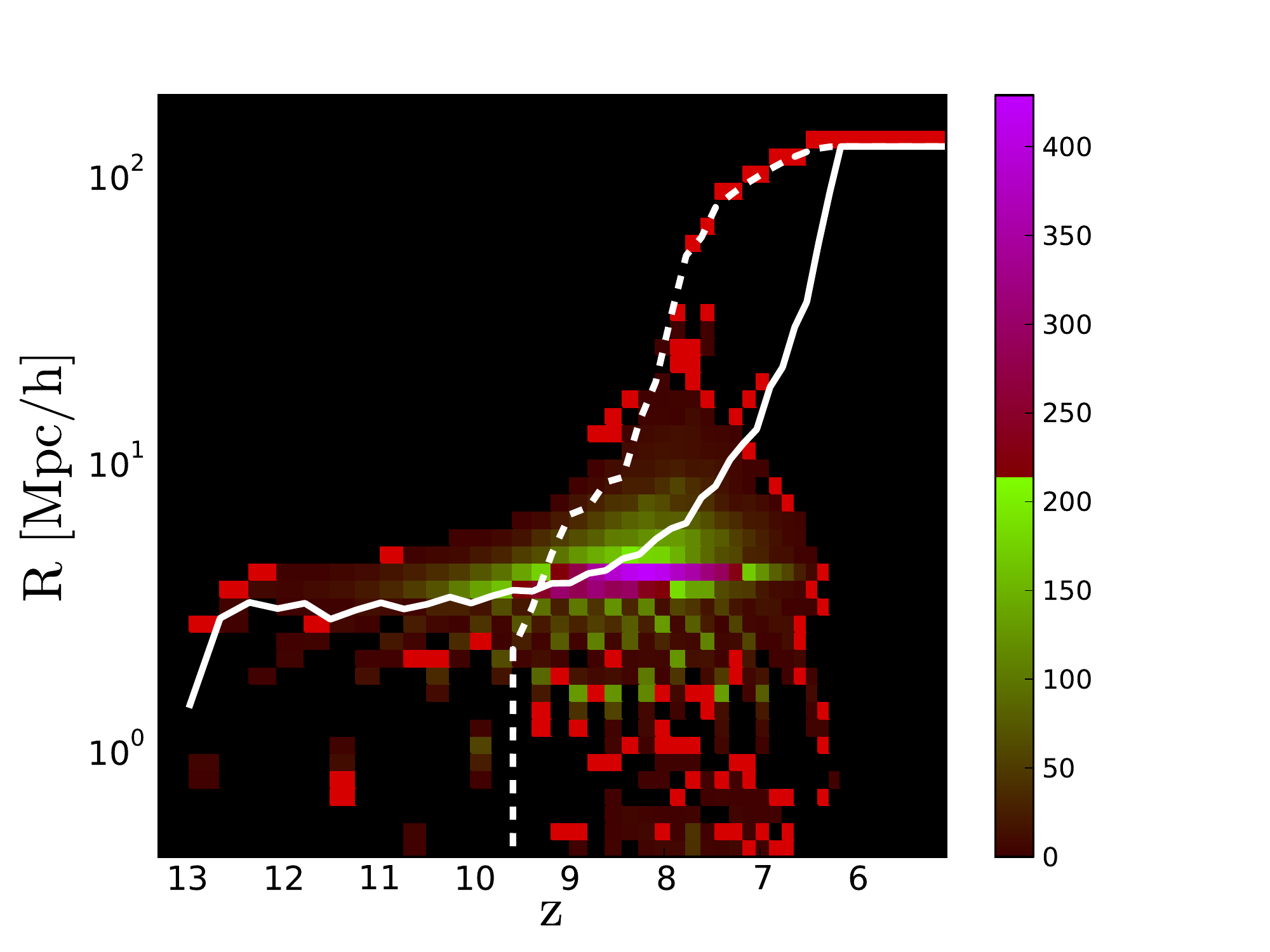}&
      \includegraphics[width=6cm,height=5cm]{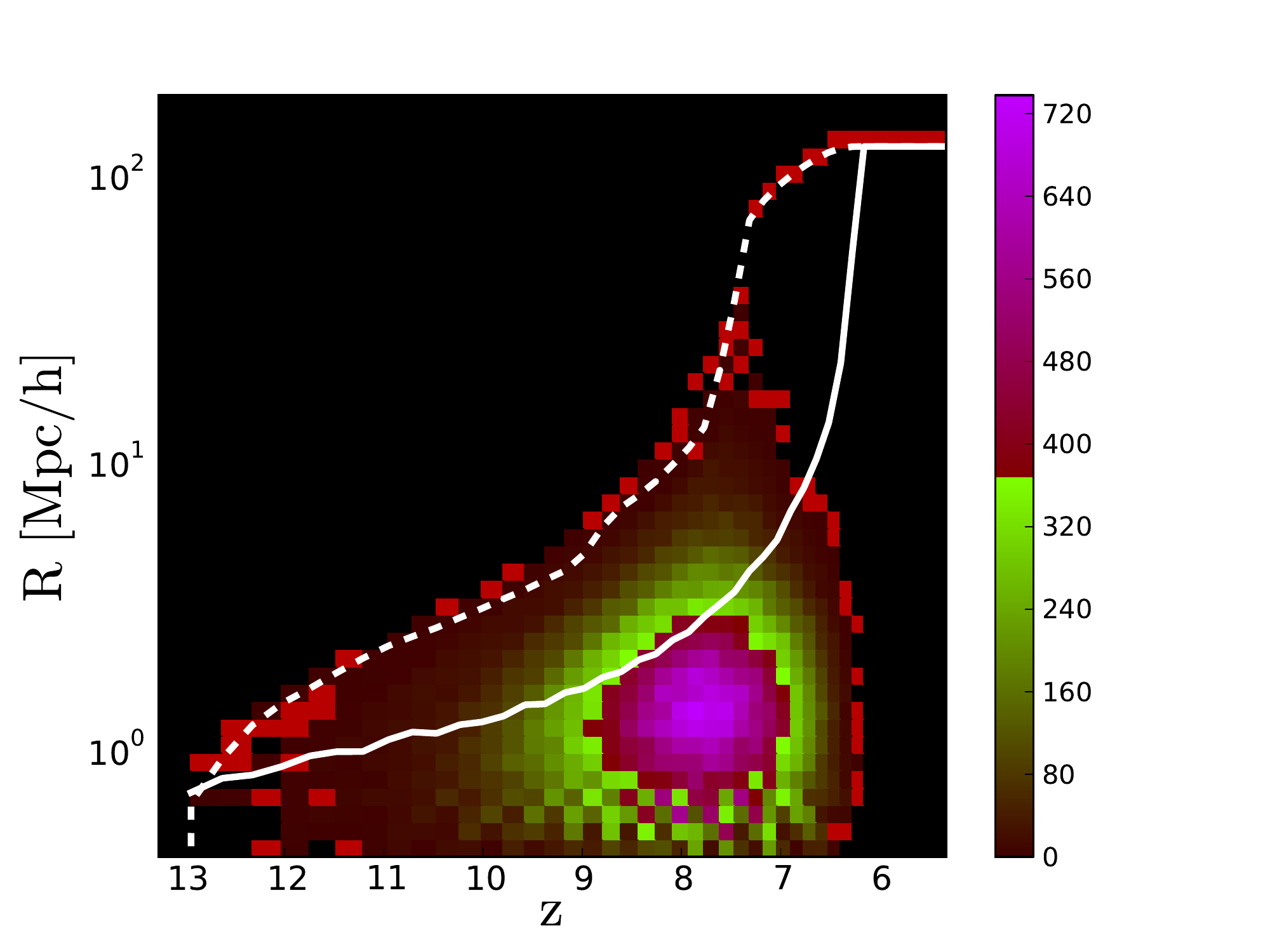} \\
       (a) Boosted Star 200 & (b) Star 200 & (c) Halo 200 \\
      \includegraphics[width=6cm,height=5cm]{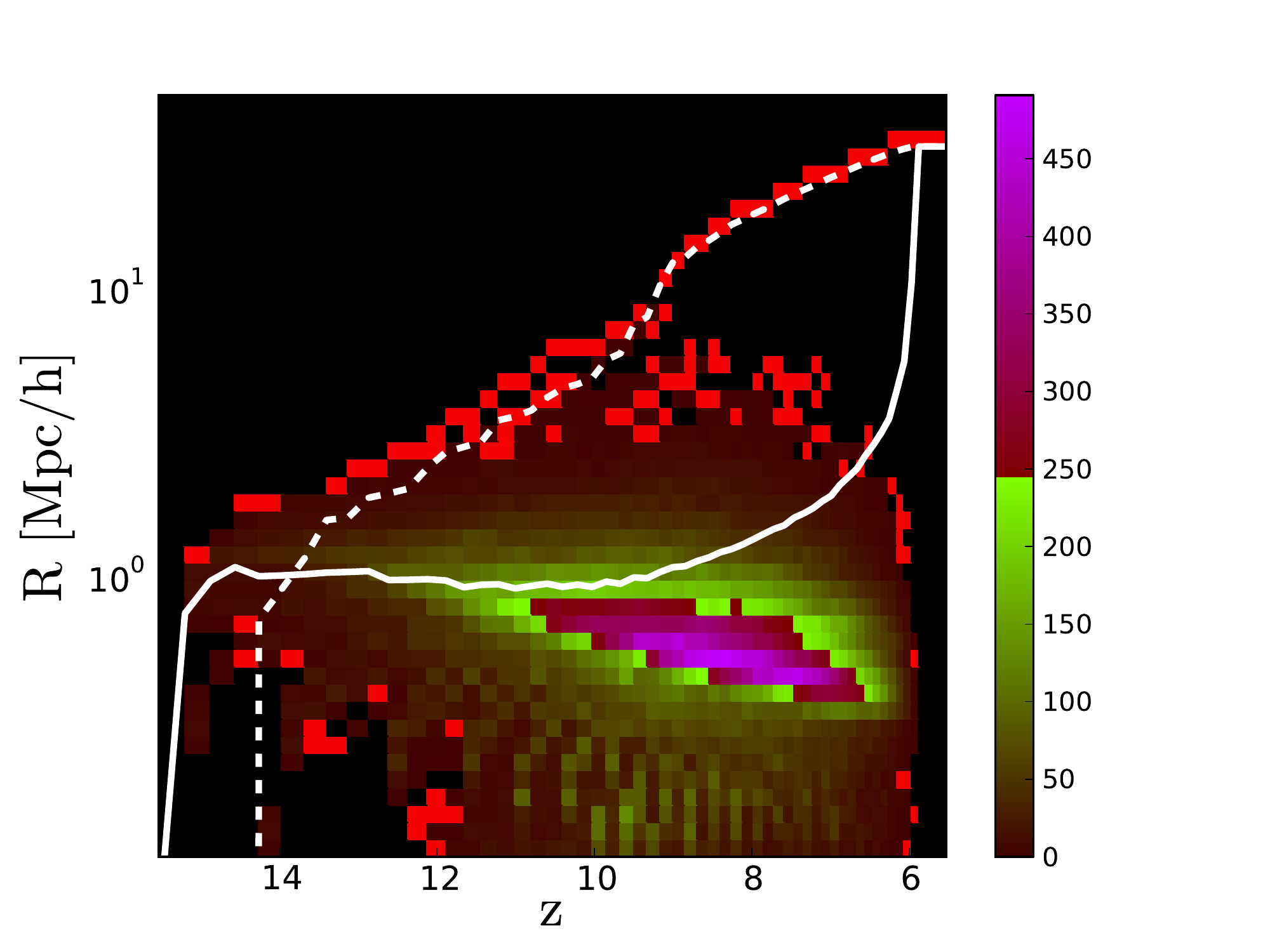} &
      \includegraphics[width=6cm,height=5cm]{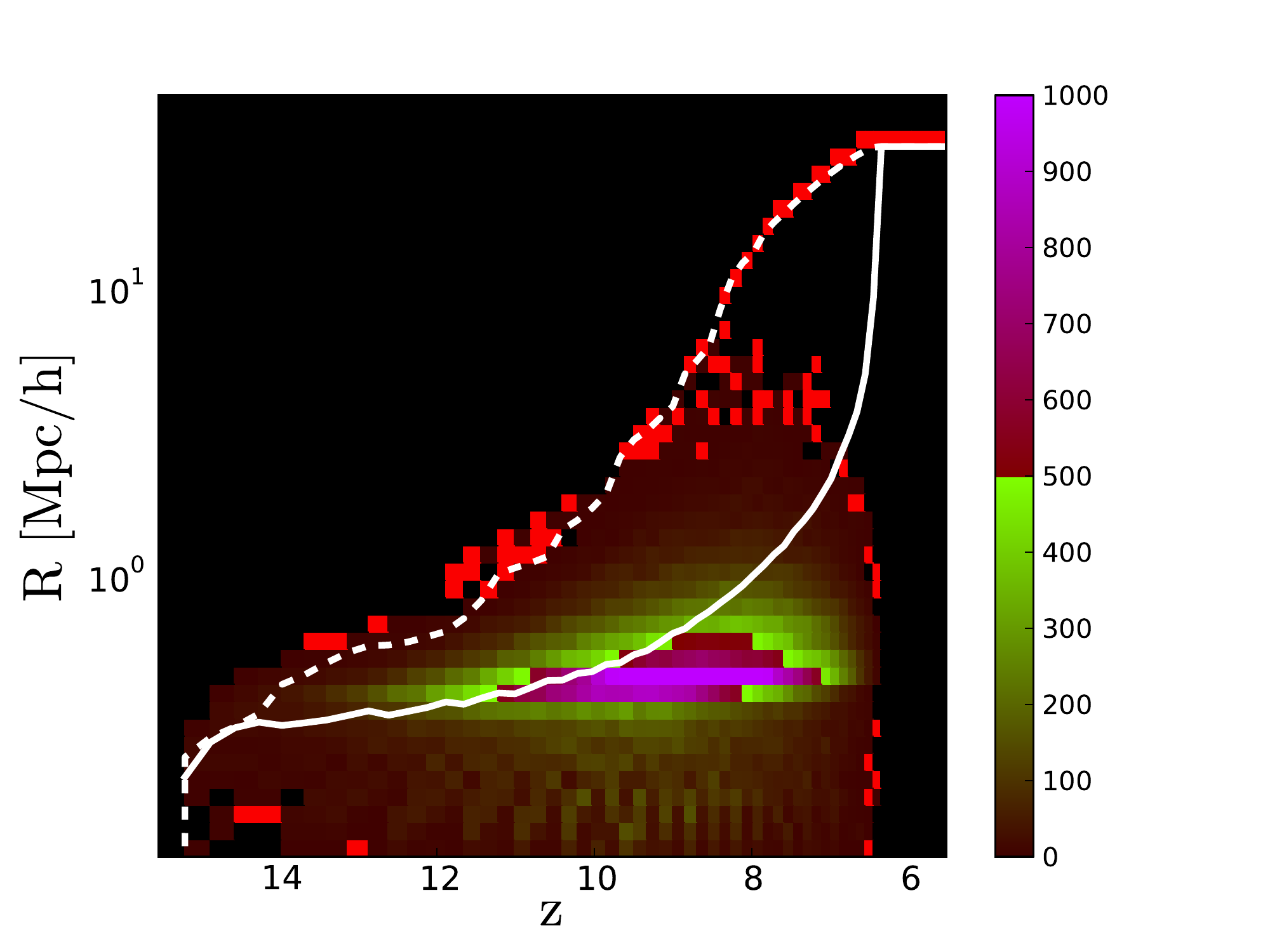}&
      \includegraphics[width=6cm,height=5cm]{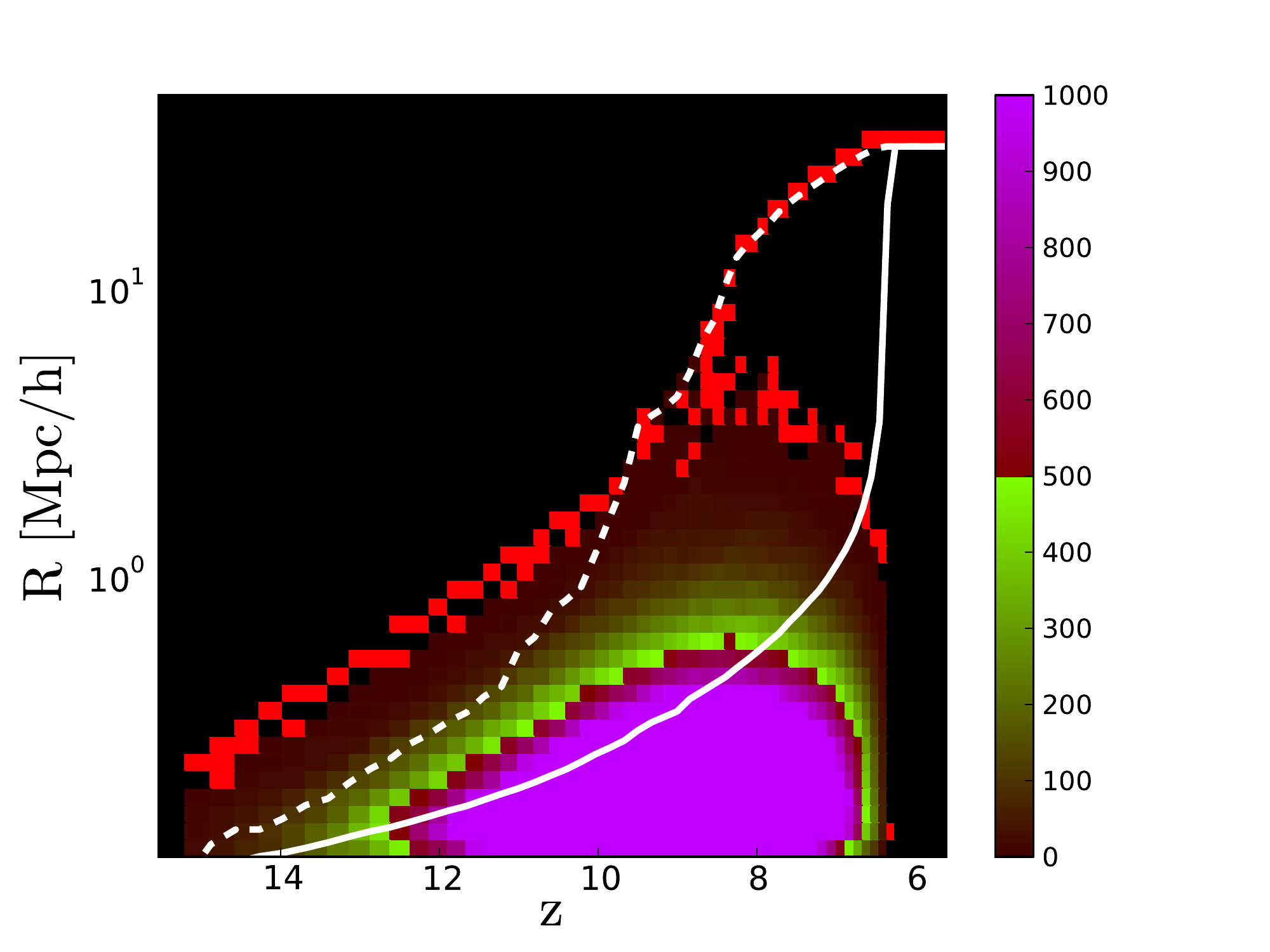} \\
     (d) Boosted Star 50 & (e) Star 50 & (f) Halo 50 \\
\end{tabular}    
   \caption{Evolution of the HII regions radius distribution as a function of redshift for the three models of ionizing sources formation and
for both boxes of 200 and 50 Mpc/h. Panels (a), (b) and (c) respectively represents the distribution for the Boosted Star, 
the Star and the Halo model for the 200 Mpc/h box, while the panels (d), (e) and (f) are for the same models but for the 50 Mpc/h box.
The color code stands as follows: the brightest red cells represent the
location in the distribution populated by \textit{a single} HII region, other
red tones up to the brightest green cells spans distribution densities from a
couple HII regions up to a value corresponding to half the maximal value of
the distribution. Finally the purple tones (from the darkest to the brightest)
encode the maximal values.  The evolution of the average radius and the
evolution of the radius of the main region are shown respectively with the solid and dashed white lines.}
    \label{rayon}
  \end{center}
 \end{figure*}

Fig. \ref{rayon} shows the evolution of the distribution of the HII regions radii 
at each instant as a function of redshift for the 200 and 50 Mpc/h box. 
Like \cite{2011MNRAS.413.1353F}, we compute the volume of the HII region and then 
we derive the effective radius corresponding to a sphere of equal volume through the following expression $R=[3/(4\pi)V]^{1/3}$.
The white dashed curve shows the radius evolution of the last HII region which remains at the end of the simulation.
With the help of the merger tree we follow this region back in time and calculate the radius of its main progenitor
at each instant. In addition the solid white line represents the evolution of the average radius of HII regions as a function of redshift.
The color code in the distributions is made so that the brightest red cell
contains only \textit{a single HII region}. Other red tones 
up to the brightest green designate cells 
populated by a couple HII regions up to a value corresponding to half the
maximal value of the distribution. Finally the purple tones
(from the darkest to the brightest) encode the largest values of the distribution.

Typically, each distribution seems to trace the underlying ionizing source
prescriptions related to the model  considered.  The Star model has a constant
star emissivity  with redshift.  This  is reflected in the radii distributions where
ionized regions are concentrated with a constant radius about $r \sim 4$ Mpc/h
over the whole range of redshift in the 200 Mpc/h box.  We find a gap at about
$r \sim 2$ Mpc/h  in the size distribution where there are  only a few regions
with  radii under this  constant radius.   This would  indicate that  the time
sampling used here cannot entirely capture the fast tracks followed by the HII
regions in the  radius-redshift space with strong inner  sources.  This effect
is attenuated for the 50 Mpc/h box (that contains typically weaker sources and
 HII regions with smaller growth rates) where we observe a more continuous
distribution even  if there remains  a typical cutoff  radius of $r  \sim 0.4$
Mpc/h, with smaller  occupation numbers at low values. 

On  the other hand, the
Halo model  implies that each  of them has  an emissivity proportional  to its
mass.  As these  masses cover a large  range, we find a large  range of radii
for the resulting  HII regions in the distribution of the  200 Mpc/h box.  The
regions are concentrated in larger intervals of radii than in the other models
that typically  trace the underlying  mass range of  halos.  The shape  of the
distribution is almost the same when we consider the 50 Mpc/h experiment, with smaller
regions   as  expected  since   smaller  halos   are  available   at  higher
resolution.  

Finally,  the  boosted  Star  model  has a  boost  for  the  star
emissivity that decreases with  redshift.  Indeed the distributions show, at early times, large
HII regions detected without small counterparts.  This gap would be the result
of the very  powerful boost for ionizing sources at  high redshift combined to
the time sampling  of the simulation which allows to  detect only regions when
they have  a large  radii early in  the reionization.   Then we can  observe a
decreasing gradient for the typical radius of the regions as the boost for the
emissivity   decreases  with   time.   Surprisingly   we  find   a  bimodal
distribution from a  redshift of $z \sim 9$ until the  reionization at $z \sim
6$.  Some HII regions are concentrated with radii below or under $\sim1$ Mpc/h
with a gap  in the distribution for  this value. To a lesser  extent, the same
effect could  be seen  in the Star  200 model.  This regions with  radii under
$\sim1$ Mpc/h  could be the recombining  regions that we have  found in figure
\ref{proportion}.  This could also be combined with the fact that around these
redshifts the  boost emissivity falls down  under a certain  value.  Then, the
time sampling of the simulation would become fine enough to detect some new or
expanding regions  with these radii.  We  will further investigate this in the
following section \ref{size_distribution_contour}. For the SB50 model, this
bimodal  distribution  disappears:  a  weaker boost  amplitude  and  variation
combined to a more stationary production of stars promotes a better tracking
of regions sizes as they grow and reduce the contribution of recombination, as
seen earlier.

Considering again the S and H  models, for both box sizes, it is interesting to
note that  the S model  produce a  truncated version of  the H one.  Above the
cutoff radius (0.4 Mpc/h (resp. 4 Mpc/h)  in the S50 (resp. S200)) the S and H
distributions are quite  similar. At smaller radii, S  has less objects than H. 
It is indicative that there is a scale above which the
clustering  of halos  combined to  their own  mass-proportional emissivities
produce a similar radii distribution to the one  provided by the clustering
of stellar sources  and their own constant emissivity.  At smaller scales, the
stellar sources  are too  scarce to reproduce  the regions created  by halos,
resulting in overpowered individual sources with large radii. At larger scales
the two approach produce equivalent size distributions and with an appropriate
calibration,  hydrodynamically  created sources  can  match  the halo result. 
It should  be noted that this typical scale  appears at smaller values
in S50 models as sources are  more numerous and are less prone to stochasticity
and therefore converges toward the halo model behavior on smaller volumes.

Finally, all the experiments present a single main HII region that dominates in size.
It appears at $z \sim 8$ in the S200 and H200 models and much earlier ($z \sim
9$) in the SB200. As suggested by the previous study on merger populations, it is likely the consequence of the boost that creates early
large regions that merge at high z to give the dominant one. Interestingly the
dominant region of the late reionization is not dominant in size at every
redshift: this is indeed the case for the H200 model and to some extent for
the S50 model too, but in all the other case this region started as a non
special one inside the population, at least from the `radius' point-of-view. Furthermore, the first progenitor of the dominant region is among the
very first ionized regions of the H200 and H50 models but appears later in the
S and SB experiments. It is noteworthy that even though we argued that S and H
model produce similar populations (for regions larger than the cutoff radius)
it seems that for individual cases, differences can persist, especially for the
buildup of the dominant region. As we will see in next sections, the channel
through which this specific region grows (through expansion rather than merger) also differs as hinted by
the difference between the smooth evolution of the dominant region of H models
and the sharp, kinked and late rise of its equivalent in e.g. the S200 model.

\subsubsection{Radius distribution of the different kind of HII regions}
\label{size_distribution_contour}

\begin{figure*}
   \begin{center}
    \begin{tabular}{ccc}
      \includegraphics[width=6cm,height=5cm]{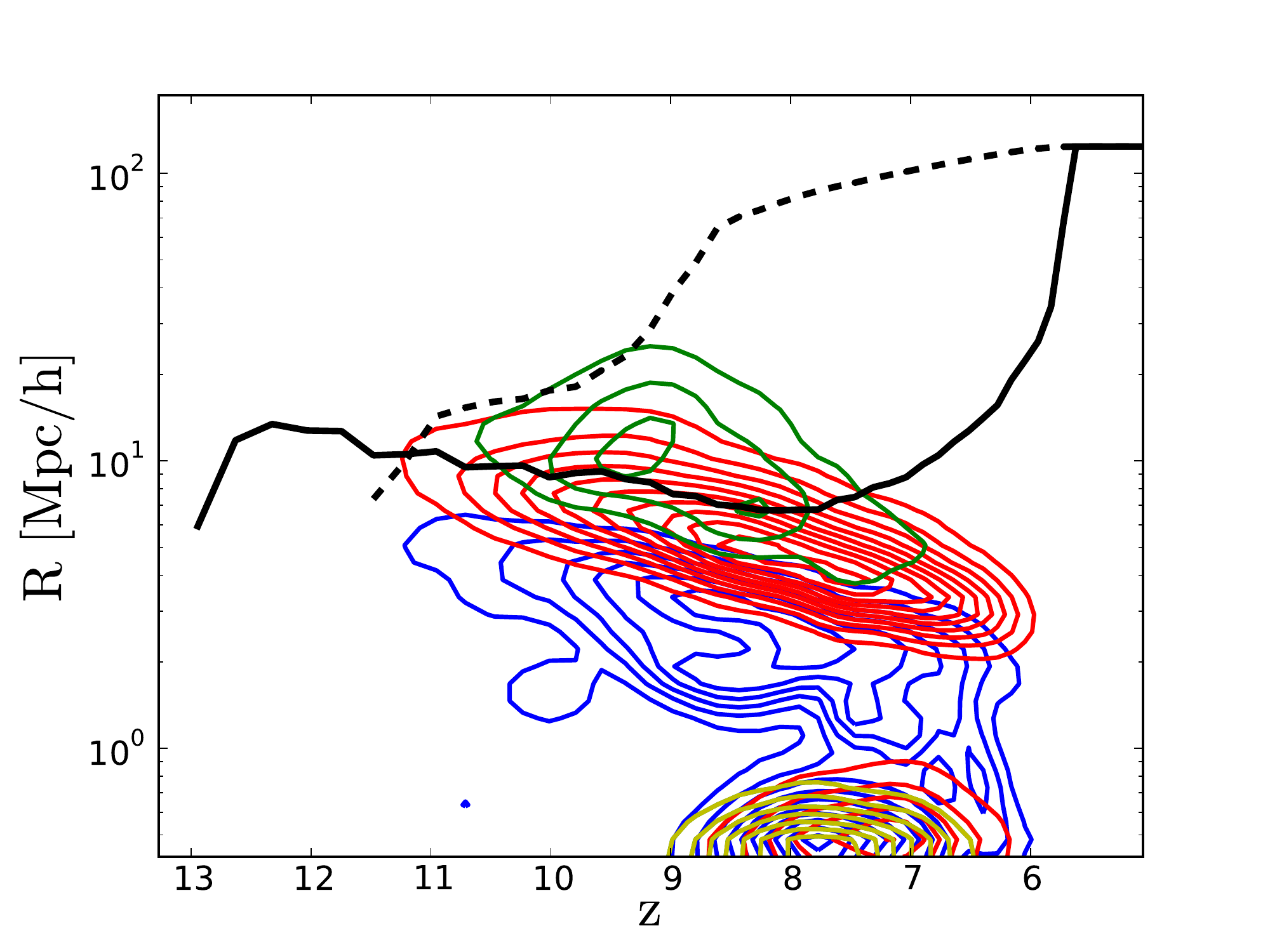} &
      \includegraphics[width=6cm,height=5cm]{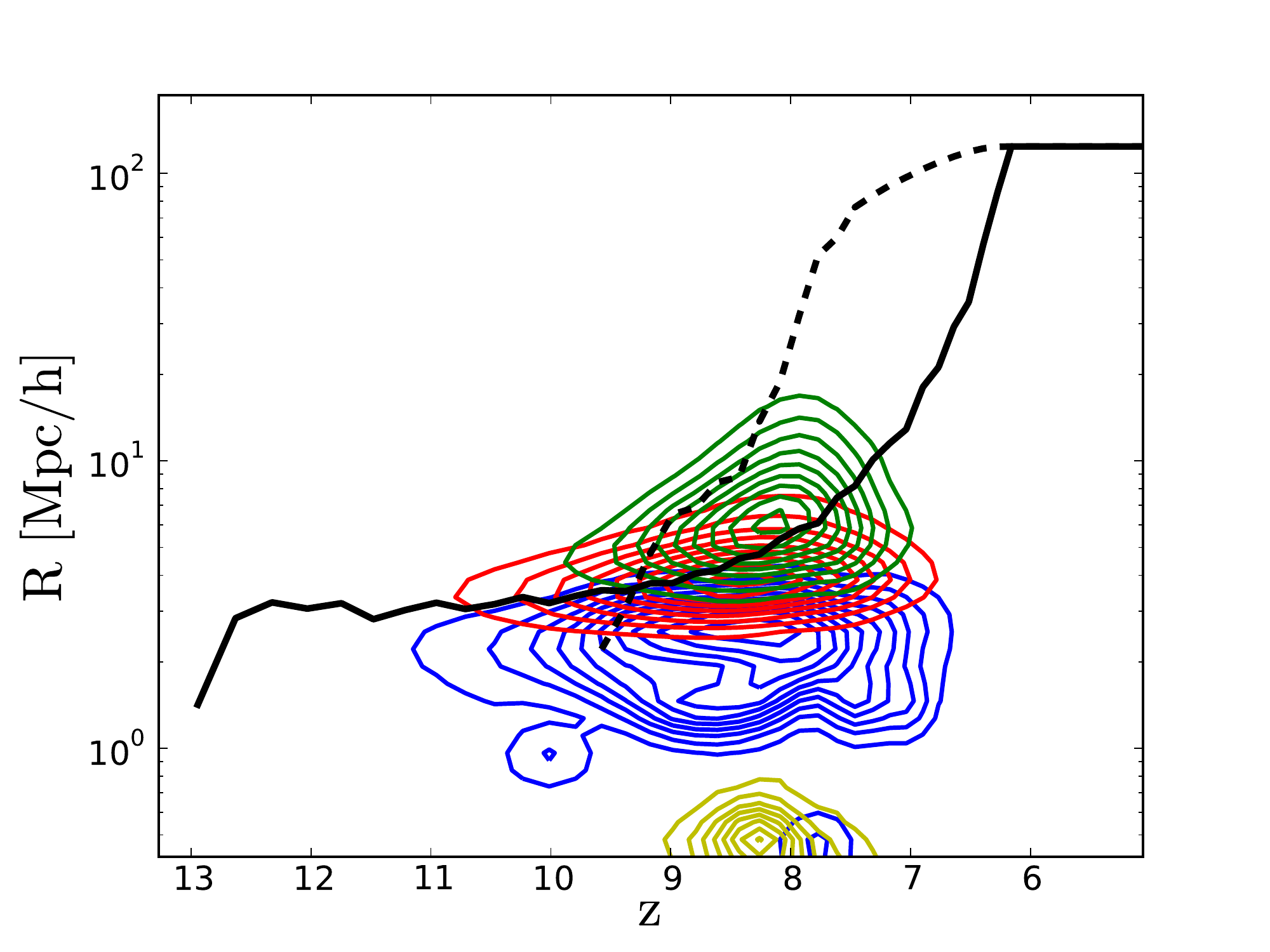}&
      \includegraphics[width=6cm,height=5cm]{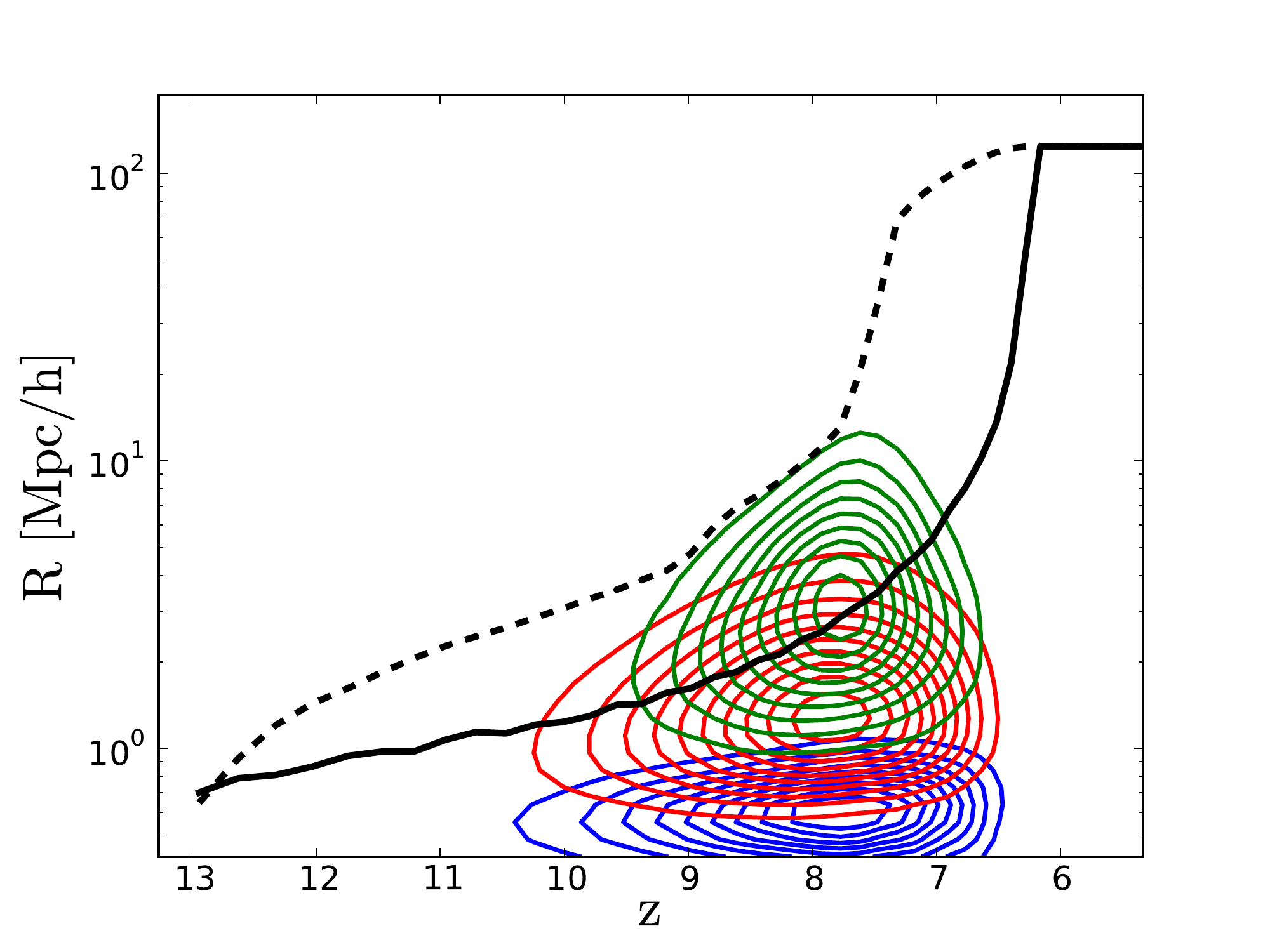} \\
     (a) Boosted Star 200 & (b) Star 200 & (c) Halo 200 \\
      \includegraphics[width=6cm,height=5cm]{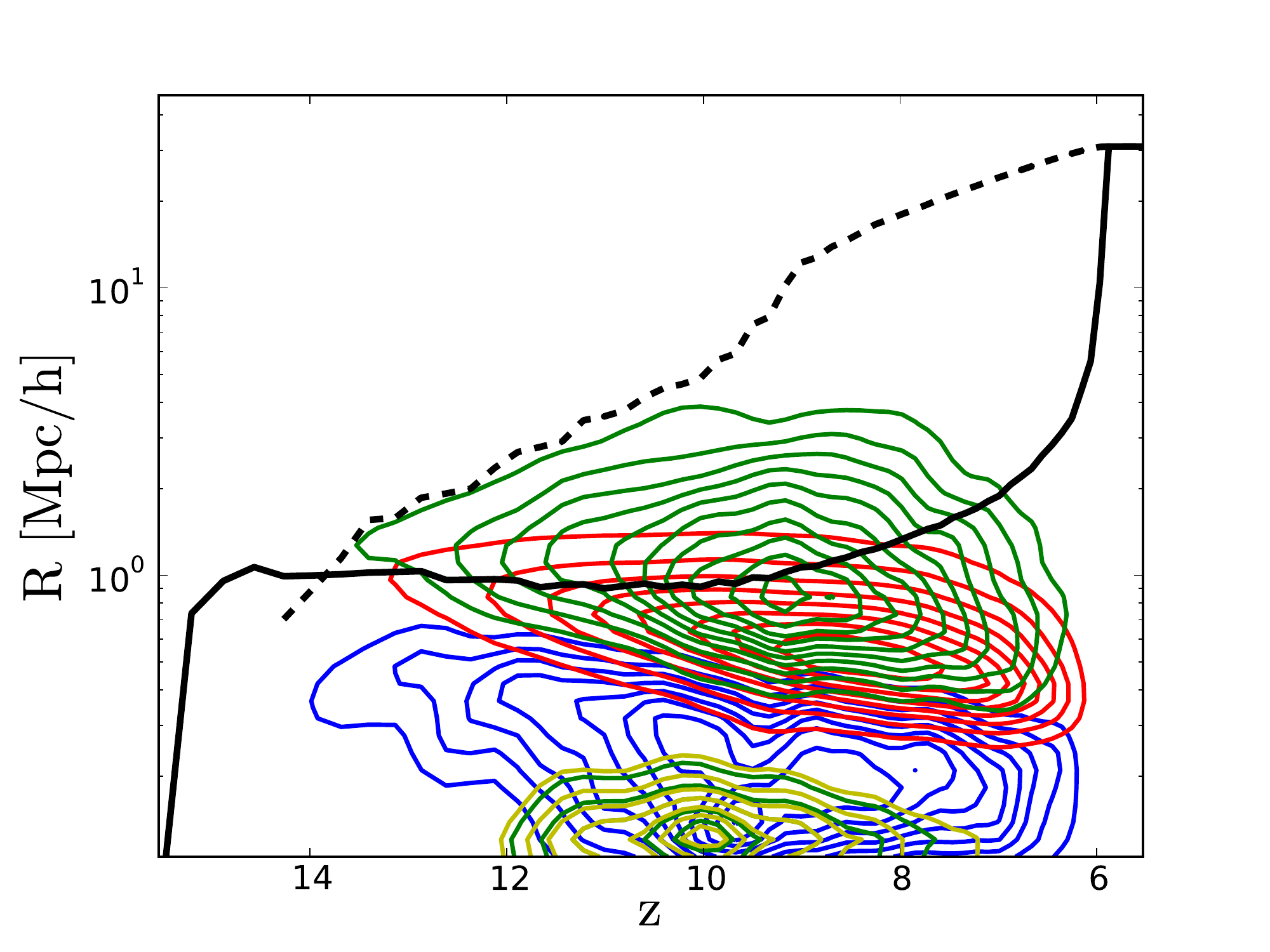} &
      \includegraphics[width=6cm,height=5cm]{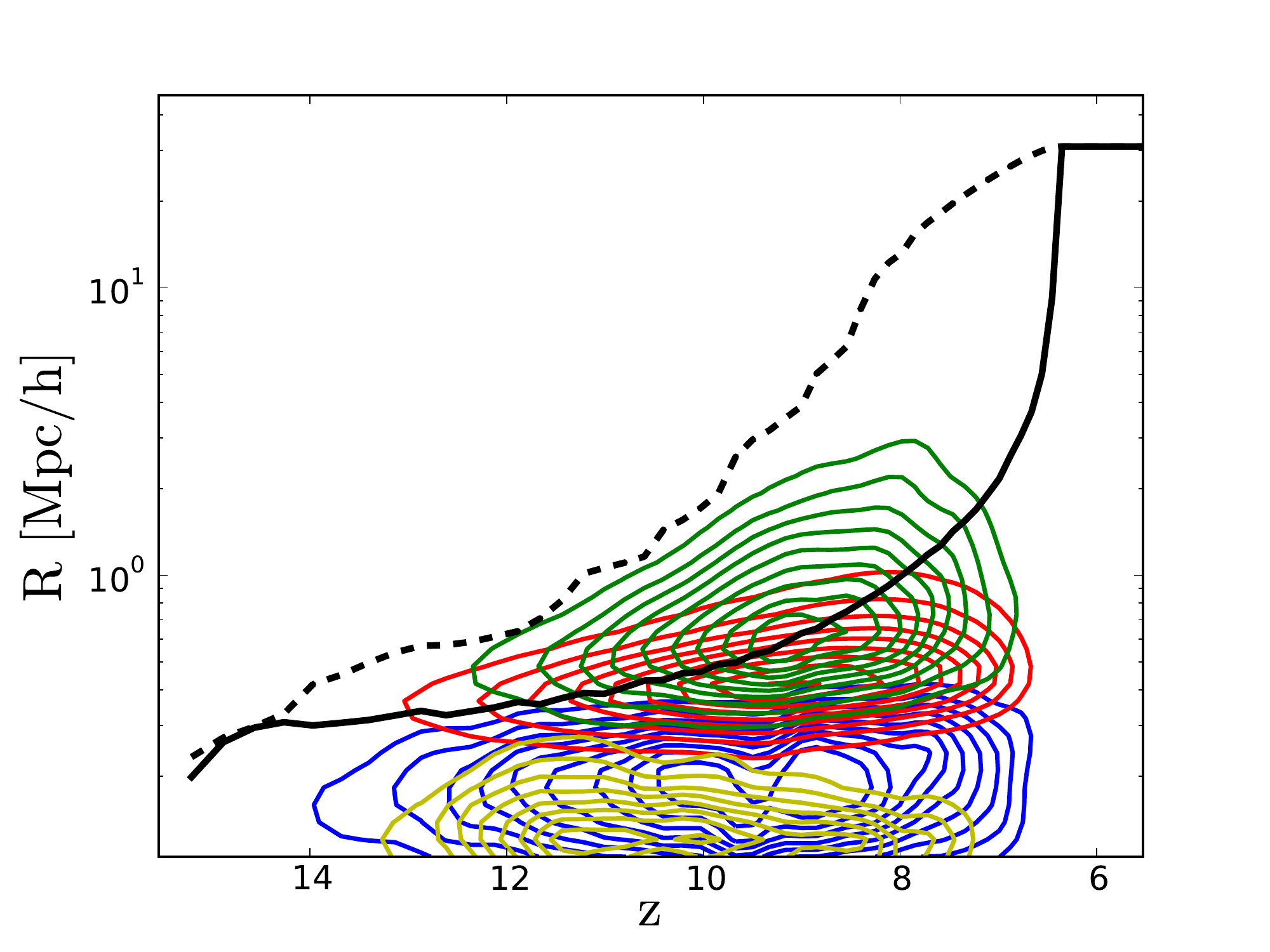}&
      \includegraphics[width=6cm,height=5cm]{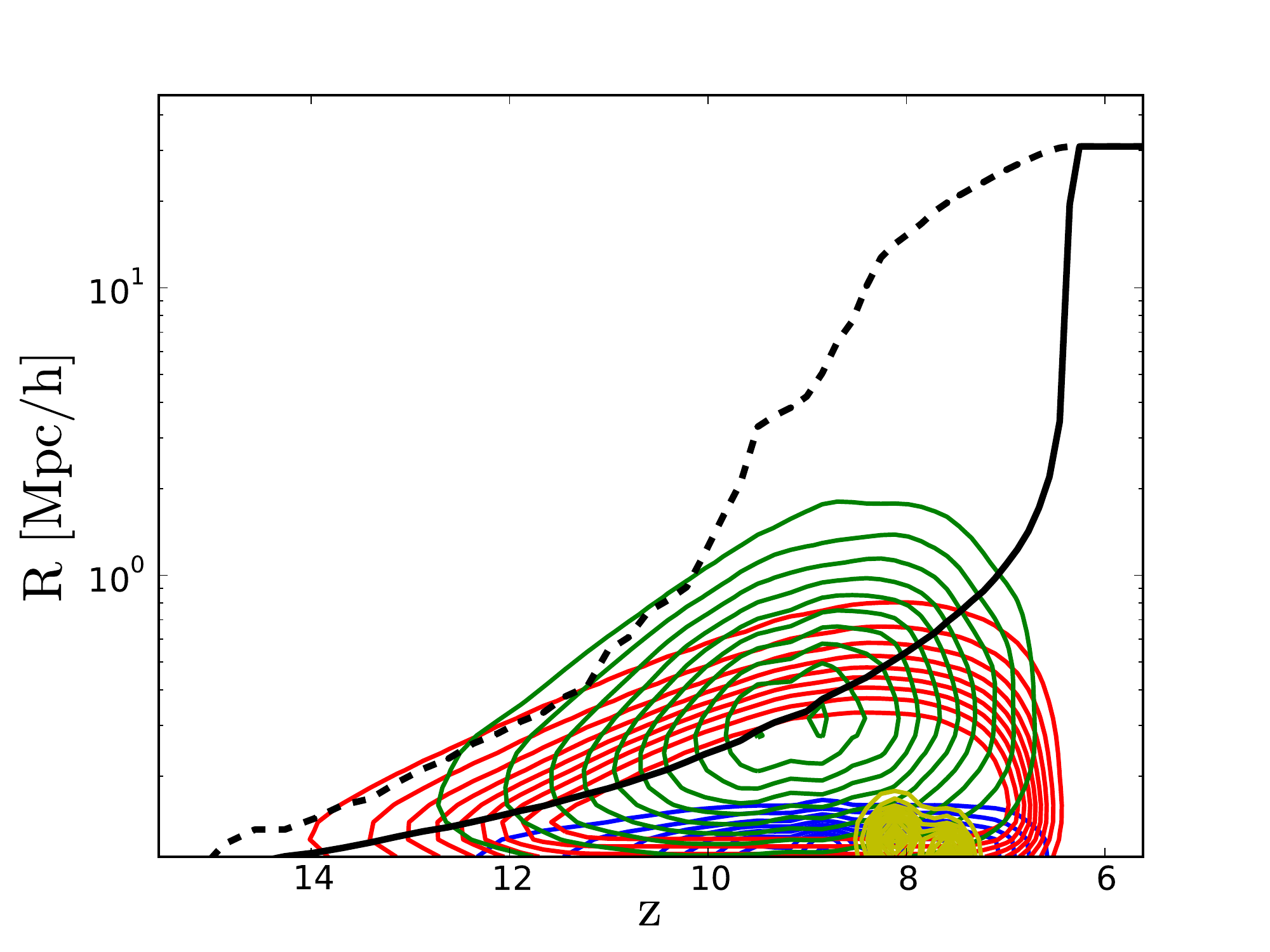} \\
      (d) Boosted Star 50 & (e) Star 50 & (f) Halo 50 \\
\end{tabular}    
   \caption{ Evolution of the radius distribution of each kind of HII regions as a function of redshift for the three models of ionizing source formation 
and for both boxes of 200 and 50 Mpc/h. Panels (a), (b) and (c) respectively represents the distribution for the Boosted Star, 
the Star and the Halo model for the 200 Mpc/h box, while the panels (d), (e) and (f) are for the same models but for the 50 Mpc/h box.
The colours stand as follows: Blue for the new HII regions, Red for the expanding regions, 
Yellow for the regions that will recombine and Green for the regions resulting from mergers. In addition the solid and dashed black lines represent respectively
the radius evolution of the main region end the evolution of the average radius for the HII regions.}
    \label{isocontour_rayons}
  \end{center}
 \end{figure*}

Fig. \ref{isocontour_rayons} presents the size distribution for each kind of HII regions as a function of redshift 
for all three models and for both 200 and 50 Mpc/h boxes.
Here we show the same distribution seen in figure \ref{rayon} but by plotting the contours related to each type of HII regions that we can discern.
The colour code stands as follows: Blue corresponds to the new HII regions, red to the expanding ones, green to those resulting from mergers 
and yellow to regions that will recombine. Once again we show the evolution of the average radius of the region with the black curve 
and the evolution of the radius of the last HII region with the black dashed
curve as in the figure \ref{rayon}.   

Initially, we observe that each kind  of HII region occupies a dedicated range
of radii  in the whole distribution  for all models  and for both 200  and 50
Mpc/h box sizes.   As expected new regions occupy the  smallest range of radii
while the regions resulting of mergers are those which populate the top of the
distribution.  The expanding HII regions  are in the middle with radii greater
than the new ones and smaller than those that will merge.  
We can  also see in each  model that the peak of  the distribution of
the figure  \ref{rayon} correspond  to a radius  range that  is simultaneously
covered  by the  three kinds  of HII  regions.  Alternatively,  we  could have
imagined a  clear separation between the  different range of  radii covered by
the  different kind  of HII  regions instead  of the  observed overlap  of the
different  distributions.  This  tells us  first that  the time  sampling here
allows us  to detect HII  regions with same  radii but belonging  to different
kind of HII regions with different paths in the radius-redshift space. It also
indicates that  the peak of  the radii distribution  must be understood  as the
most likely radii of detection, with  a mix of HII regions at different stages
of their evolution (newborn, growing or merging).
  
Interestingly,  for  each kind  of  regions, we  find  the  properties of  the
underlying ionizing source models in  the evolution of the radius distribution
with redshift.  In other words the R(z) distribution for the size evolution of each kind
of regions  as function  of redshift  is dictated by  the R(z) distribution of  the new
regions and  is propagated to the  other kind of regions.  We thus can
see a  similar R-z relationship for  the expanding region  and those resulting
from merger with a simple translation.  In the Star models S50 and  S200 the new regions appear with a relative
constant range of radii during the whole reionization; again, it reflects the
typical constant emissivity of the stars appearing during the whole simulation
combined  to the  time sampling  used here.  The expanding  regions  and those
resulting from mergers also show  this constant radius range but translated to
greater radius ranges.  The S50 allows to track smaller  new regions thanks to
its higher resolution in terms  of sources with smaller individual emissivity:
as a consequence  all subsequent types show smaller typical  radii than in the
S200 model.

The H200  model presents  new regions that  appear with sizes  more distributed
than in the  S50/S200 models and typically result from the  mass range of the
dark matter  halos that are assumed  as ionizing sources.   Then the expanding
regions cover a much greater radius range compared to the two other models and
we find the  same trend for the regions resulting from  mergers. The last point
indicates that the merger process is not dominated by a few large regions as
in the two other  models but also involves sets of small  HII regions.  In the
H50 model  the mixing is  more pronounced and  the ranges of radii  covered by
each kind of regions overlap. Highly clustered regions are more represented at
high resolution  where dense set of  small halos with small  regions can also
merge. In  H200, mergers occur  at later stage  of radii evolution, in  a more
homogeneous distribution of halos at low resolution.

Finally,  SB200  shows  that  new  regions appear  with  decreasing  radii  as
reionization progresses  and as the  boost for the star  emissivity decreases.
The  expanding regions  and  those resulting  from  merger show  the same  R-z
decreasing trend, seeded  by the new regions population  properties. It should
be noted that at some point ($z<9$), new regions can be detected at small radii:
the declining boost leads to individual regions that  appear with smaller size
and growth  rate, increasing the  likelihood to spot  them at small  radii.  A
weaker R-z  slope is found in  the SB50, as  a consequence of a  smaller boost
correction at higher resolution. As expected, new regions with small radii are
found earlier (for $z<12$) and globally share a greater number of broad features
(like the range of radii of each type) with the S50 model.  


Let  us  finally  focus  on  the  HII  regions  that  recombine  in  the  size
distributions   of   each   model.    We   find   as   expected   in   section
\ref{size_distribution} that regions that  will recombine are those with radii
below  the radius  of $\sim  1$ Mpc/h  in both  SB and  S models  and  at both
resolution.  As  already mentioned,  the  recombination  is  driven either  by
density or  a modification of  the sources.  In the case  of SB models,  it is
fairly obvious that strong early sources  create large HII regions that may be
unsustainable by later generation of sources which are  individually  weaker by construction of the
boost. Furthermore,   the  boost  guarantees  a  global
convergence  but locally  a strong  early source  can be  replaced by  too few
weaker  later  sources,  especially  if  the  source  renewal  is  subject  to
stochasticity. Once sources become too  dim, it leads to the fragmentation of
regions and recombinations at small  scales. S models also present some degree
of  recombination, even  though they  do not  contain declining  sources. These
recombinations can  be  explained by  local  dense  patches  and it  could  be
supported by the fact that such regions only appear at later time ($z<9$) in the
S200 model, when local clustering is high enough. Furthermore, in both S50 and
SB50 models,  these regions can be found  over extended period  compared to their
200  Mpc/h equivalent, related  to the  fact that  dense patches  that trigger
recombination are more likely to be found at high resolution.

 
\subsection{Mergers of HII regions}
\label{merger_HII_regions}

\subsubsection{General evolution of the merger process of HII regions}
\label{parents_distribution}

\begin{figure*}
   \begin{center}
    \begin{tabular}{ccc}
      \includegraphics[width=6cm,height=5cm]{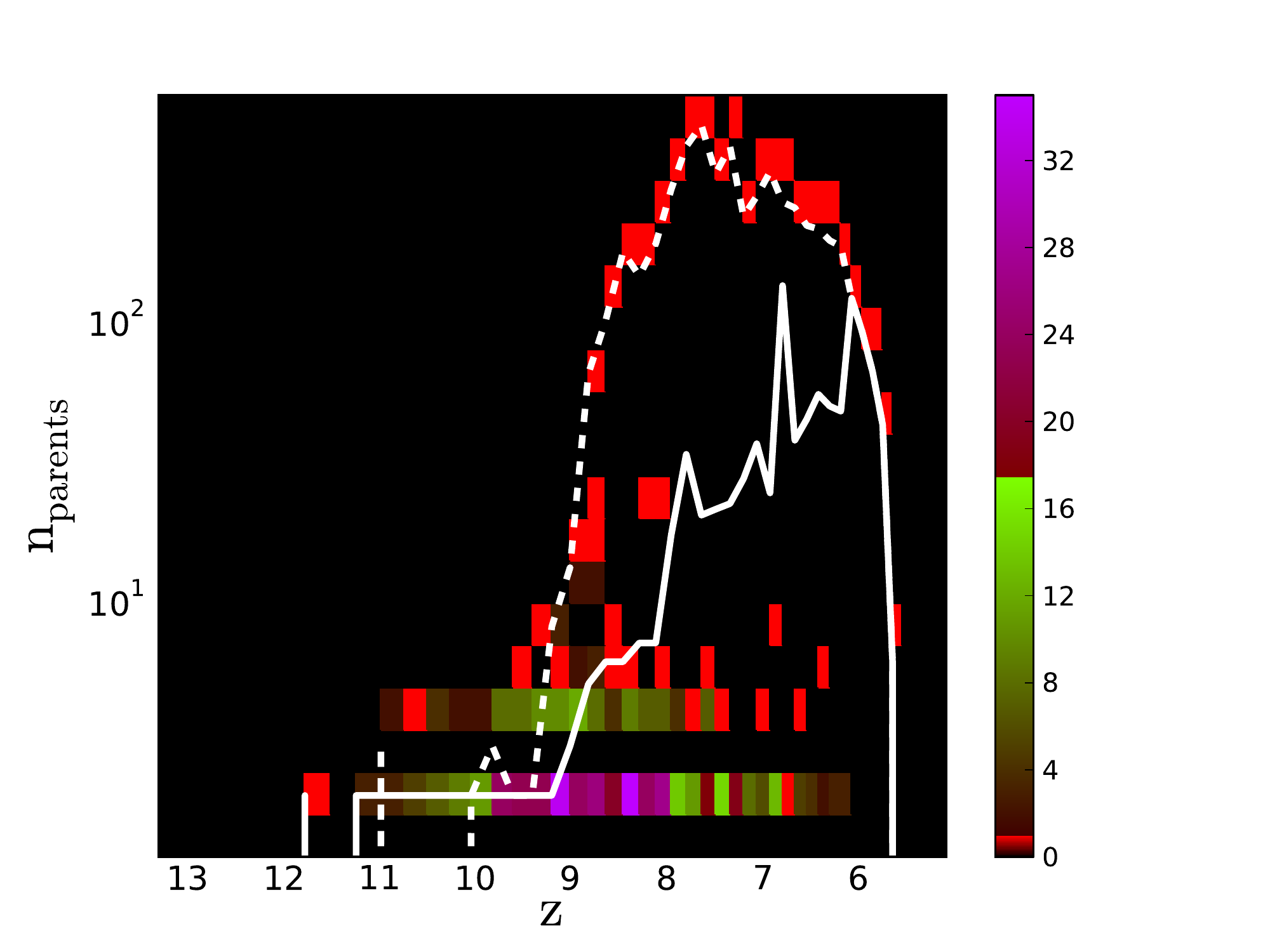} &
      \includegraphics[width=6cm,height=5cm]{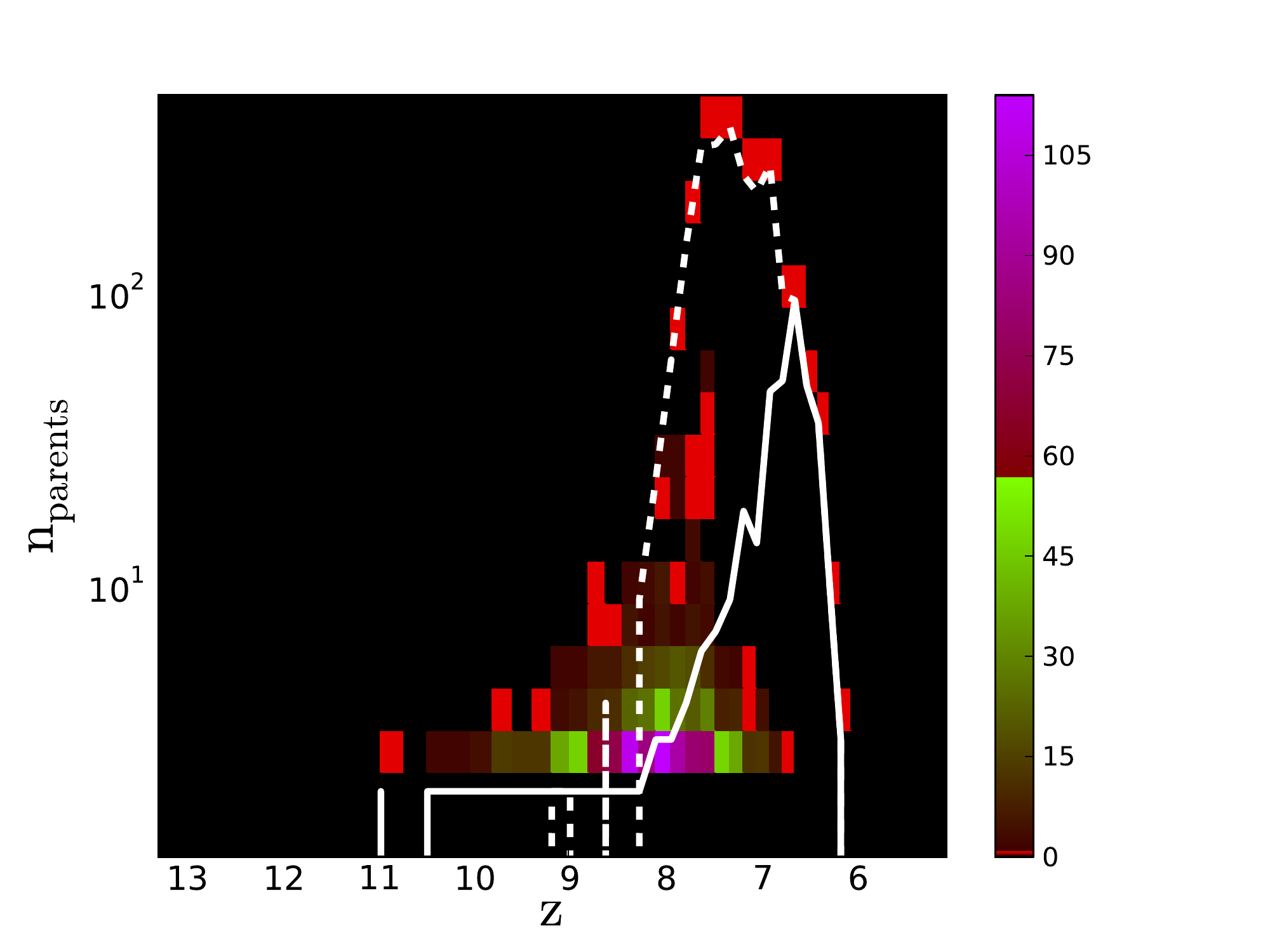}&
      \includegraphics[width=6cm,height=5cm]{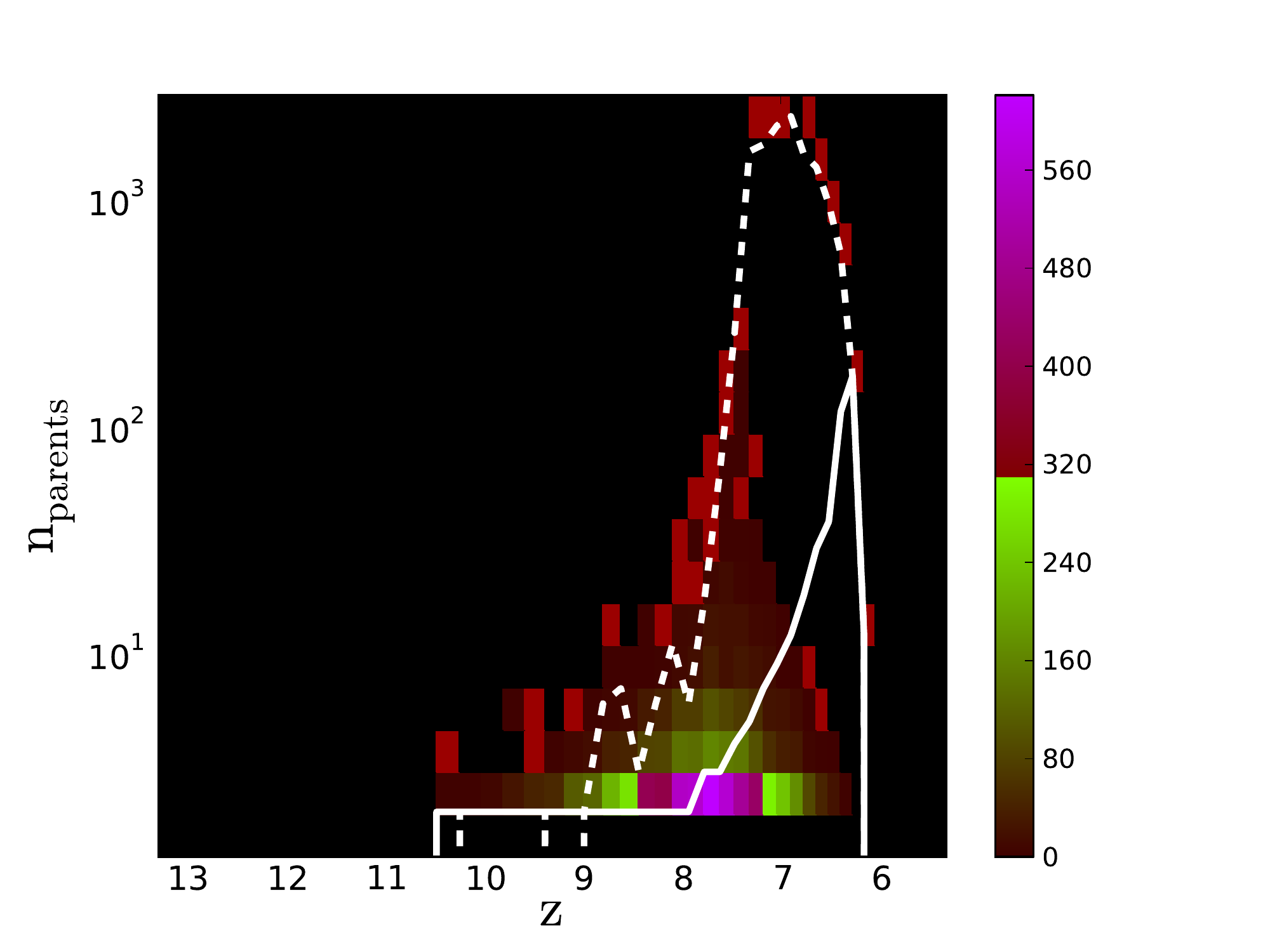} \\
     (a) Boosted Star 200 & (b) Star 200 & (c) Halo 200 \\
      \includegraphics[width=6cm,height=5cm]{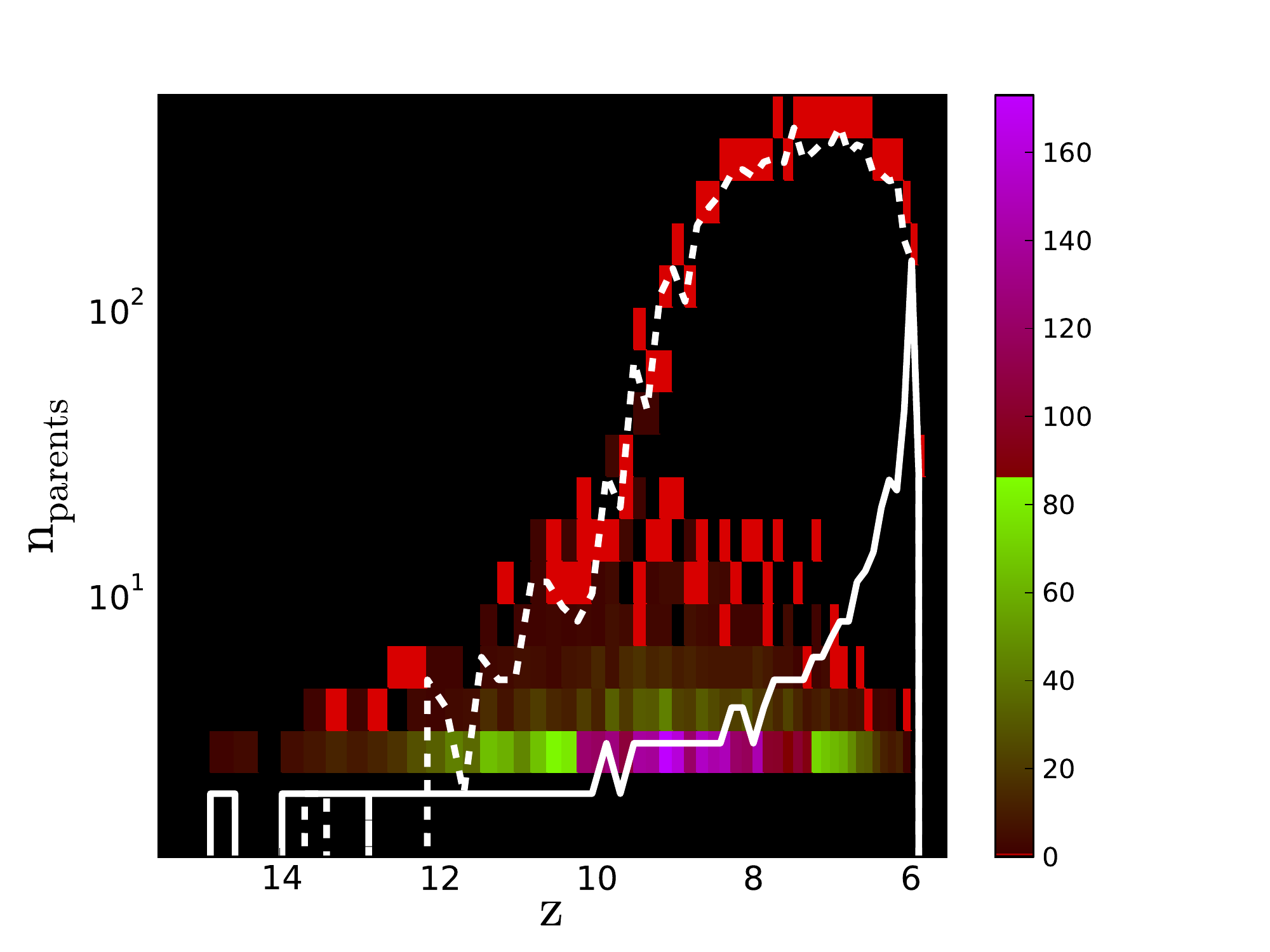} &
      \includegraphics[width=6cm,height=5cm]{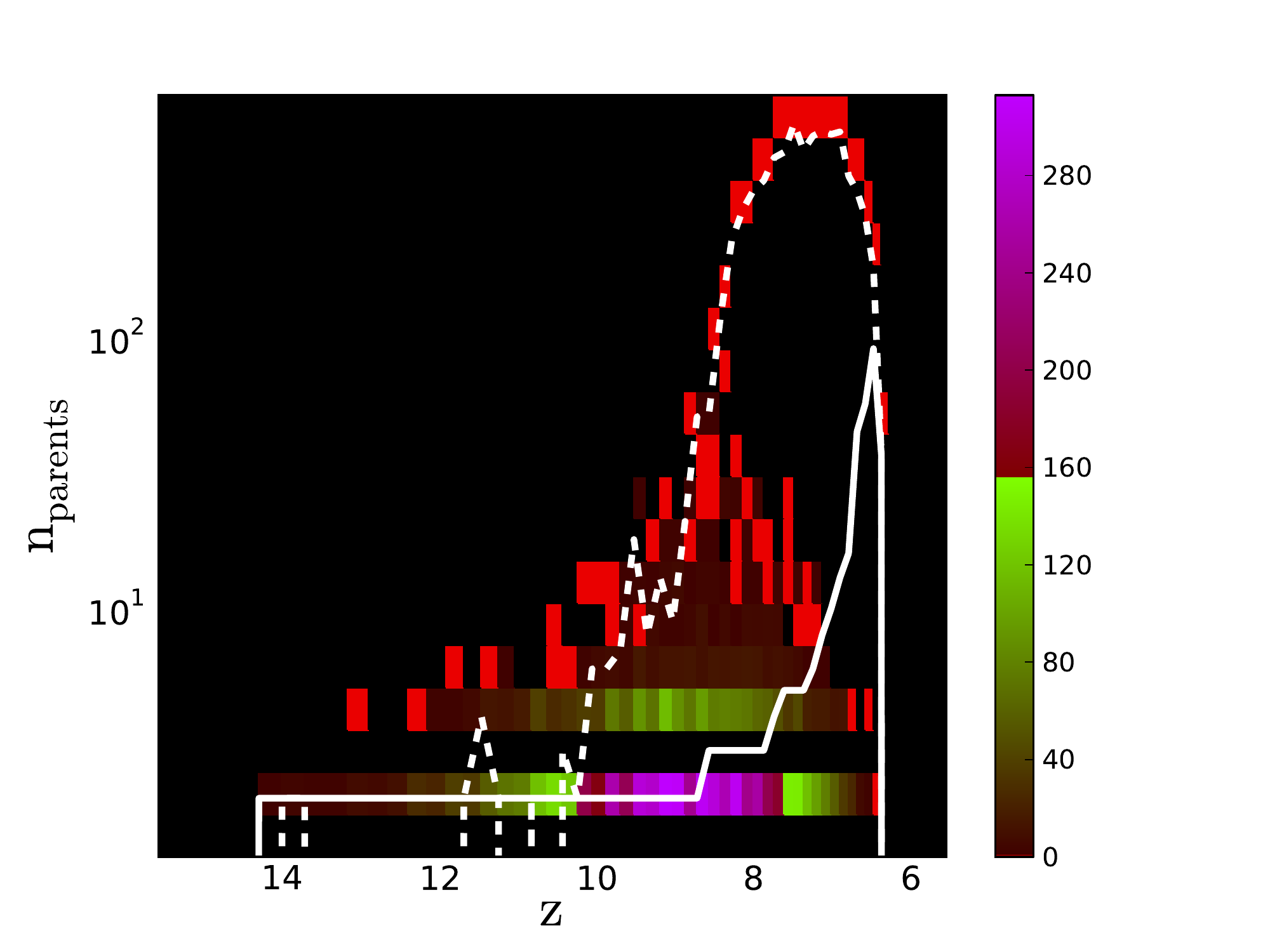}&
      \includegraphics[width=6cm,height=5cm]{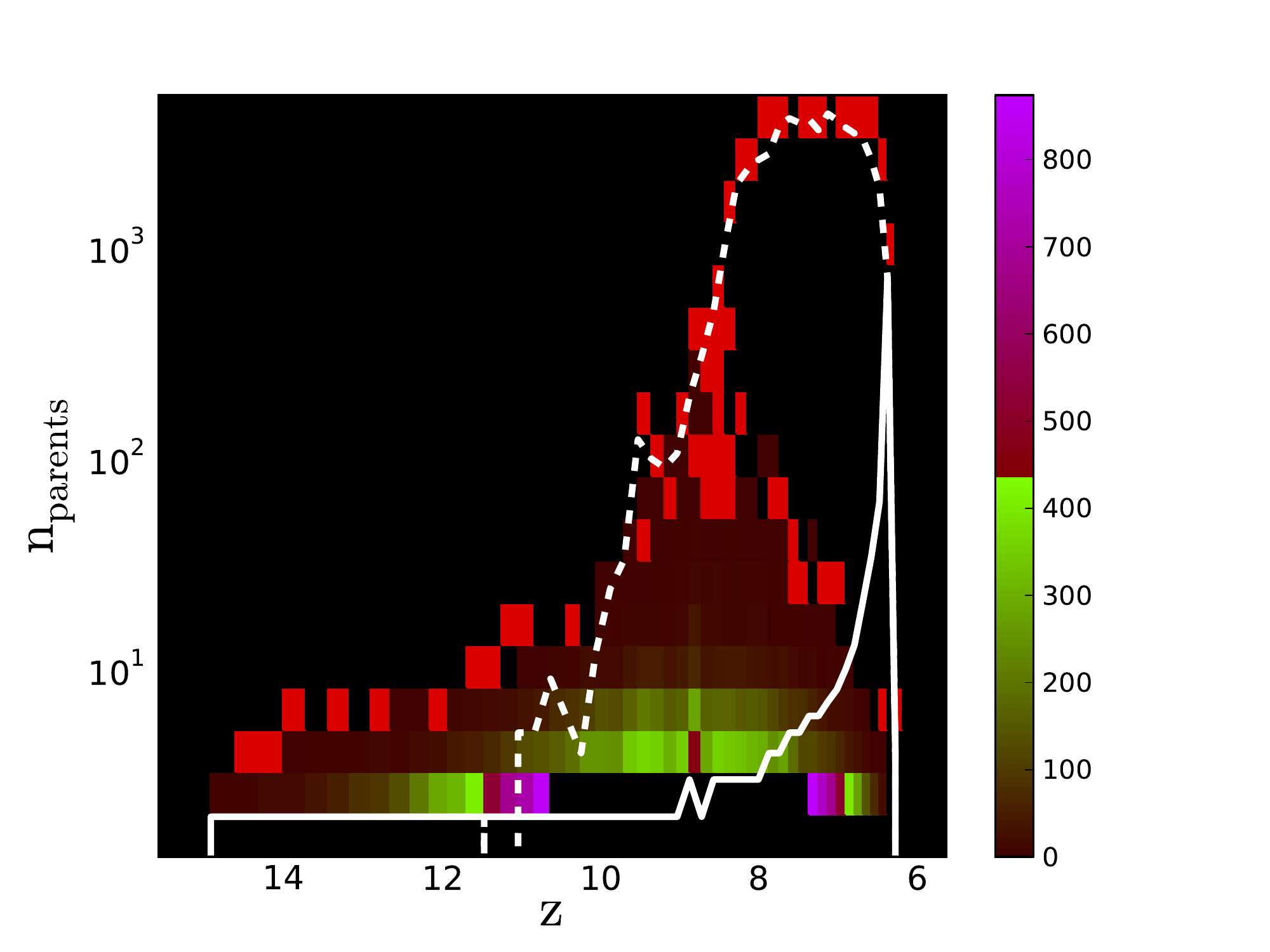} \\
    (d) Boosted Star 50 & (e) Star 50 & (f) Halo 50 \\
\end{tabular}    
   \caption{Distribution of the number of parents for the HII regions resulting from merger as a function of redshift for the three kinds of ionizing sources
and for both box sizes of 200 and 50 Mpc/h. Panels (a), (b) and (c) respectively represents the distribution for the Boosted Star, 
the Star and the Halo model for the 200 Mpc/h box, while the panels (d), (e) and (f) are for the same models but for the 50 Mpc/h box.
The color code stands as follows: the brightest red cells represent the
location in the distribution populated by \textit{a single} HII region, other
red tones up to the brightest green cells spans distribution densities from a
couple HII regions up to a value corresponding to half the maximal value of
the distribution. Finally the purple tones (from the darkest to the brightest)
encode the maximal values.
 We represent in addition the evolution of the average number of parents and the evolution of the number of parents of the main region respectively with the solid and dashed white lines.}
    \label{parents}
  \end{center}
 \end{figure*}

Fig. \ref{parents}  is restricted to regions  that have more  than one parent,
i.e.  that result  from  mergers. It  presents  the redshift  evolution of  the
distribution of  the number of  parents. It  can be seen  as a measure  of the
patchiness  or granularity  of the  overlapping process  and is  also  a first
indication  of the inner  complexity of  the regions:  an ionized  volume that
result from  tens of progenitors is  shaped (like e.g.  the front propagation,
its  growth rate,  or the  inner  distribution of  UV flux)  by the  different
properties of  at least tens  of sources whereas  an HII region coming  from a
binary merger is  likely to be, maybe naively,  more straightforward to relate
to its inner sources.  In addition,  the solid white line in figure \ref{parents} shows the
evolution of  the average  number of parents and the white  dashed line
stands  for the  evolution of  the number  of parents  for the  last remaining
region at the end  of the simulation. Again, with the help  of the merger tree
we follow this region back in time  and evaluate its number of parents at each
instant.
The color code in the distributions is identical to the one used in the figure \ref{rayon}.

Firstly,  in most  of the  models the  mergers between  regions operates  in a
binary-tertiary  manner  as   seen  with  the  peak  at   $\sim  2-3$  in  the
distributions. It is  essentially an indication that the  temporal sampling of
the tree  is fine enough  to track  individual mergers  whereas a  coarser grained
sampling would have presented a typical larger value. One exception is the H50
model, where  clearly such  mergers are missing  from $z  \sim 10$ to  $z \sim
7$.  In this  case,  the abundance  of  clustered individual  regions at  high
density is much  higher than in other  models and leads to multiple mergers during a single time step
for a given region.

Then, in  every model we find  the emergence of  a single region which  is the
result of more mergers than other  regions.  In the S200 and H200 models, this
region appears at $z \sim 8$ and earlier  in the SB200 one, at $z \sim 9$. The
same discrepancy can be found at higher resolution with a formation at $z \sim
9$ for the S50 and H50 models  and $z\sim 10$ in the SB50. Interestingly, this
merger-dominant  regions appear  at  the same  moment  as the  radius-dominant
region discussed  in the previous  section.  In Fig. \ref{parents},  the white
dashed  line   shows  the   evolution  of  the   number  of  parents   of  the
radius-dominant  region  and  unsurprisingly   it  matches  the  path  of  the
merger-dominant  one. Thus,  a  moment exists  when  an HII  region begins  to
monopolize the  merger process in  the box.  Then  the region grows  faster by
successive  mergers  and will  quickly  dominate  other  regions in  terms  of
size. From this instant (called $\mathrm{z_{\,BKG}}$), the probability that a location is part of the global
UV \textit{background} instead of being irradiated  by a local source increases and can
be considered as the onset of the overlapping process.  This is typically the moment
when on average the local  information about the reionization process start to
be lost to the benefit of a single reionization history.

Size is an important factor for regions to get high numbers of progenitors as a larger volume
naturally promote encounters with distant ionized patches. However it can be noted that the radius-dominant region is only one among the
merger-dominant regions. Clearly the distribution shows regions with larger
number of progenitors and this is especially the case for H200 and H50 models:
the high density of sources and their clustering can induce high merger rate
in smaller volumes. More generally, high resolution experiments present
greater merger rate thanks to a higher number of sources and higher local
clustering. Also H models can achieve number of progenitors that are tenfold
greater than the S and SB equivalent as another consequence of large number of
halos. The two stellar models differ also in the number of regions with a
large number of progenitors: close to $\mathrm{z_{\,BKG}}$, S models present
clear detection of regions with a number of progenitors close to those found
for the radius-dominant one (10-50 progenitors), whereas in SB models such regions are hardly
found. It indicates that the early buildup of a dominant region in the SB
model tends to prevent the formation of separate regions that aggregate large
number of sources on their side, probably because they were incorporated early
in this dominant volume. At higher resolution the discrepancy is weaker, but
in the S50 models, the dominant region share with more other regions the
property of having a large ($>10$) number of progenitors.

\subsubsection{Growth of the main HII region}
\label{merger_main_HII_region}

\begin{figure*}
   \begin{center}
    \begin{tabular}{ccc}
      \includegraphics[width=6cm,height=5cm]{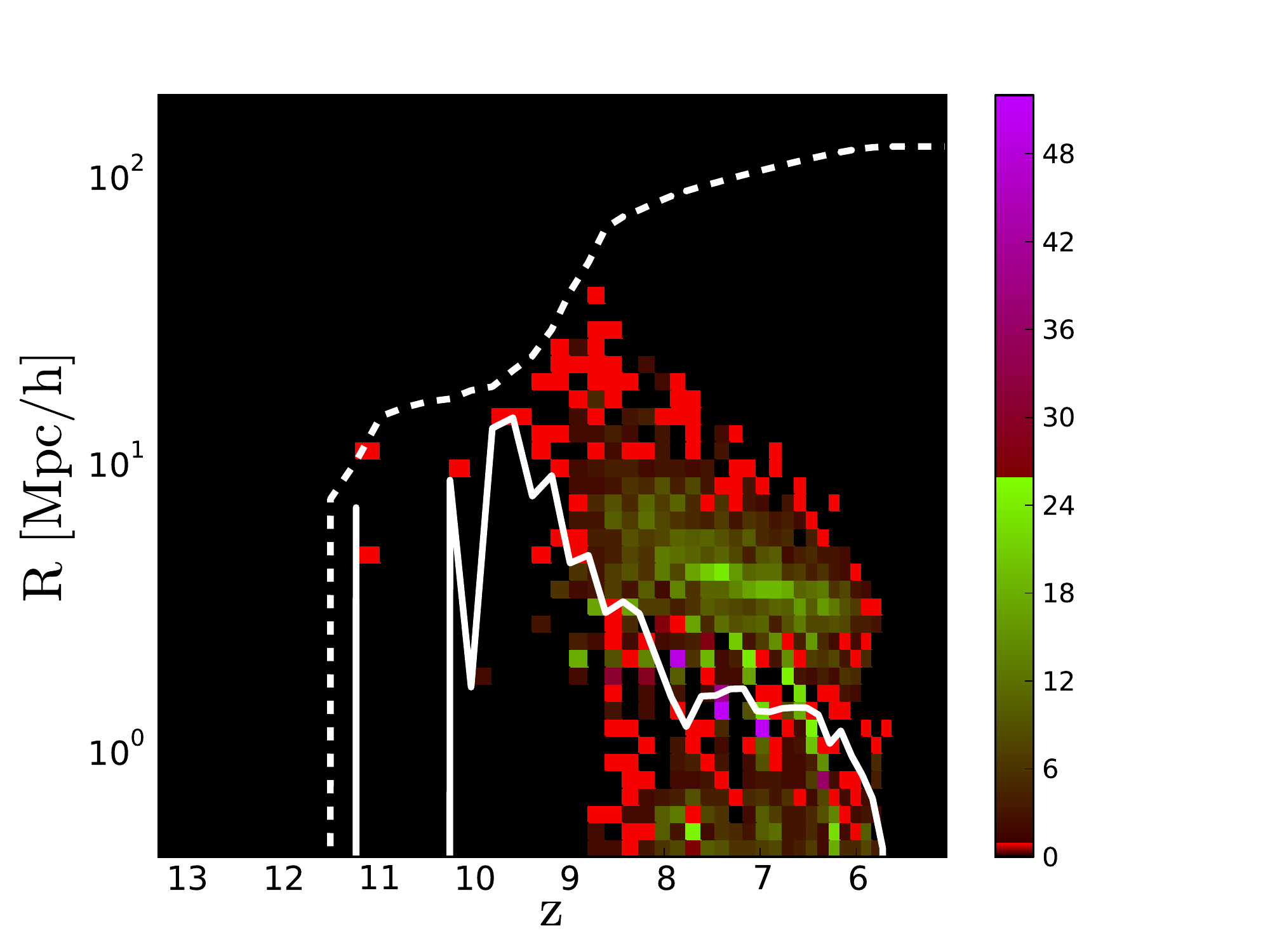} &
      \includegraphics[width=6cm,height=5cm]{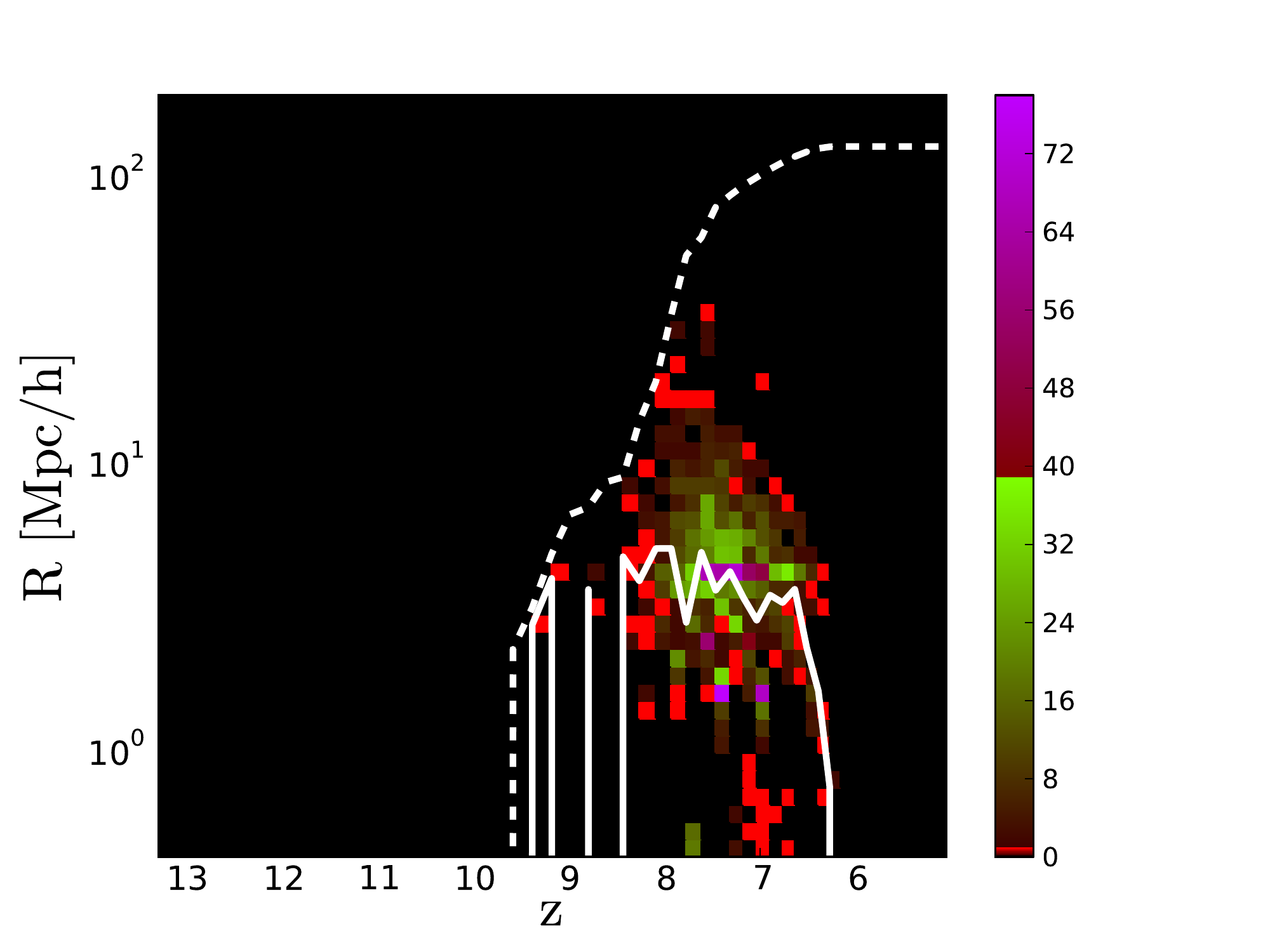}&
      \includegraphics[width=6cm,height=5cm]{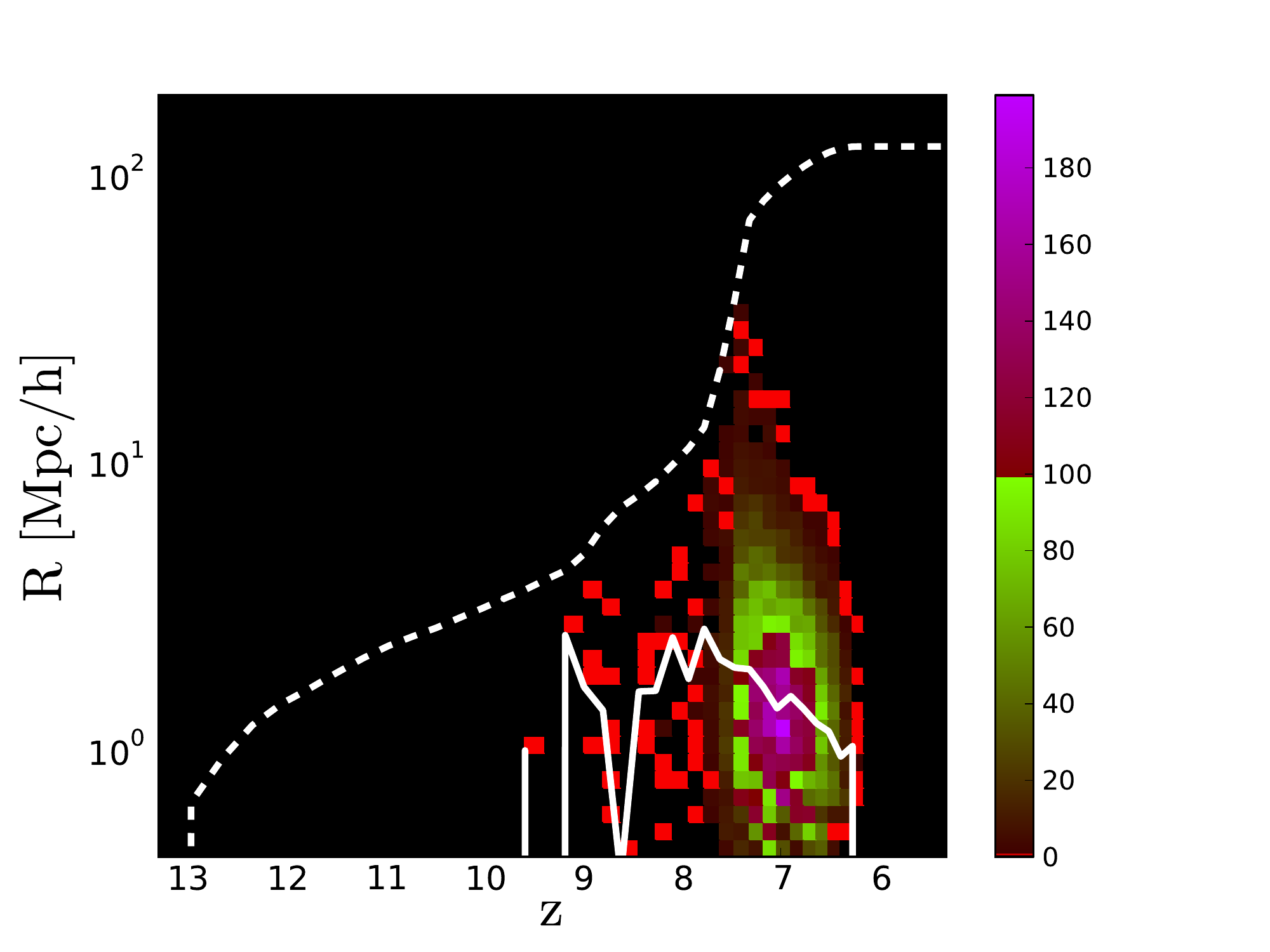} \\
      (a) Boosted Star 200 & (b) Star 200 & (c) Halo 200 \\
      \includegraphics[width=6cm,height=5cm]{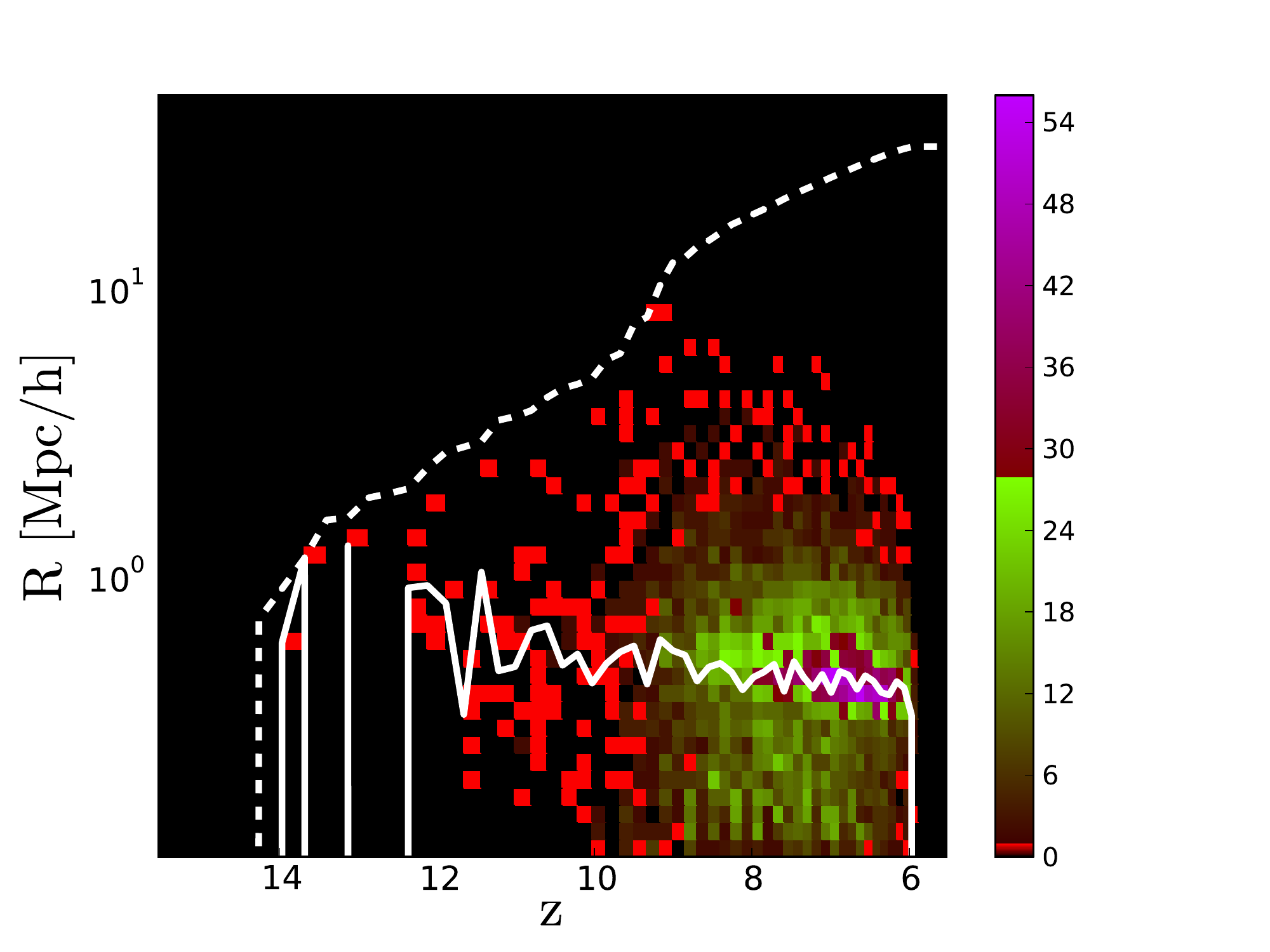} &
      \includegraphics[width=6cm,height=5cm]{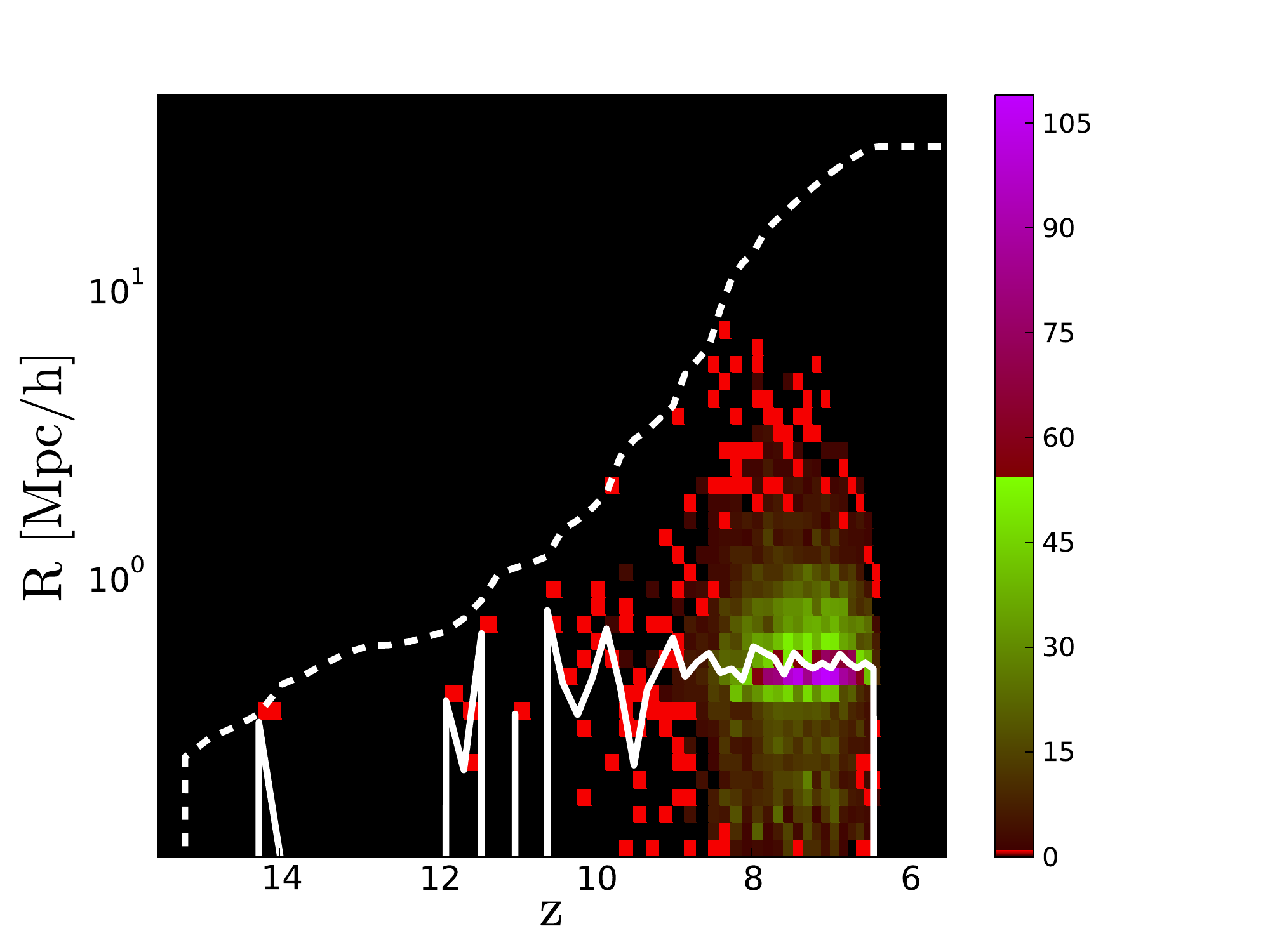}&
      \includegraphics[width=6cm,height=5cm]{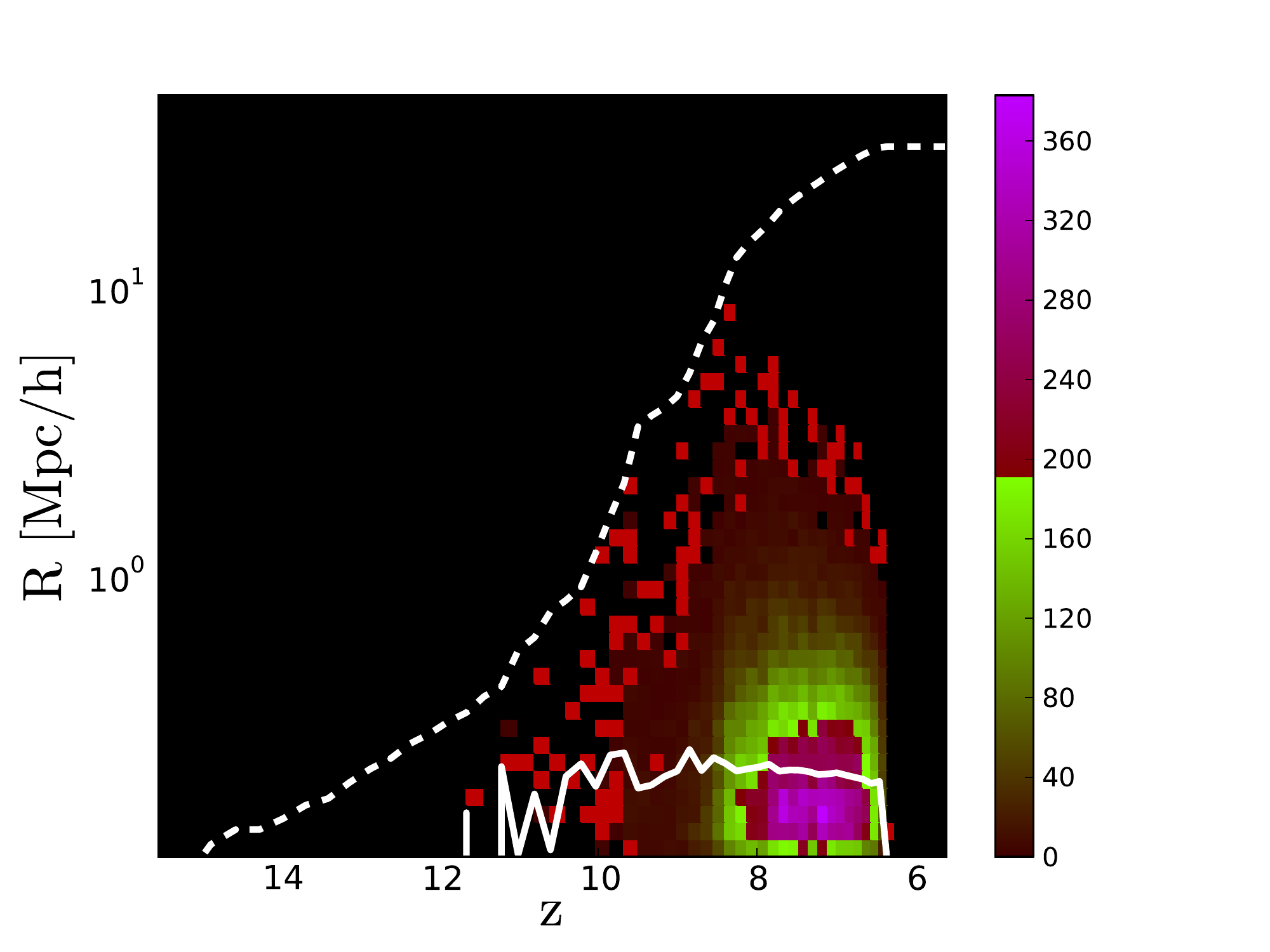} \\
      (d) Boosted Star 50 & (e) Star 50 & (f) Halo 50 \\
     \end{tabular}    
  \caption{Distribution of the HII regions sizes for the regions that merge with the dominant region as a function of redshift for the three kinds of ionizing sources
and for both box sizes of 200 and 50 Mpc/h. Panels (a), (b) and (c) respectively represents the distribution for the Boosted Star, 
the Star and the Halo model for the 200 Mpc/h box, while the panels (d), (e)
and (f) are for the same models but for the 50 Mpc/h box.
The color code in the distributions is made so that the brightest red cell
contains only a single HII region. Other red tones 
up to the brightest green designate cells 
populated by a couple HII regions up to a value corresponding to half the
maximal value of the distribution. Finally the purple tones
(from the darkest to the brightest) encode the largest values of the
distribution. The evolution of the average radius for the regions that merge
with the dominant region and the evolution of the radius of the main region
are represented respectively with the solid and dashed white lines.}
    \label{rayon_fusions_grosse}
  \end{center}
 \end{figure*}

We aim with this section at understanding how the dominant region leads to a loss of information about the expansion process of others regions. 
By looking at the evolution of the properties of the regions that merge with
the dominant one, we will evaluate how each model is resistant to the emergence of this main HII region.

Fig. \ref{rayon_fusions_grosse} presents the evolution of the radius distribution for HII regions that merge with the 
radius-dominant one for all models and both 200 and 50 Mpc/h boxes. With the
white dashed curve we show again the evolution of the radius of the dominant
region and with the solid white line we show the evolution of the averaged
radius of the HII regions that merge with it \textit{only}.
The color coding of the distributions is identical to the one used in the figures \ref{rayon} and \ref{parents}.

Immediately we see that, from a certain redshift, every model presents a
radius distribution which is representative of the whole distribution of all
HII regions, as discussed previously and shown in figure  \ref{rayon}.
In other words, from a certain moment onwards, the major part of the present HII regions are regions that will merge with the main region in the next snapshot.
Thus this moment can be seen as the time when the main region imposes its domination.  
Then the later this moment appears, the longer the individual expansion
process of ionized region can be tracked. In all models and at both
resolution this moment is broadly coincident with $\mathrm{z_{\,BKG}}$: from
this redshift any region (in terms of size) can be incorporated into the background.

 
It is also noteworthy to see how this main region is built before it begins to impose its domination.
For instance both H50 and H200 models, the earliest stage of the buildup does
not include any merger and the growth is purely driven by the inner source
until z=9.5 (resp. 11.8)  for the H200 (resp. H50) model. The equivalent
region in the S200 model is detected later, with a large radius ($\sim 3
$ Mpc/h instead of $0.5$ Mpc/h for H200) and incorporates immediately other
regions with similar sizes. At higher resolution, the radius dominant region
of S50 is always larger than its H50 counterpart and also incorporate other
relatively similar region quite early (at $z\sim 14$ for instance). Clearly
the stronger emissivity of individual stellar source produce larger
progenitors for the S models than for the halo-based ones, promoting a
merger-driven growth sooner. Meanwhile the SB models exhibits the expected and
quite different evolutions for the dominant region: the large emissivity correction at early
times creates a large dominant region at the earliest stages and it quickly
incorporates other regions also `puffed up' by the boost and therefore
belonging to the same class of large radius. In particular, SB50 present a
whole succession of large regions, close to the white evolution track of the
dominant one. For the latter, it implies that its radius is always larger than
found in the S50 and H50 models and is driven by mergers much
sooner. Previously, we found that high resolution tend to diminish the
discrepancies of the SB model but in the current case, the higher level of
clustering in the 50 Mpc/h box leads to a buildup scenario of the main region that is arguably more
even more distinguishable compared to H and S models than in the 200 Mpc/h experiments.

\section{Summary and conclusion}
\label{Discussion}

We developed a new methodology based  on merger trees of HII regions to study
the history of reionizations  in cosmological simulations with full radiative
transfer post-processed  by the GPU-driven code ATON  on RAMSES hydrodynamical
snapshots. We  applied systematically the technique  in two sets  of 200 Mpc/h
and 50 Mpc/h simulations, where each set involved three kinds of source model:
\begin{itemize}
\item a halo  model where halos act as sources of  photons with an emissivity
  proportional to their mass,
\item a stellar model where `star' particles produced by the cosmological code
  act  as  the   sources,  with  an  emissivity  corrected   to  complete  the
  reionization at the same redshift as the halo model,
\item a  boosted stellar model, where  the same `star' particles  are used but
  with time-dependent and decreasing correction on the emissivity to reproduce
  the converged emission of UV photons at each instant.
\end{itemize}
Corrections  are   necessary  as  the  self-consistent   production  of  `star
particles' by  the hydrodynamical code  is resolution-dependent and  not fully
converged  at these  resolutions. The  emissivities  were tuned  to produce  a
similar ionization  fraction evolution at  both resolutions. HII  regions were
detected thanks to a FOF algorithm  and linked along time within a merger tree
structure.

Firstly, the three  models present an evolution comparable  in terms of global
features.  Indeed,  the evolution  of the optical  depth and  ionized fraction
present  similar  shapes  in  all  experiments.  

The evolution  of the number  of present HII  region reaches a maximum  at the
same similar redshift for all models and for both box sizes (z=8.5 and z=9 for
the  200  and  50 Mpc/h  experiments).   However  the  study of  the  separate
evolution  of  the different  kinds  of HII  regions  shows  that the  intense
episodes  of mergers occur  earlier in  the boosted  Star model  while it
occurs later  and at the same  moment in the Star  and Halo models  in the 200
Mpc/h box.  Also the boosted Star  model is sensitive to recombination, due to
sources unable to sustain the existing HII regions while
only few recombinations occur  in the Star model and no one  in the Halo model
in  the 200  Mpc/h.  In the  50  Mpc/h simulations,  episodes  of mergers  are
similarly  distributed in the  three models,  whereas increased  densities are
likely to produce the larger population of recombining patches in the Star experiments.

We also investigated  the evolution with redshift of  the size distribution of
HII regions.  All models present the emergence of a main region in size from a
certain redshift onwards.  We also have found that the
evolution of the  radii distribution is typically related  to the evolution of
the ionizing source emissivities in each model: a large dispersion of possible
radii with  small structures in the halo-based models, a  similar distribution
for the stellar model with a lack of small regions due to stronger and sparser
emitters and a distribution strongly
skewed toward  large regions at early times in the boosted  stellar experiments. The
discrepancies decrease at higher resolution but remain, especially for the last
model. We have then be able to show
that this radius-redshift correlation, imprinted in the distribution  of new regions is
somehow  memorized and  kept in  the distribution  of growing  and merging ionized patches.

The evolution  of the  number of  parents involved in  the percolation  of HII
regions has also  been studied. In all experiments,  merger-dominant (with 10
or  more progenitors)  regions exist  and among  them one  corresponds  to the
radius-dominant one and drives the  reionization. Its domination is delayed in
both halo and  stellar model compared to the boosted  one. As an illustration,
we investigated the  size distribution of HII regions  assimilated by the main
one and found that indeed  it becomes representative of the whole distribution
first in the boosted stellar model, than  in the star model and finally in the
halo one. Hence individual reionizations can  be tracked on a longer period in
the  two   latter  models.  Increasing   the  resolution  reduces   again  the
differences, thanks  to an  increased convergence between  the three  types of
sources. 

Globally, we can say that the  Star and the Halo models have similar histories
even  though  the  Star model  lack  small  scale  power and  therefore  local
merger events. Meanwhile  the boosted  Star model  presents  its own
reionization  history, where  we see  the early  emergence of  a  dominant HII
region in  size that  concentrates rapidly the  merger process.  On  the other
hand,  both other  models show  individual growths  for HII  regions  that can
be tracked individually  on a longer  time and resist  to the domination  of the
main region later  on during the reionization.  In the 50  Mpc/h box, when the
spatial  resolution is  increased, the  histories of  the three  models become
comparable even if the boosted Star  model still presents the domination of an
early main region that concentrates the merger process.

Clearly, the large scale experiments show that the lack of convergence in the
self-consistent formation of star particles leads to substantial differences in
the HII regions properties and their evolution during the percolation process,
even  though  they  present   satisfying  properties  regarding  their  global
`integrated' features.  At higher resolution (corresponding to  50 Mpc/h boxes
in our cases),  we find that overall the discrepancies are  reduced and can be
seen as  the result of  smaller corrections applied  to sources and  a greater
match  between the  halo population  and the  `stellar' one.  A  question also
remains on the quantitative impact of the rare `emitters' more likely in the large
boxes or the lack of density field variance in the 50 Mpc/h. Future experiments will
allow to assert  them even though we can predict e.g. that rare emitters should be
relevant for the rise of the dominant regions whereas density field variance on small
scales should impact the rate of local mergers.

We aim at systematically applying  this technique in  the future. For
instance,   it  provides   a  reproducible   analyzes  framework   to  compare
simulations, numerical  techniques and the  impact of physical  ingredients on
the morpho-chronology  of the reionization. Furthermore, it  puts the emphasis
on   individual  HII   regions,   leading   to  an   analyzes   in  terms   of
\textit{reionization-s} instead  of \textit{The Reionization}.  In the context
of galaxy  formation, it may provide  insights on the  proximity effects which
impacts the properties of the reionization as  being seen by one or a group of
galaxy (see \citealt{2012MNRAS.000..000F} submitted).

\section*{Acknowledgments}

The authors would like to thank B. Semelin, R. Teyssier, and
H. Wozniak for valuable comments and discussions. This work is supported by the ANR grant LIDAU.

%
\bibliographystyle{aa}
\bibliography{biblio}

\appendix
\section{FOF algorithm}
\label{fof}

\subsection{Implementation}

\begin{figure}
\hskip 1.5truecm
\includegraphics[width=6cm]{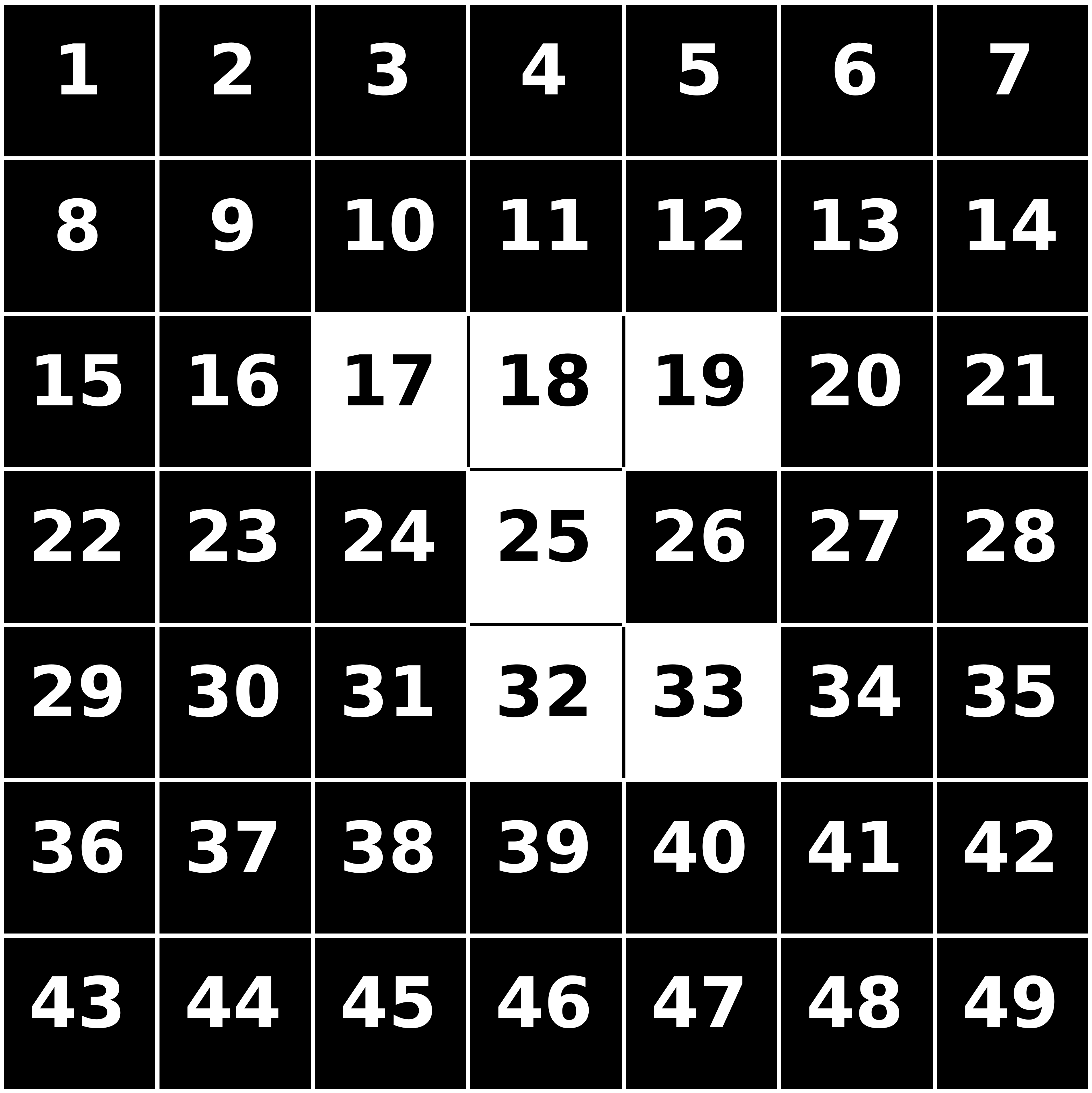}
\caption{Illustration of the \textit{friend-of-friend} algorithm (see Section
  \ref{fof} for details).
White and black cells correspond to ionized and neutral cells respectively.}
\label{identification_algorithm}
\end{figure}

Our FOF algorithm proceeds by scanning all cells from the box and testing 
at each iteration the ionization status of the cell being explored.
The algorithm simply proceeds as follow:
\begin {itemize}
\item{If an ionized cell is encountered, it is given an ID corresponding to the 
ionized region being tested and the exploration of its neighbors begins~:}
\begin{enumerate}
\item{If a neighbor is not ionized, the cell is marked as visited. The parent
  cell is then reconsidered and the ionization status of the other near
  neighbors not already marked is tested.}
\item{If the neighbor is ionized, it is given an ID corresponding to the
  ionized region being tested and we mark the cell as visited. The exploration
  of its own neighbors begins. The HII regions IDs diffuse from near neighbor to near neighbor.}
\end{enumerate}
\item{ When all the neighbors of an ionized cell have been explored, we go
  back to its parent cell and continue the exploration of its near neighbor
  which were not previously visited. The identification stops when the
  starting cell is visited again and all its near neighbors are marked as visited.}
\end {itemize}    
  
The exploration of the cosmological box proceeds until another starting
ionized cell is found and which is not already marked as visited. At this
moment, a new identification starts with a new identification number.

In Figure \ref{identification_algorithm}, an illustration of the algorithm in 2 dimensions is presented.
In this figure, black cells correspond to neutral cell and
white cells to ionized cells with a ionized fraction $ x \ge 0.5 $.
In this example, 6 cells belongs to the same HII region, with numbers 17, 18, 25, 32 and 33
and all of them should receive the same identification number. Following the
algorithm described above, the cells would be visited in the following order :
17-18-19-18-25-32-33-32-25-18-17. At this stage, the algorithm recover the
starting cell with all its ionized neighbors marked as visited, it stops.

To implement this method, a recursive scheme could be used as well as linked
lists. We chose the second option as it limits the use of large stacks of memory.
Let us also mention that for boxes of $1024^{3}$, we have parallelized
the identification on the large box by cutting it in 16 sub-boxes. 
We thus do the identification on this 16 sub-boxes in parallel to accelerate 
the calculation. We reassemble the entire box after and merge the IDs.

\subsection{Effect of the neutral ionization threshold}
\label{studfof}
\begin{figure*}
   \begin{center}
    \begin{tabular}{ccc}
      Boosted Star & Star & Halo \\
      \includegraphics[width=6cm,height=5cm]{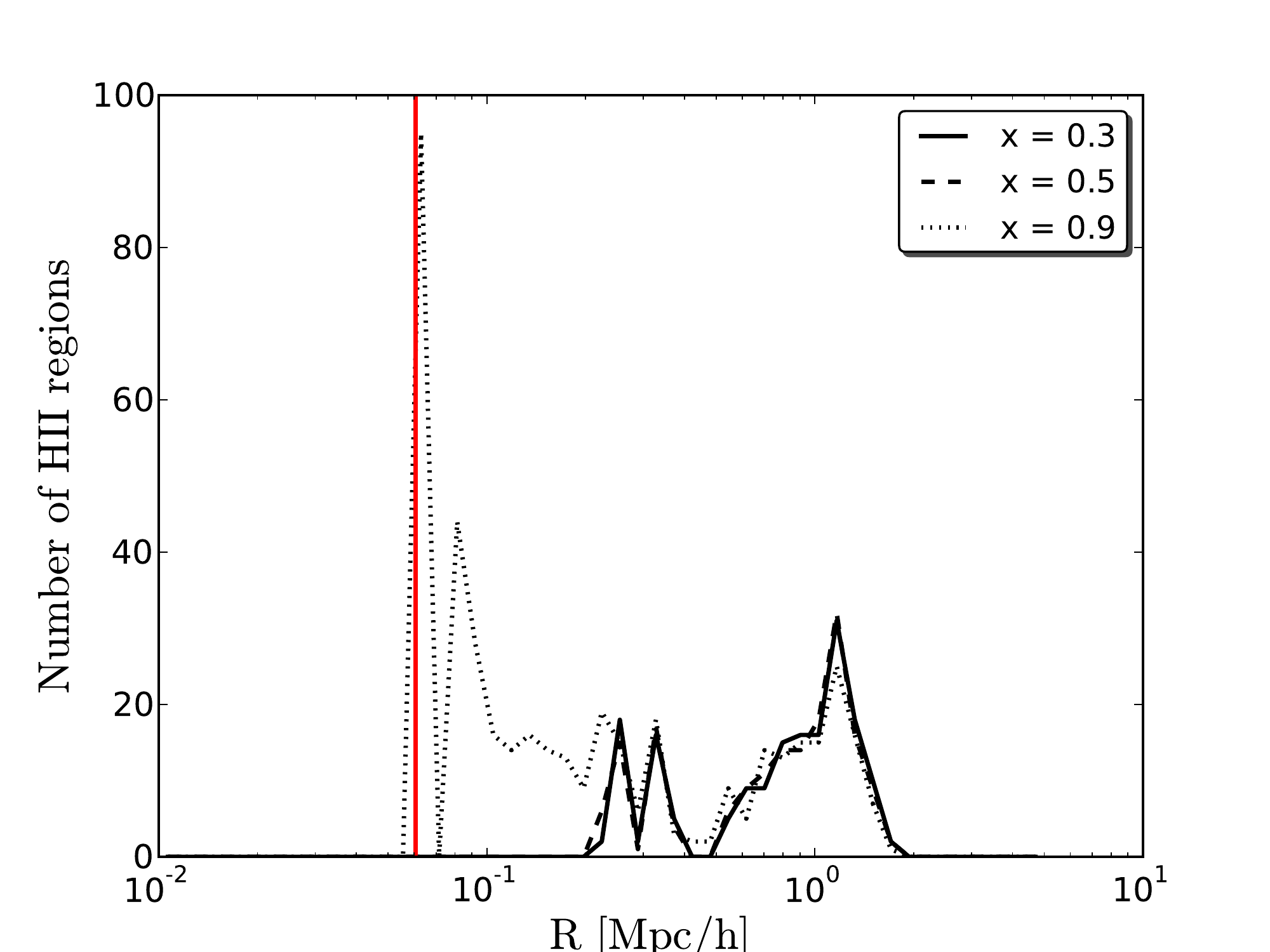} &
	\includegraphics[width=6cm,height=5cm]{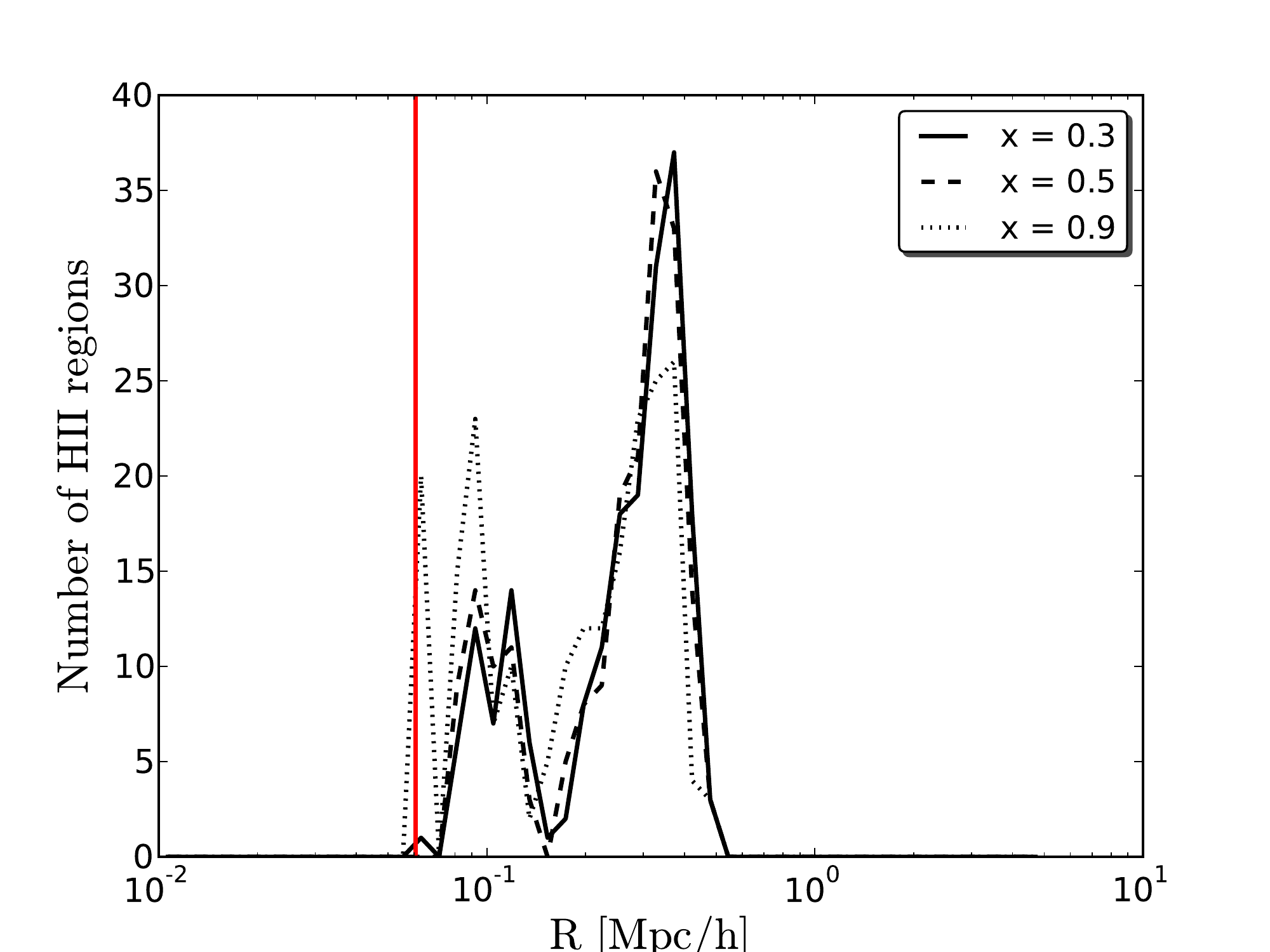} &
	\includegraphics[width=6cm,height=5cm]{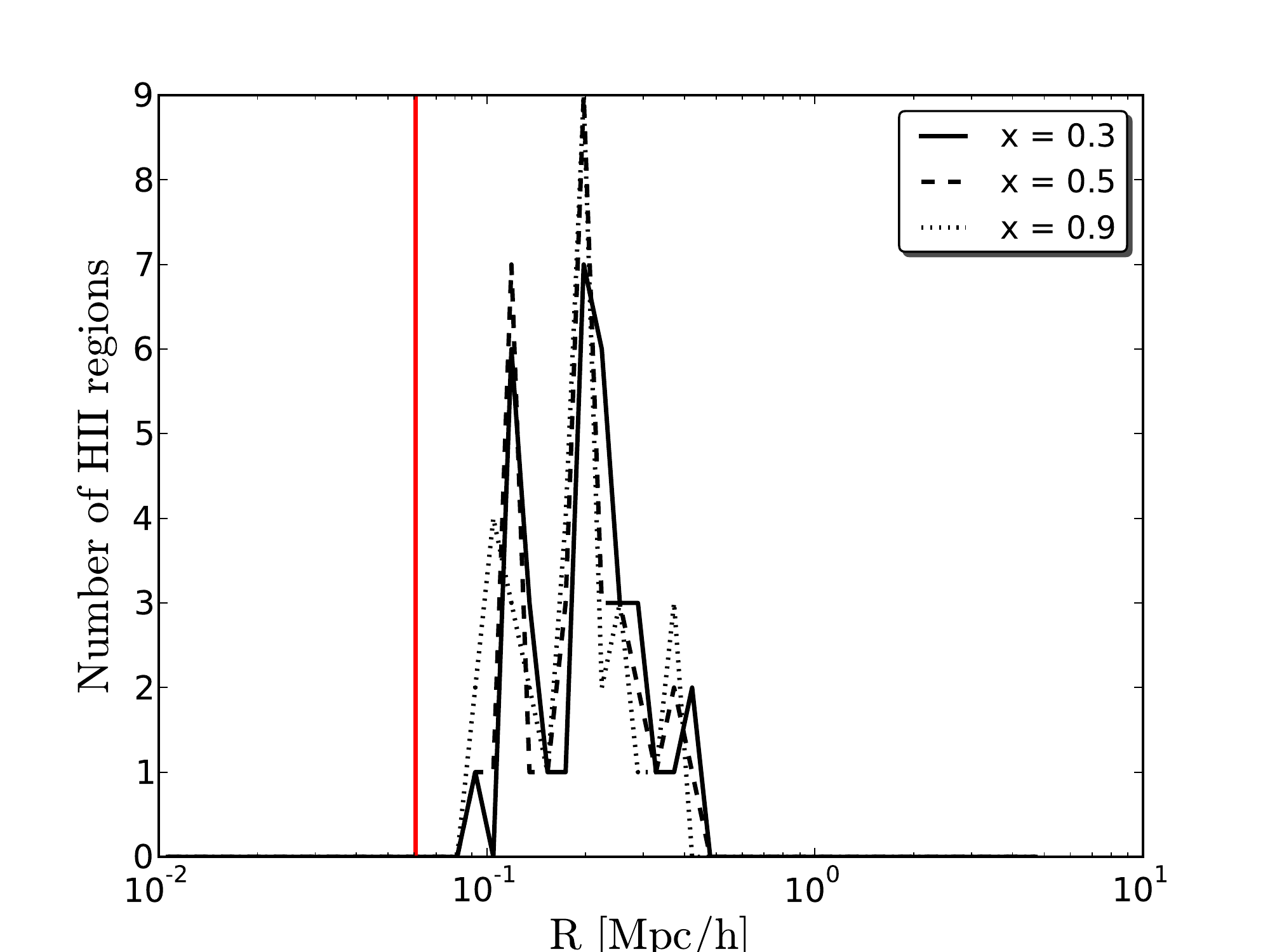} \\
     $\mathrm{z = 14.0}$ & $\mathrm{z = 14.0}$ & $\mathrm{z = 14.0}$ \\
	\includegraphics[width=6cm,height=5cm]{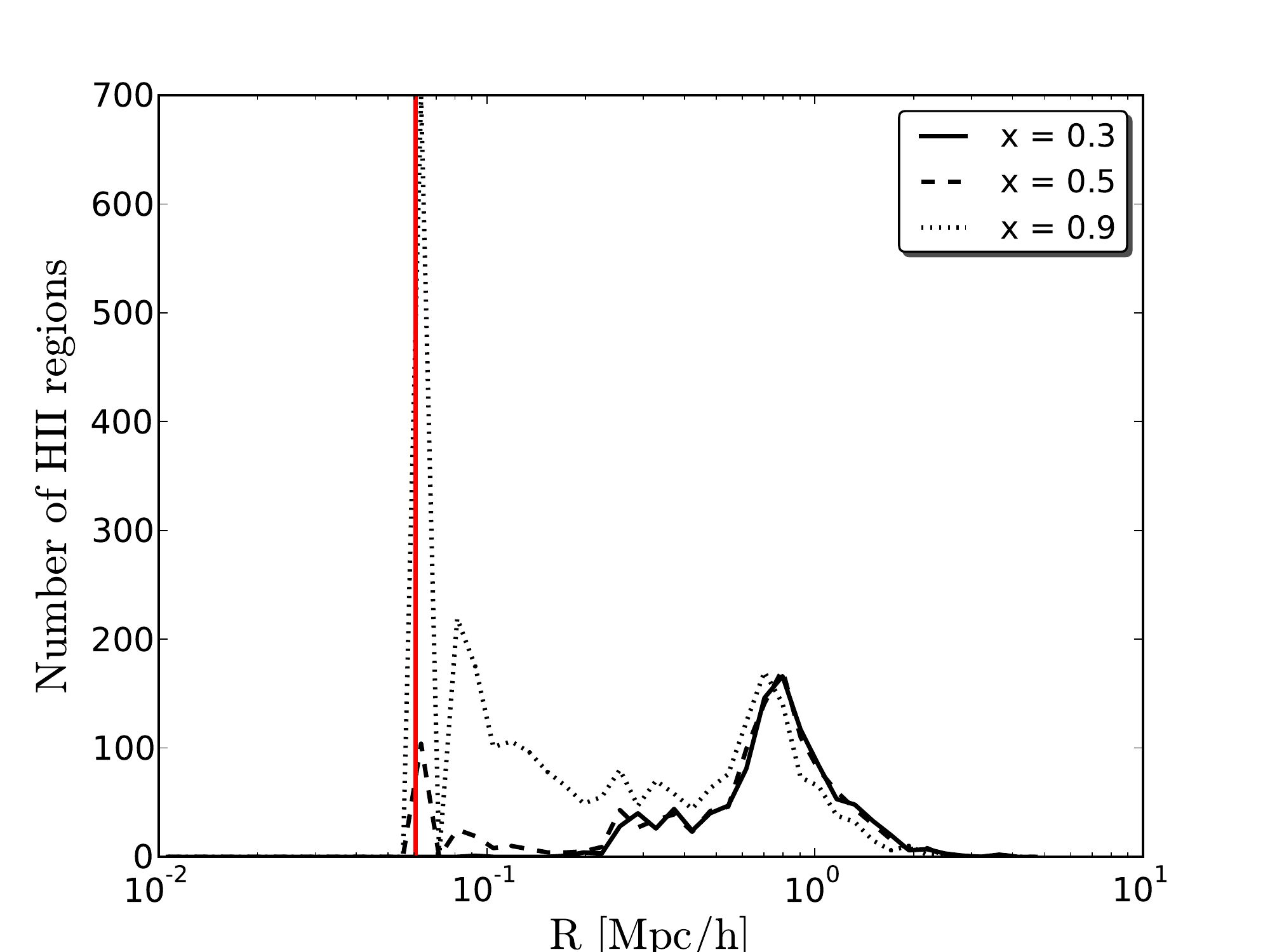} &
	\includegraphics[width=6cm,height=5cm]{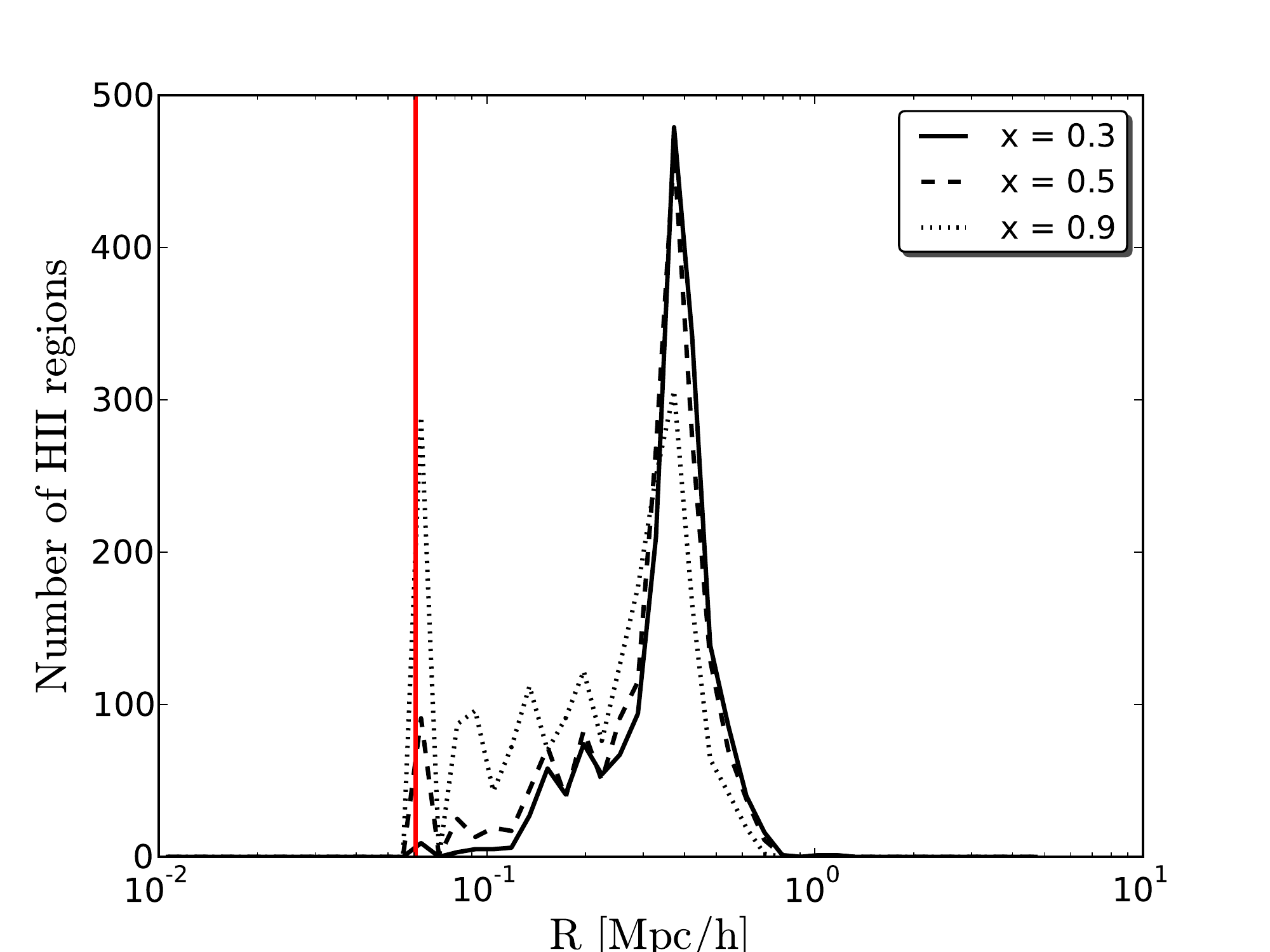} &
	\includegraphics[width=6cm,height=5cm]{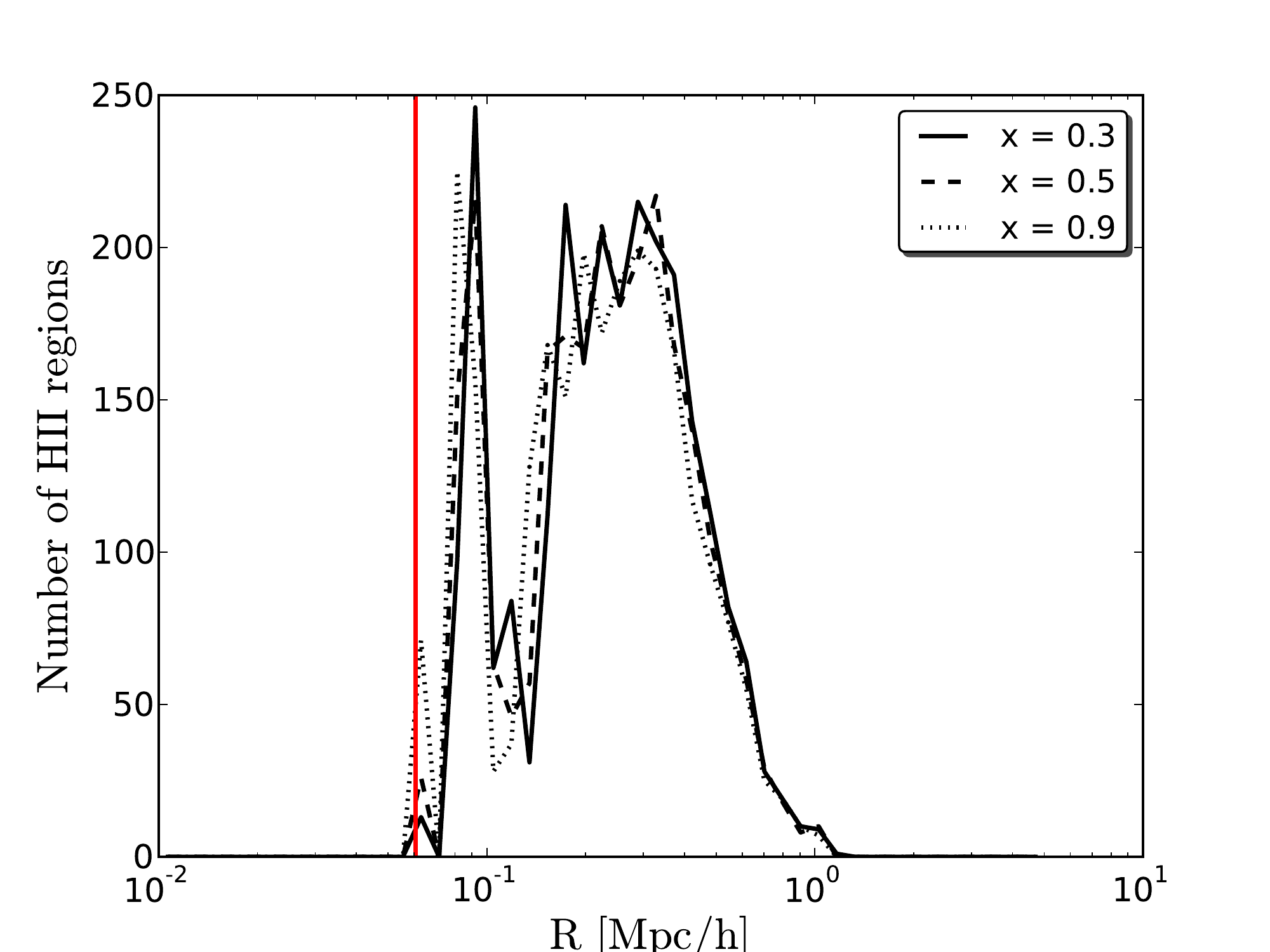} \\
      $\mathrm{z = 11.4}$ & $\mathrm{z = 11.4}$ & $\mathrm{z = 11.4}$ \\
	  \includegraphics[width=6cm,height=5cm]{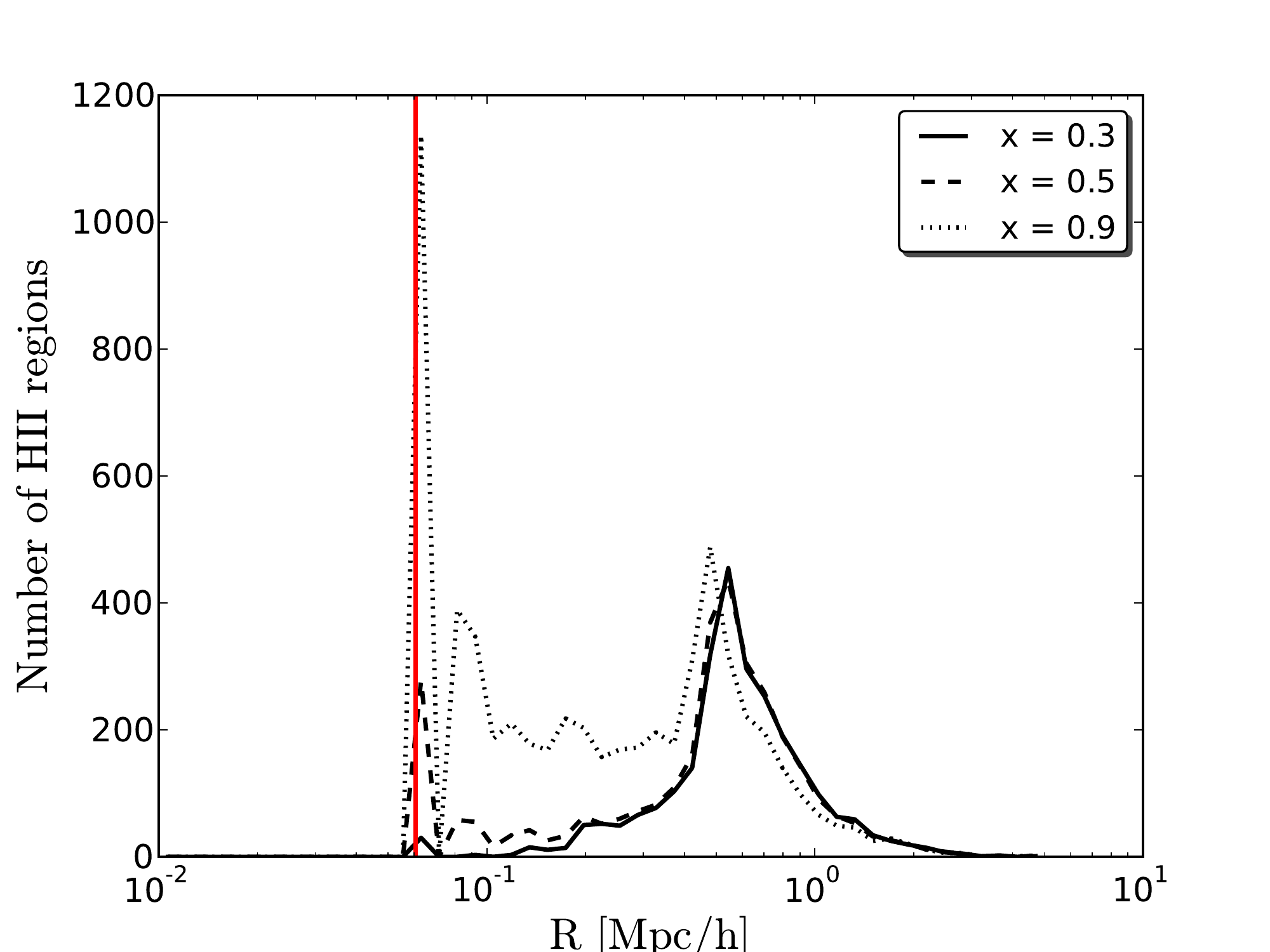} &
	  \includegraphics[width=6cm,height=5cm]{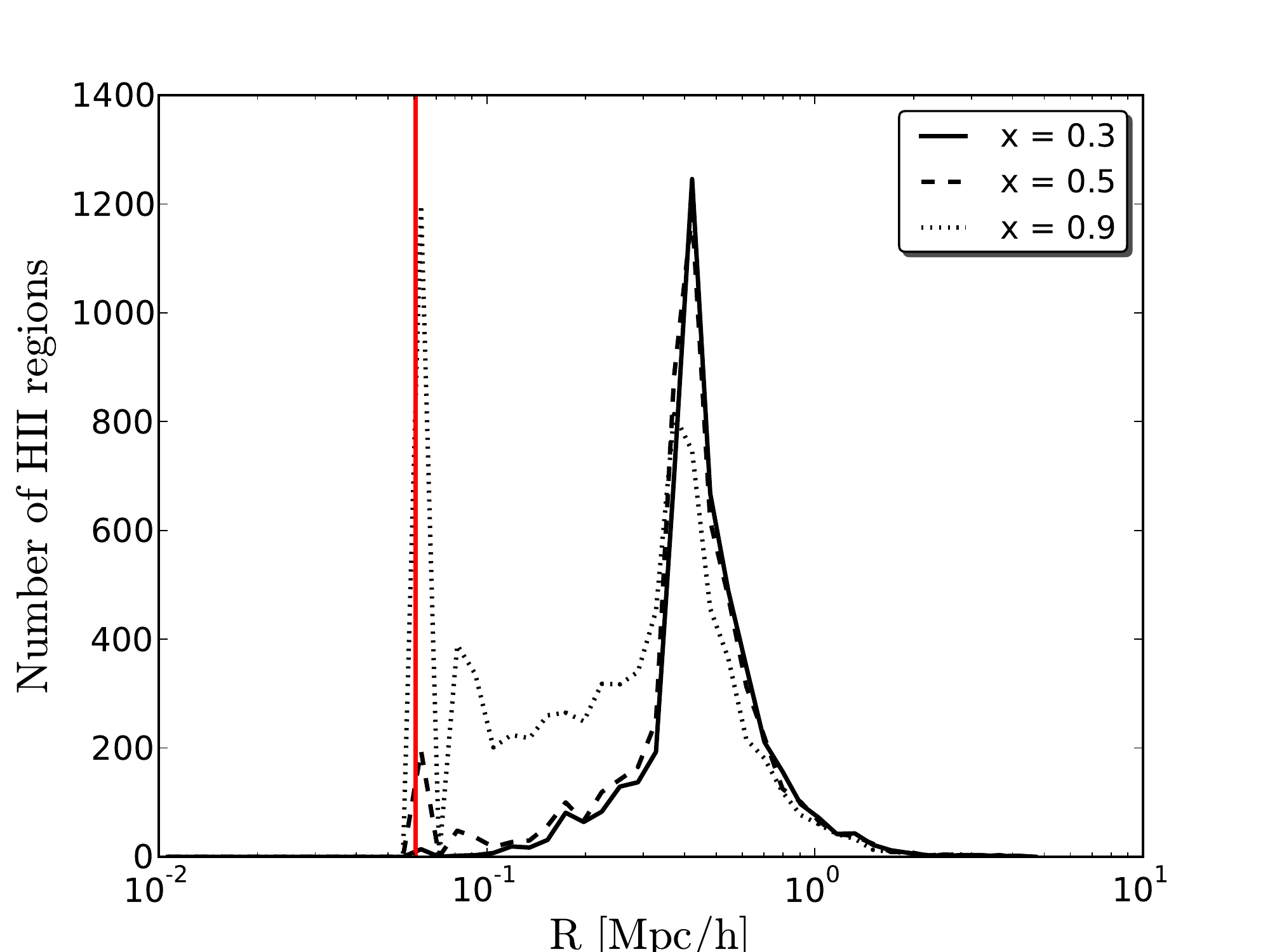} &
	  \includegraphics[width=6cm,height=5cm]{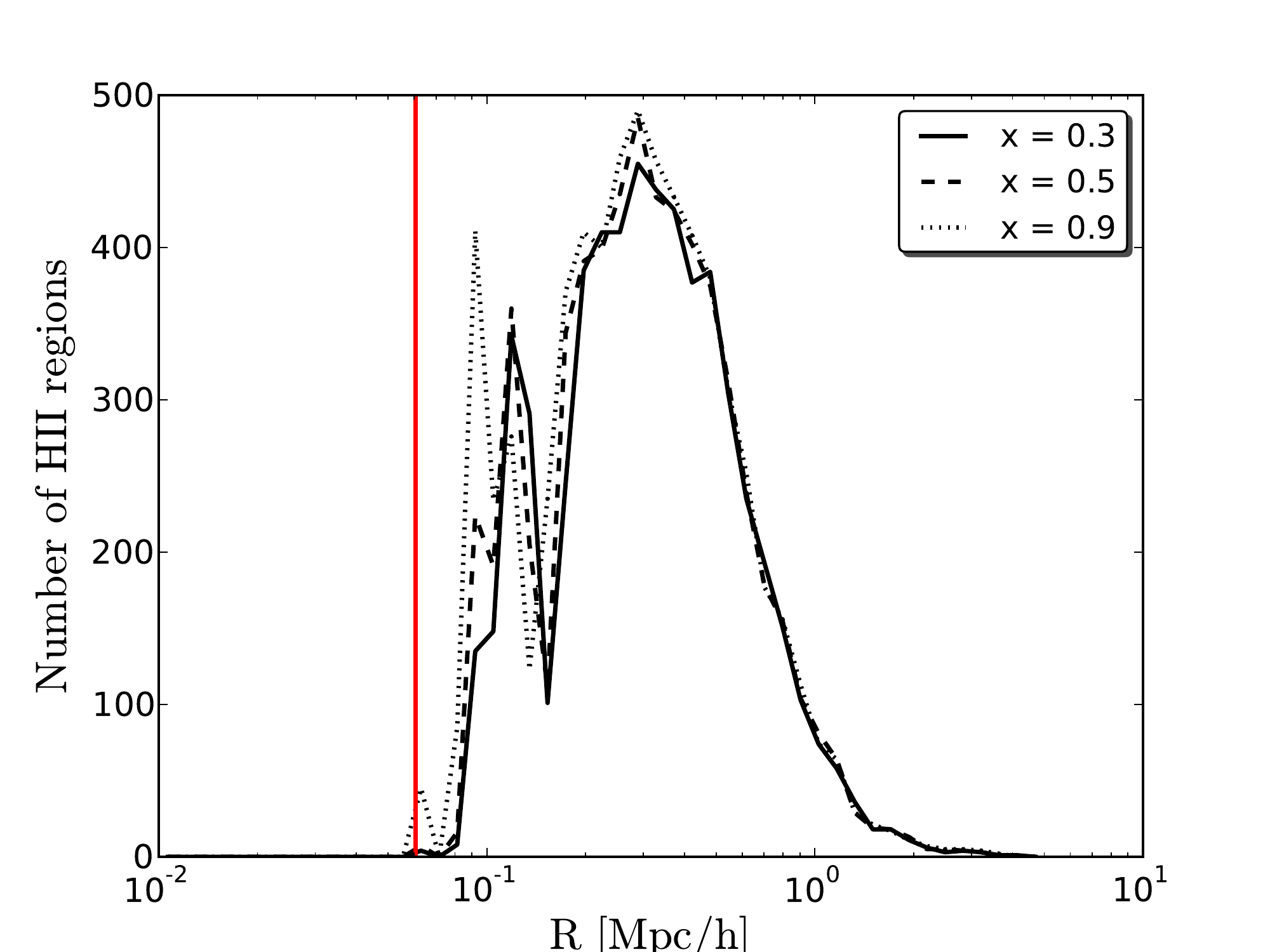} \\
     $\mathrm{z = 9.0}$ & $\mathrm{z = 9.0}$ & $\mathrm{z = 9.0}$ \\
\end{tabular}    
  \caption{Representation of the radius distribution of all HII regions at three different redshifts for the three simulations for the 50 Mpc/h box. 
We compare the difference in the radius distribution according to the ionization threshold for the detection of the HII region with the FOF procedure.
We represent the volume of one single cell with the vertical red line.}
    \label{variation_neutral_fraction}
  \end{center}
 \end{figure*}

\begin{figure*}
   \begin{center}
    \begin{tabular}{ccc}
      Boosted Star & Star & Halo \\
      \includegraphics[width=6cm,height=5cm]{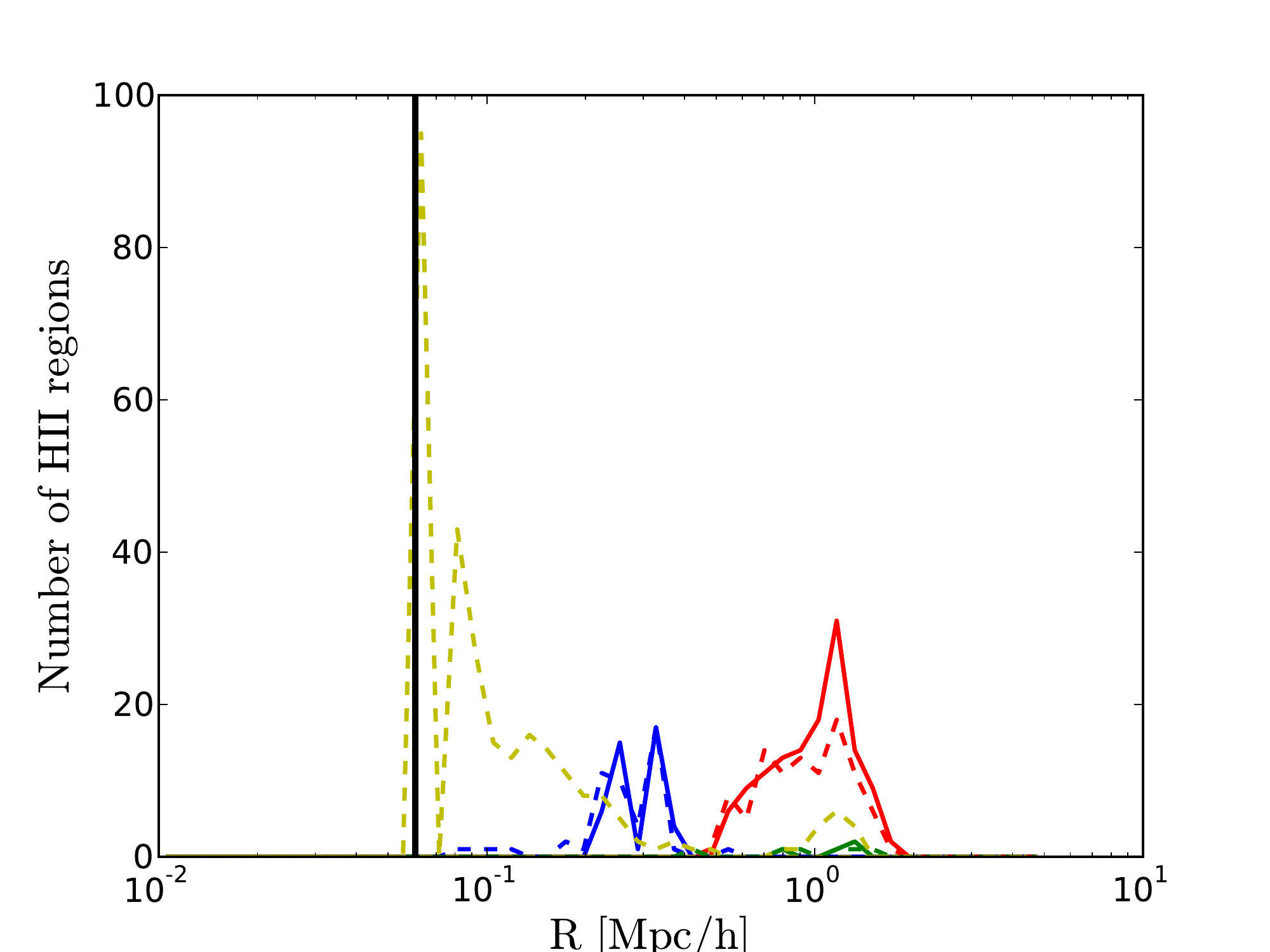} &
	\includegraphics[width=6cm,height=5cm]{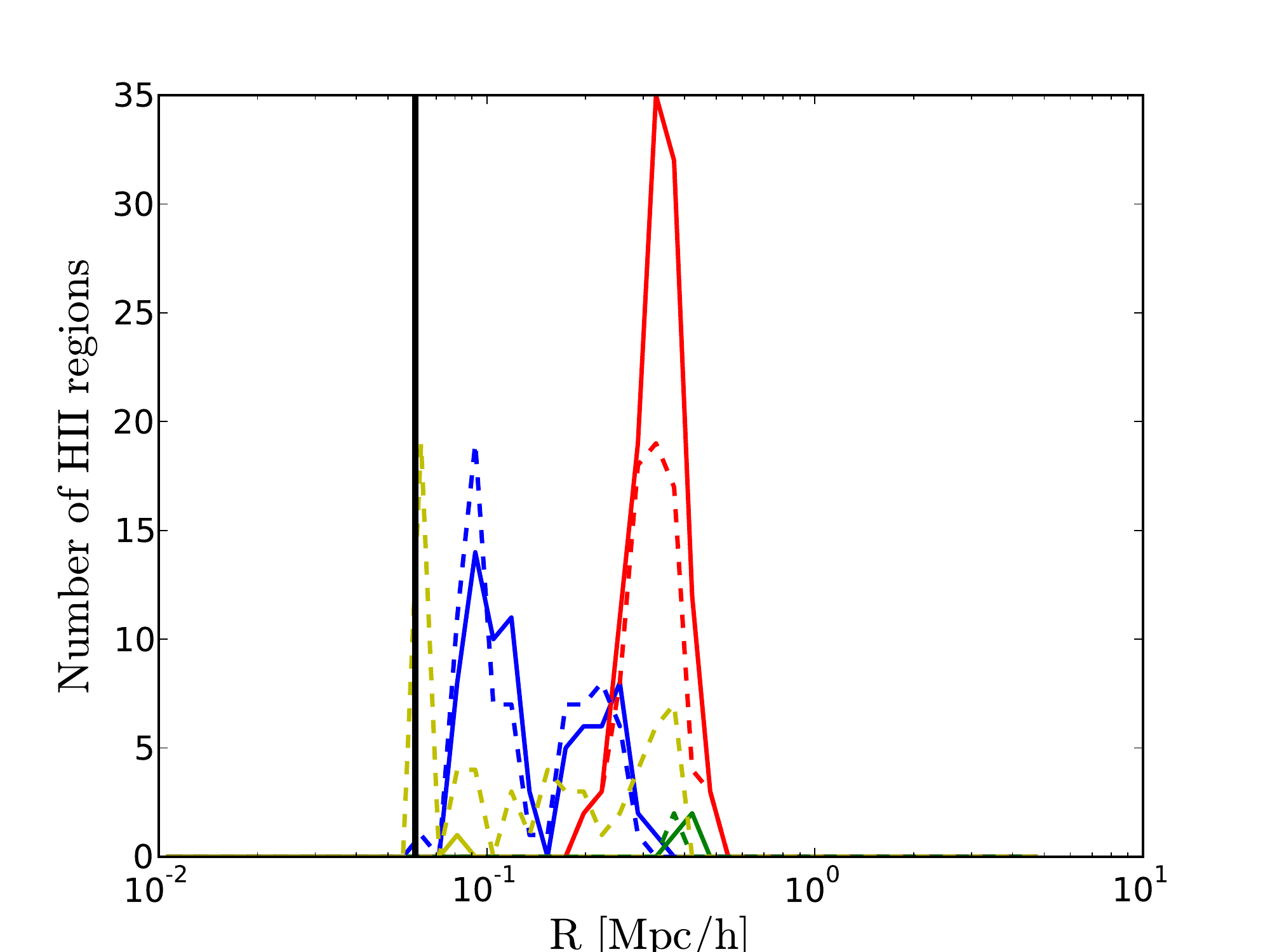} &
	\includegraphics[width=6cm,height=5cm]{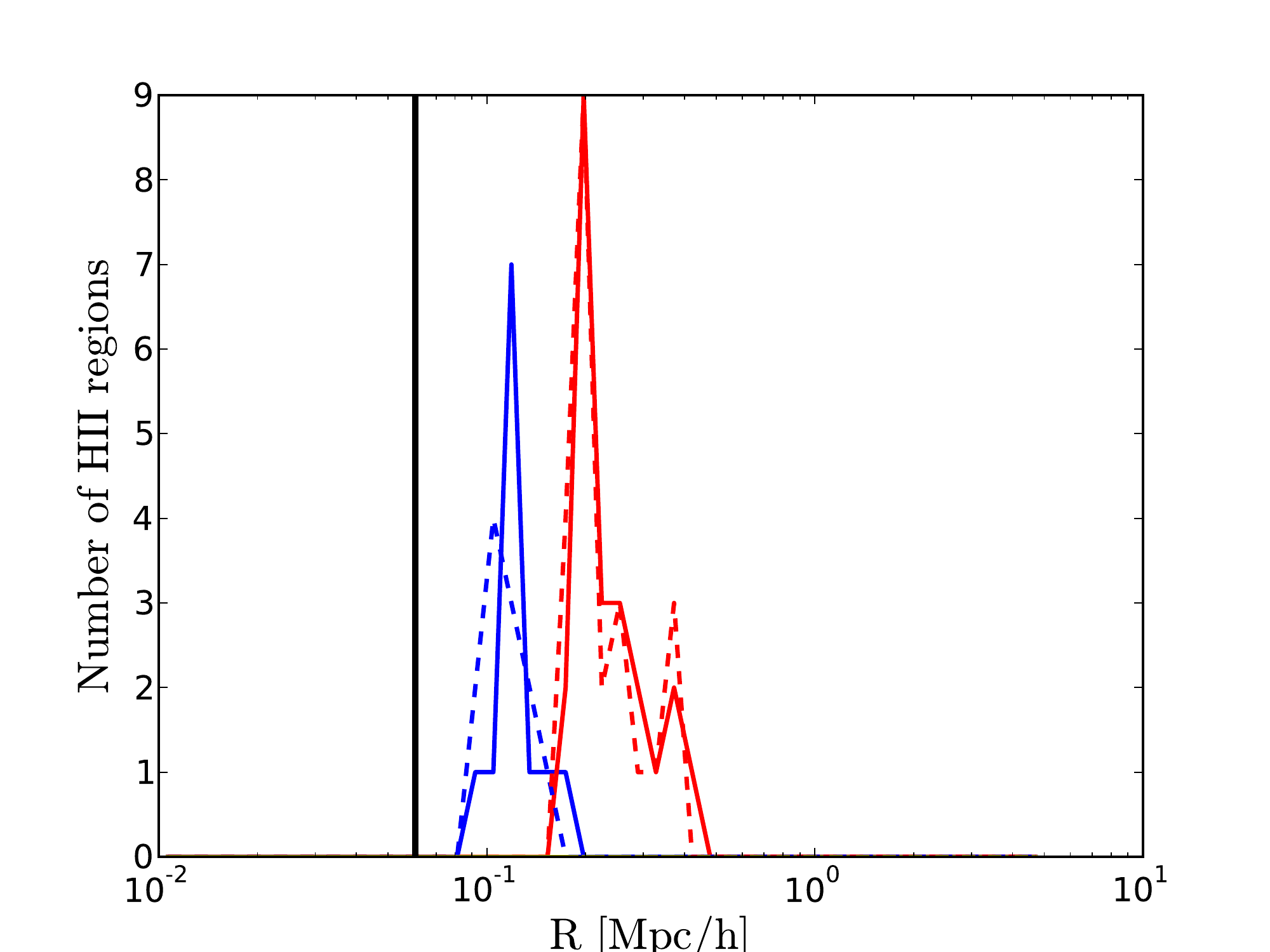} \\
     $\mathrm{z = 14.0}$ & $\mathrm{z = 14.0}$ & $\mathrm{z = 14.0}$ \\
	\includegraphics[width=6cm,height=5cm]{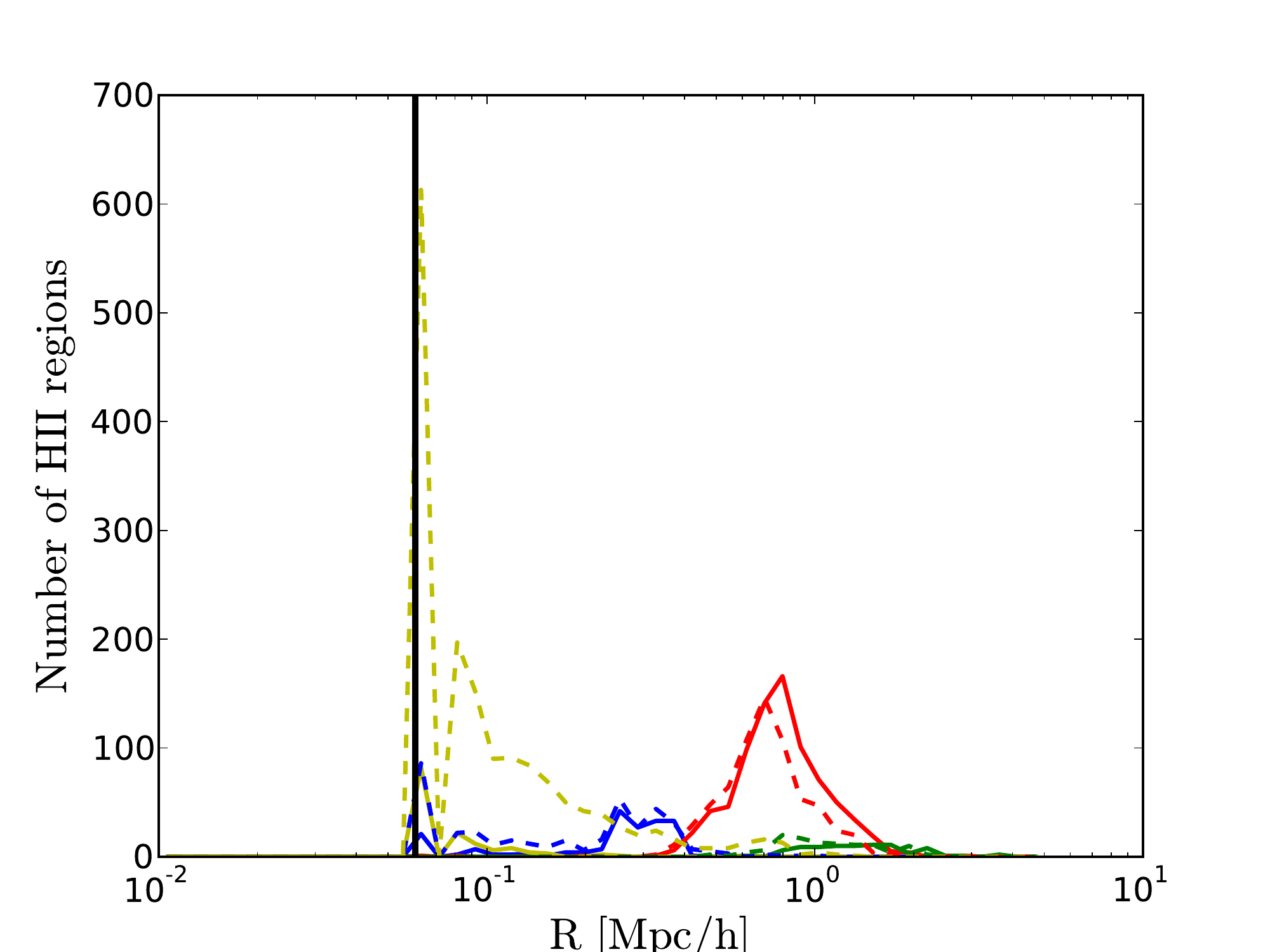} &
	\includegraphics[width=6cm,height=5cm]{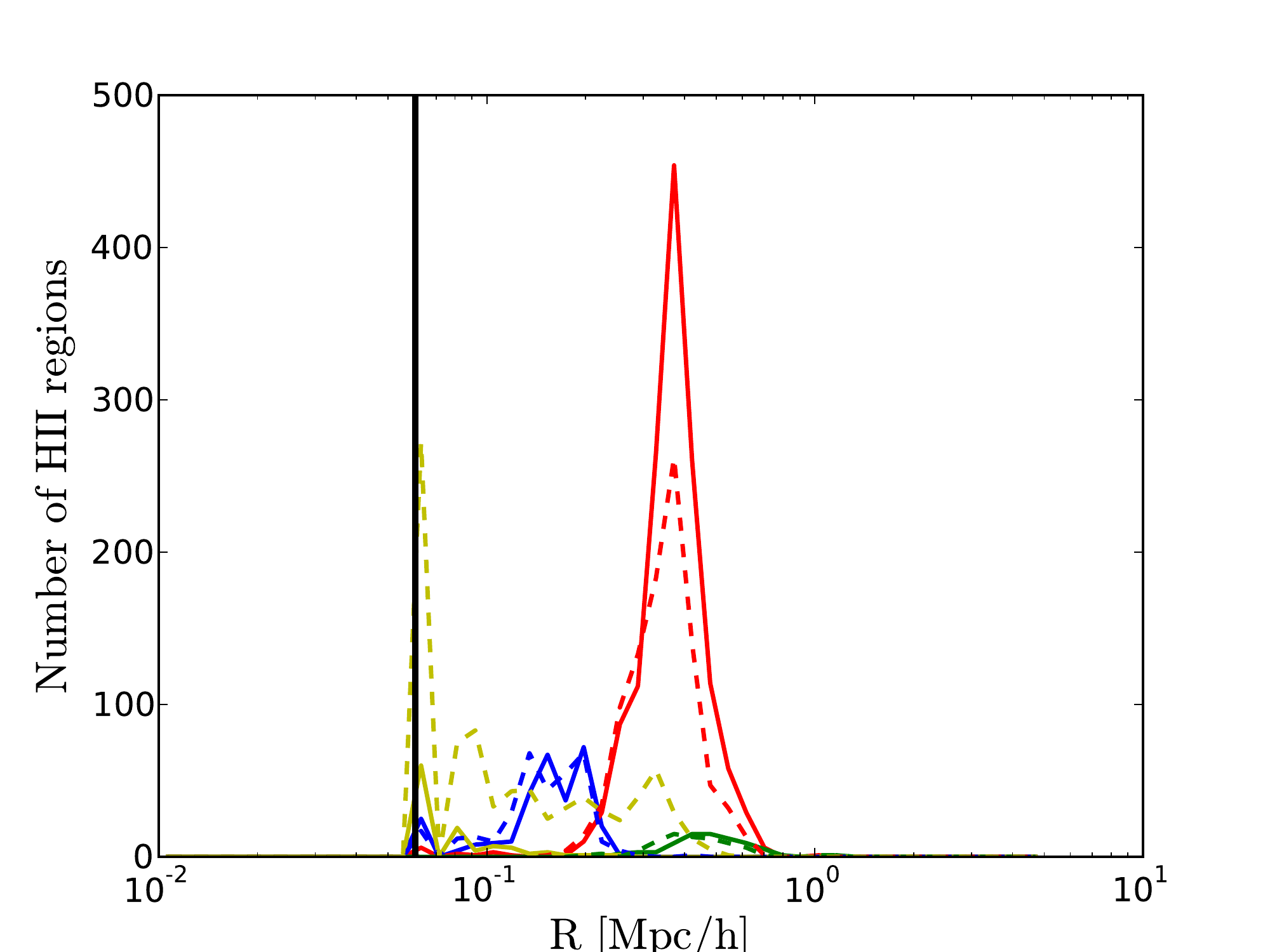} &
	\includegraphics[width=6cm,height=5cm]{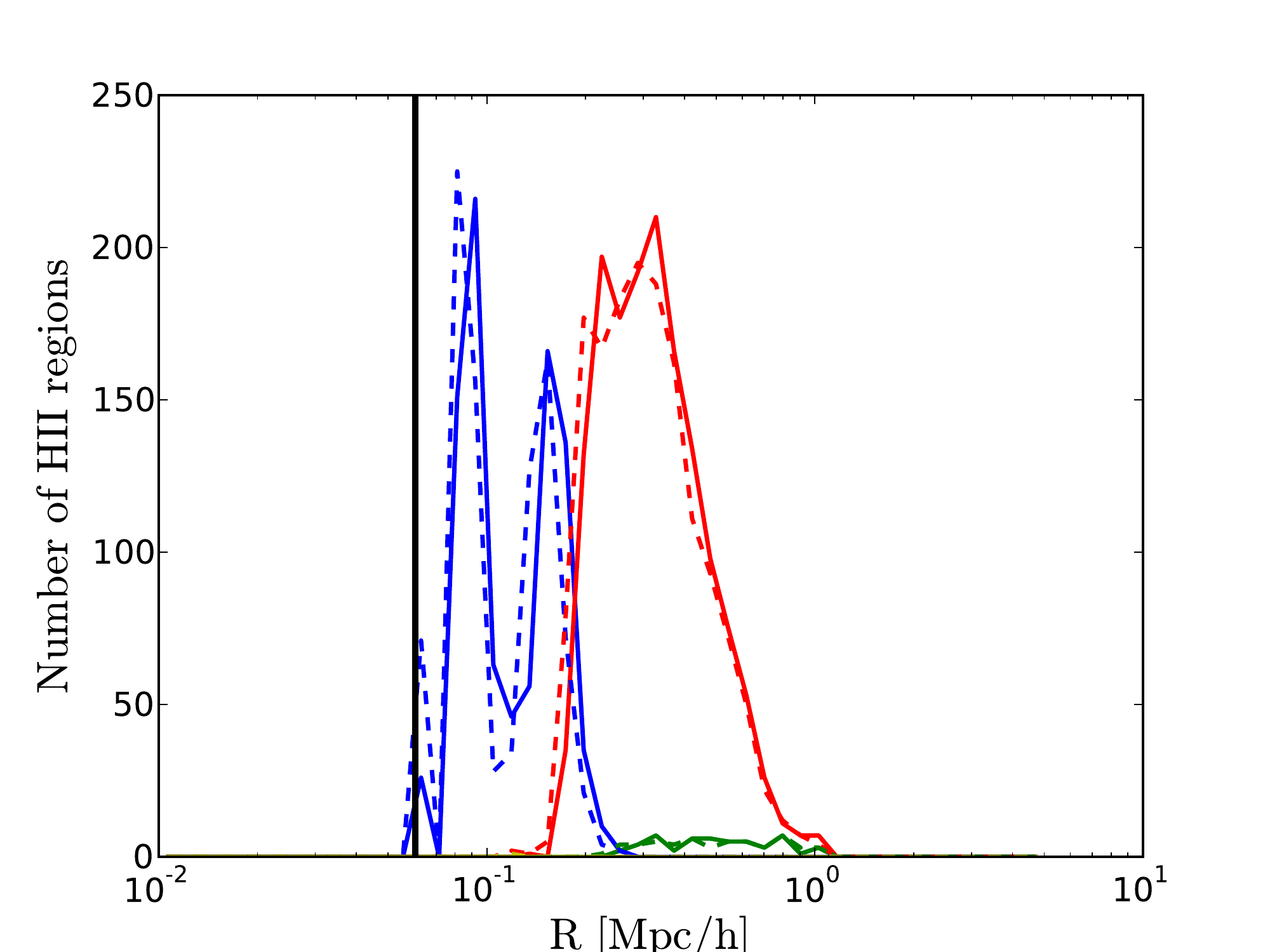} \\
      $\mathrm{z = 11.4}$ & $\mathrm{z = 11.4}$ & $\mathrm{z = 11.4}$ \\
	  \includegraphics[width=6cm,height=5cm]{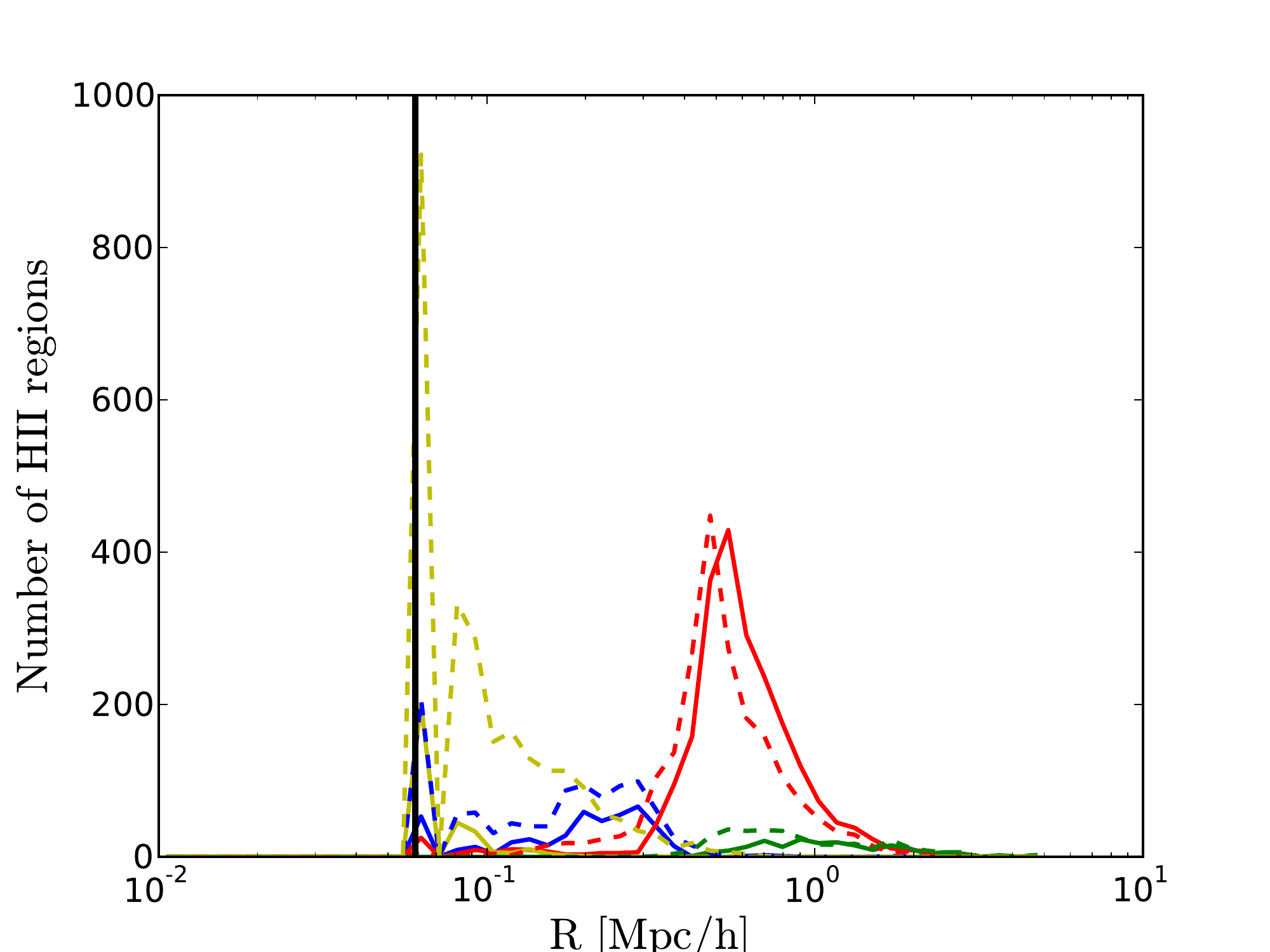} &
	  \includegraphics[width=6cm,height=5cm]{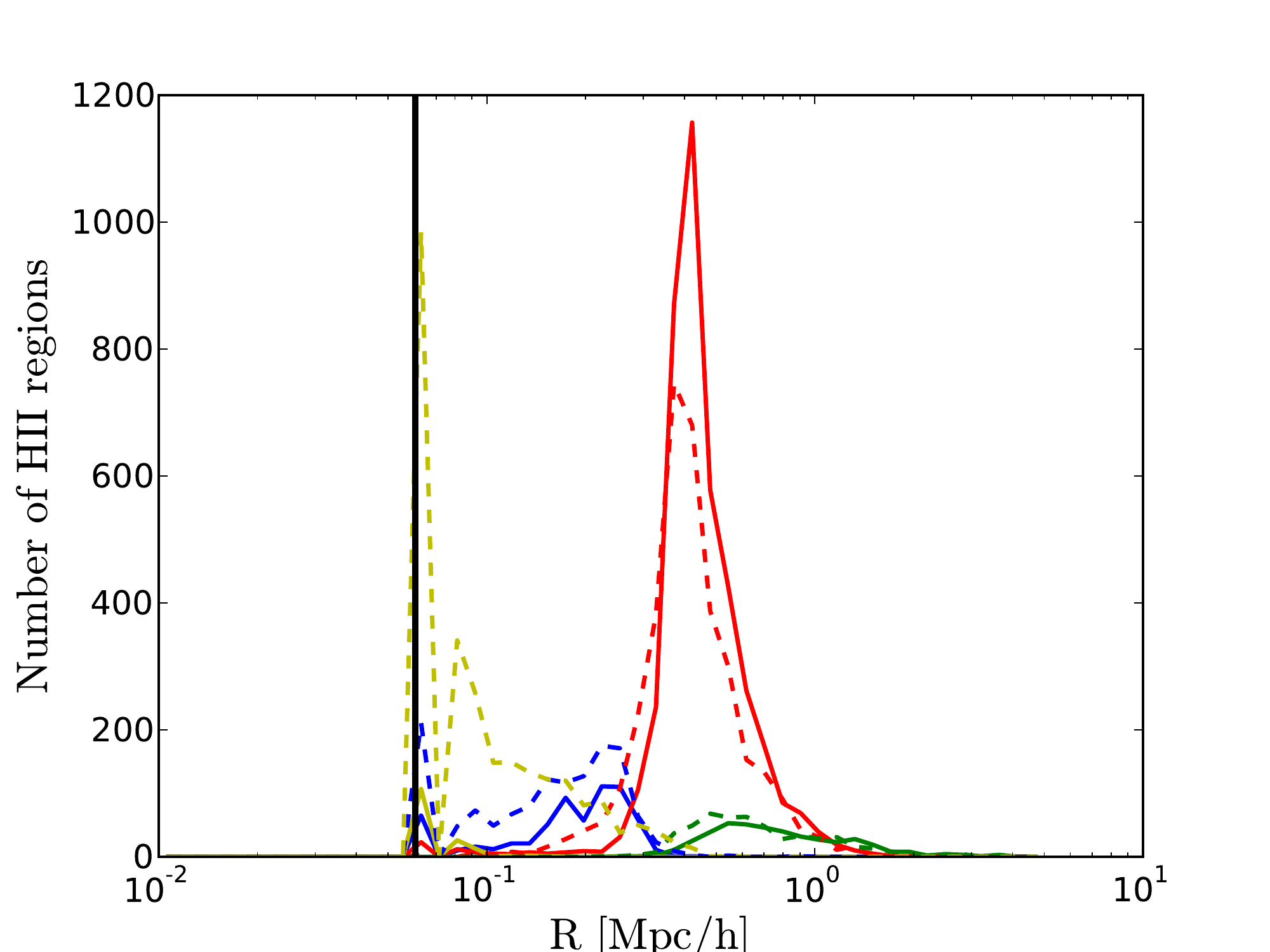} &
	  \includegraphics[width=6cm,height=5cm]{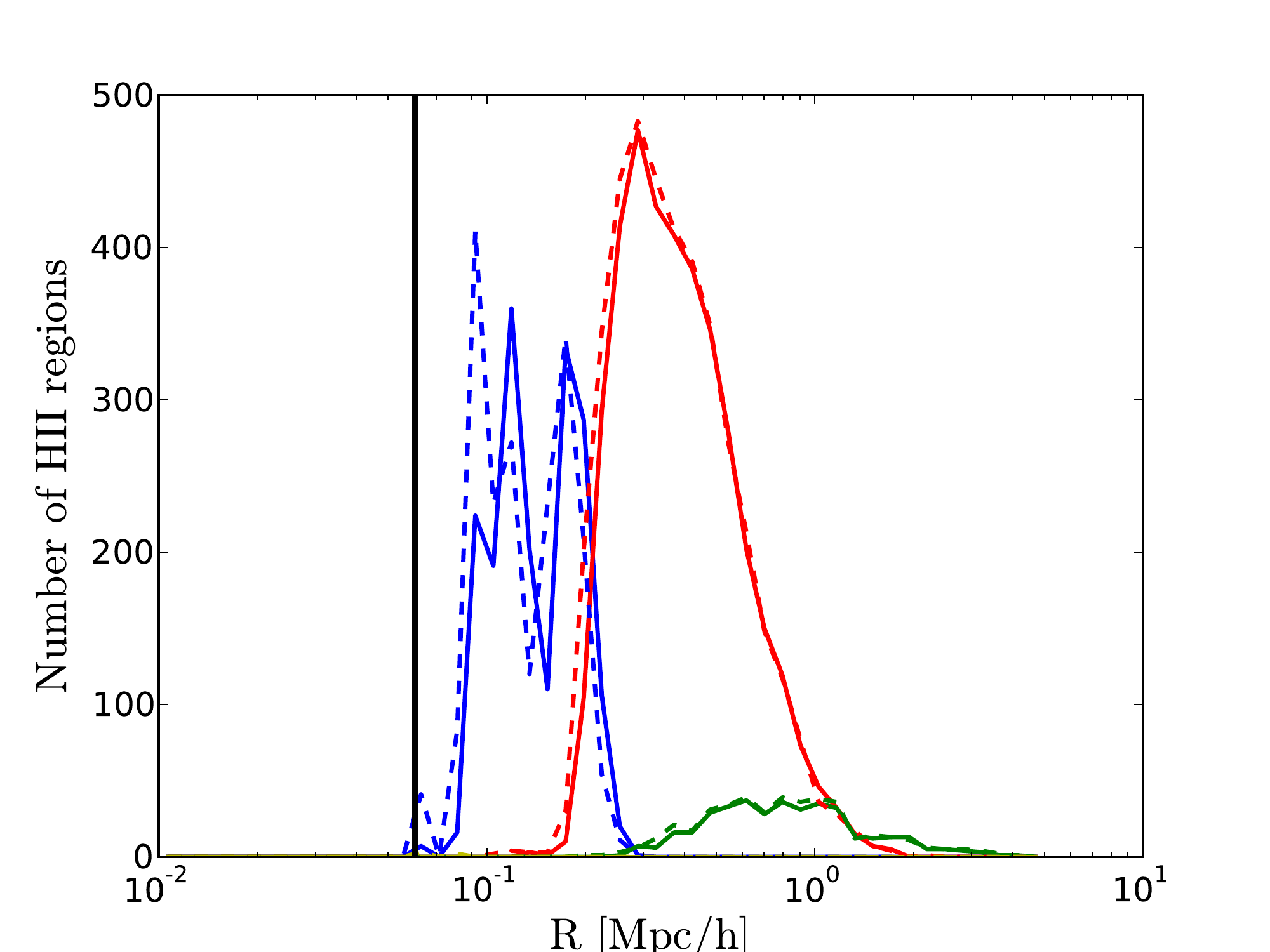} \\
     $\mathrm{z = 9.0}$ & $\mathrm{z = 9.0}$ & $\mathrm{z = 9.0}$ \\
\end{tabular}    
  \caption{Radius distribution of all HII regions at three different redshifts for the three simulations for the 50 Mpc/h box.
This time, we dissociate the distribution in the different counterpart of the type of HII region considered.
We compare the difference in the radius distribution according to both ionization threshold of $ x \ge 0.5 $ and $ x \ge 0.9 $ for the detection of the HII region with the FOF procedure.
The color codes are as follows: Blue: the new regions, Red: the expanding regions, Green: the regions resulting from merger and Yellow: the regions that will recombine. 
In addition we represent the volume of one single cell with the vertical black line.}
    \label{variation_neutral_fraction_differente_region}
  \end{center}
 \end{figure*}

The FOF algorithm assumes an ionization threshold in order to decide if a cell is ionized or not.
As often used in the literature, we have chosen a value of $ x \ge 0.5 $ to consider a cell as ionized.
In order to test the impact of the threshold's value for the identification of
the HII regions we have also tried other values for the ionization fraction in the FOF procedure.

The figure \ref{variation_neutral_fraction} represents the radius distribution of the HII regions for three different instants.
The distributions are done for the three model of ionizing sources for the 50
Mpc/h box size. We compare the radius distributions according to three different ionizing thresholds of $ x \ge 0.3 $, $ x \ge 0.5 $ and $ x \ge 0.9 $ to assume that a cell is ionized.
The three distributions are very similar for
all models and at each instant regardless of the chosen ionizing
threshold. The most significant variation stands for small regions with radius below
$\sim 2\times10^{-1}$ Mpc/h that are detected in greater number with $ x \ge 0.9 $.
Meanwhile, the distributions with $ x \ge 0.3 $ and $ x \ge 0.5 $ are almost superimposed in every cases.
Additionally, we used the merger tree properties to plot the same distributions 
in figure \ref{variation_neutral_fraction_differente_region} but dissociating the contributions of each type of regions.
Thus new regions are in blue, expanding regions in red, regions resulting from
mergers in green and the regions that will recombine in yellow. This time, we
have just represented the distribution for both  $ x \ge 0.5 $ and $ x \ge 0.9
$ ionization threshold, that previously exhibited the greater difference.
We see that globally that the families remain unchanged and therefore that the
merger tree properties are conserved even when varying the ionization threshold. The only difference being that 
with $ x \ge 0.9 $ the FOF procedure tends to detect more small recombining
regions  than with $ x \ge 0.5 $ in the boosted star and star models. 

Therefore the only remarkable effects by varying the threshold is for high
threshold values within recombining regions, that represent a tiny number and
volume fraction of the HII regions network. It does not come as a surprise
though, since recombination occur in already ionized regions and any
fluctuations within, combined with a high detection threshold lead to a
subdivision of regions in smaller ones. One might even consider that such
effect should be avoided since it increases branches in the merging tree with
regions that may be short lived or unstable. As a conclusion, we can thus
consider that the value of $x\ge 0.5$ used in the paper for the identification of the HII regions is a good compromise, in addition of being the one commonly used in the literature.

\section{Merger tree algorithm}
\label{merger}

\begin{figure*}
\hskip 4.5truecm
\includegraphics[width=10cm]{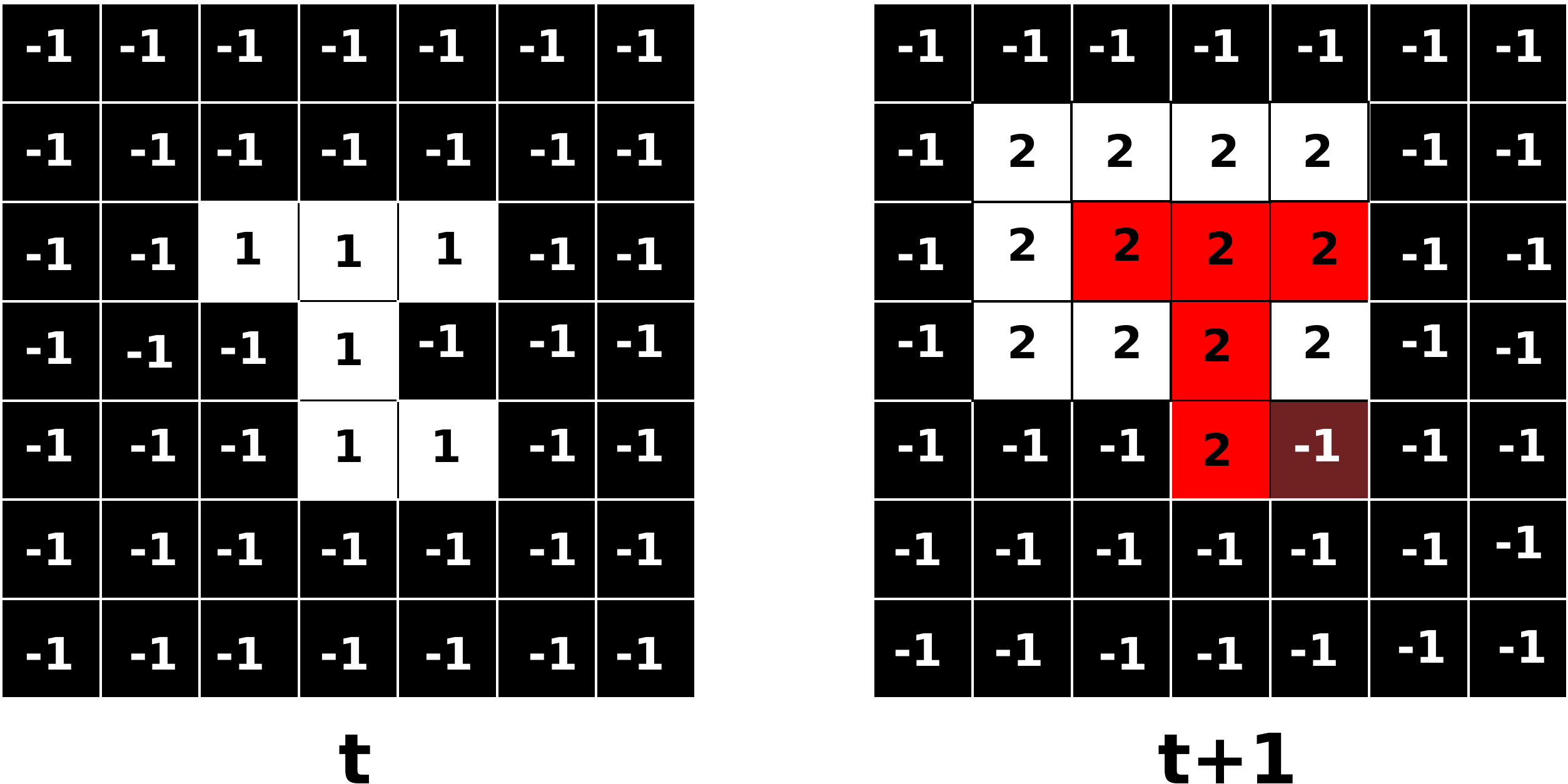}
\caption{Merger tree algorithm. All neutral cells are black with an identification number of -1.
All ionized cells are white with an identification number. At time t, we have the ionized region identified with an identification number of 1.
We look forward at time t+1 where cells of this ionized region are located
(shown in red). Here ID 1 at time t is linked to ID 2 at time t+1.}
\label{illustration_merger_tree}
\end{figure*}

The implementation of the merger tree algorithm is performed as follow (see also Figure \ref{illustration_merger_tree}):
\begin{enumerate}
 \item{We look at an instant $t$ of the simulation where are located the cells 
  of a particular ionized region. They all share the same ID.} 
 \item{Then, we extract the same cells at time $t+1$. These cells may have
   different IDs.}
 \item{The most common ID at time $t+1$ is kept and is linked to the ID
   investigated at $t$.}
\end{enumerate}
This scheme is repeated for all HII regions between the snapshot $t$ and
$t+1$. We finally reproduce this process between each snapshots.

\end{document}